\documentclass[aip,jmp,amsmath,amssymb,reprint,numerical]{revtex4-1}
\usepackage{graphicx}
\graphicspath{{Images/}}
\usepackage{dcolumn}
\usepackage{bm}
\usepackage[table]{xcolor}
\usepackage[utf8]{inputenc}
\usepackage[T1]{fontenc}
\usepackage{mathptmx}
\usepackage{etoolbox}
\usepackage{makecell}
\usepackage{enumitem}
\usepackage{textcomp}
\usepackage{multirow}
\usepackage{lineno}

\makeatletter
\def\@email#1#2{%
 \endgroup
 \patchcmd{\titleblock@produce}
  {\frontmatter@RRAPformat}
  {\frontmatter@RRAPformat{\produce@RRAP{*#1\href{mailto:#2}{#2}}}\frontmatter@RRAPformat}
  {}{}
}
\makeatother
\begin{document}

\preprint{AIP/123-QED}

\title[Numerical Investigation of the Effect of an Oblique Flow Entry on the Pressure Losses in Square Channels]{Numerical Investigation of the Effect of an Oblique Flow Entry on the \\ Pressure Losses in Square Channels}

\author{C. Samuels}
    \email{samuelsc2@uni.coventry.ac.uk}
    \affiliation{Fluids \& Complex Systems, Coventry University, United Kingdom}
  
\author{T. C. Watling}
\affiliation{Johnson Matthey plc, Reading, United Kingdom}

\author{S. Aleksandrova}
\affiliation{School of Engineering, University of Leicester, United Kingdom}

\author{H. Medina}
\affiliation{Faculty of Engineering, University of Nottingham, United Kingdom}

\author{R. Holtzman}
\affiliation{Fluids \& Complex Systems, Coventry University, United Kingdom}

\author{I. Rusli}
\affiliation{Fluids \& Complex Systems, Coventry University, United Kingdom}

\author{S. Benjamin}
\affiliation{Faculty of Engineering, Environment \& Computing, Coventry University, United Kingdom}

\date{\today} 

\begin{abstract}
Flows in square channels are common in applications, such as automotive after-treatment systems and heat exchangers. Flows with axial flow entry are well understood, but for oblique flow entry, there is no clarity on the additional pressure loss magnitude or the flow regime. Laminar flow is often assumed, even though flow separation at the channel entrance can cause a transition to turbulence. Here, the impact of oblique flow entry on the flow is investigated using LES (Large Eddy Simulation) and RANS (Reynolds Averaged Navier Stokes) models, and their advantages and limitations are identified. The LES simulations show that the shear layer at the channel entrance produces continuous shedding of eddies that persist downstream even at moderate channel Reynolds numbers ($\approx$ 2000). The predicted pressure losses mostly agree with experimental data. The differences observed for some parameters are attributed to the difficulty of accurately replicating the experimental geometry. It is shown that LES results are susceptible to the rounding of the leading edge (present in experiments). Including edge rounding improves the pressure predictions. RANS simulations predicted pressure losses within 5\% of experimental values for most cases, apart from where transitional flow was observed (resulting in differences up to 40\%). This study provides insights into the flow structure and sources of pressure losses in square channels and highlights the importance of understanding key flow and geometric features when using LES to predict complex flows involving flow separation and shear layers.
\end{abstract}

\maketitle
%\linenumbers
\section{\label{sec:intro}Introduction}

Flow dynamics within channels and pipes is a fundamental aspect of fluid mechanics, with applications ranging from industrial processes to biological systems. Devices such as catalytic converters, particulate filters, and heat exchangers, utilise an array of (often square) channels to increase surface area, promote mass and heat transfer, and/or chemical reactions. \cite{Williams} These channel arrays are commonly known as monoliths. 

\begin{figure*}
    \centering
    \includegraphics[scale=0.31]{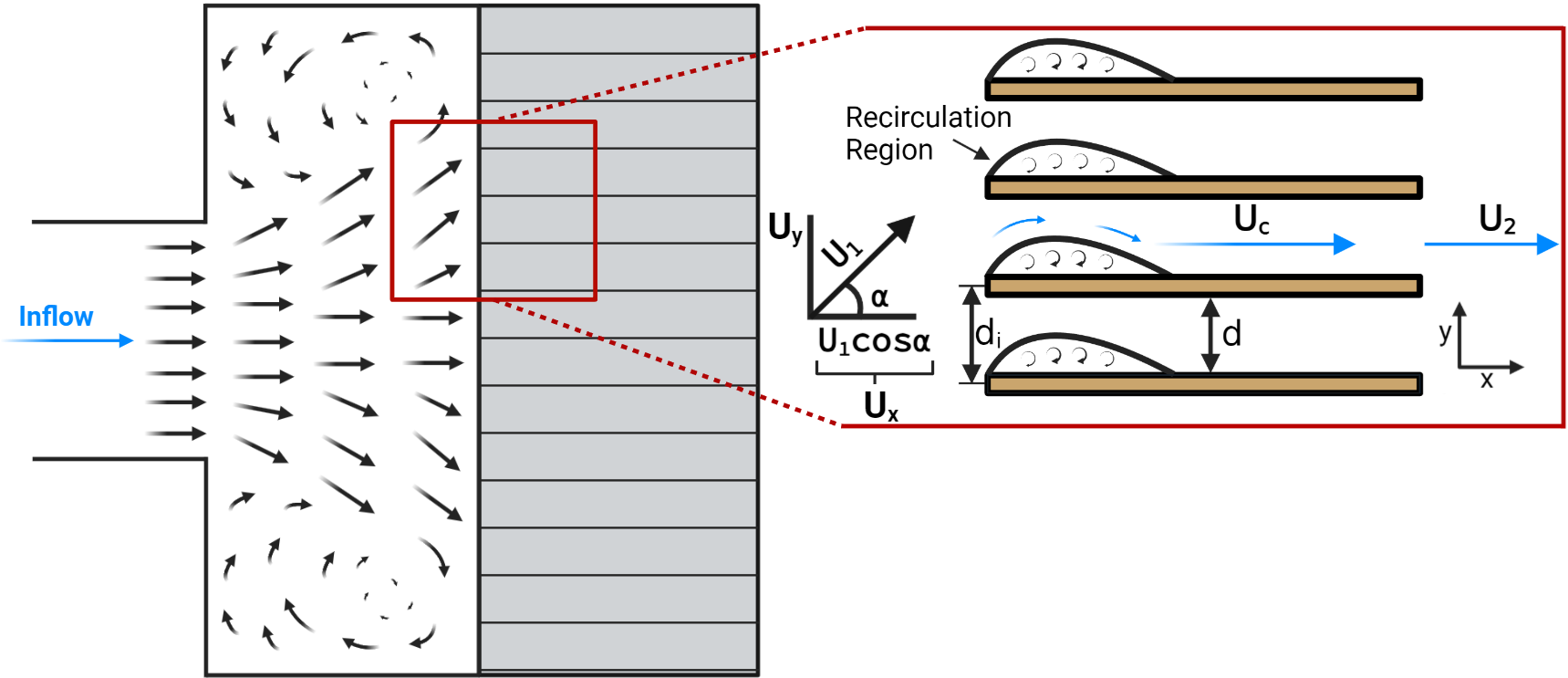}
    \caption{Diagram illustrating formation of recirculation zones and shear layers in a channel with oblique flow entry into the monolith channels caused by the sudden expansion upstream of the monolith.}
    \label{recirc}
\end{figure*}

In many applications, the flow upstream of the monolith is not always aligned axially but can enter the channels at an angle due to the upstream geometry (as illustrated in Fig. ~~\ref{recirc}). This oblique entry is significant because it increases pressure losses, which can adversely affect the efficiency and performance of the system. \cite{Koltsakis1,Fakheri} Understanding the impact of the oblique entry on pressure drop and overall system behaviour is crucial for optimising the design and operation of monolithic devices.

Flows in square channels have been studied extensively with the flow entering the channel axially, or from a larger vessel (e.g. flow with a contraction). \cite{Dullien} With axial, uniform flow entering the channel, the losses associated with the flow development and fully developed flow are well described by the correlation by Shah and London, \cite{ShahLondon} developed for channels with different cross-sections. When evaluating pressure drop in multi-channel devices, authors usually presume uniform flow distribution between the channels within the monolith, \cite{Bissett,KonstOG,Tim} assuming the pressure drop in one channel is indicative of the entire monolith pressure drop. Moreover, authors commonly assume that the flow entering the channels of a monolith does so axially, and with a flat, laminar velocity profile. \cite{Bissett,KonstOG,Masoudi,Haral} The impact of the upstream components' geometry, such as the pipes, and other elements that immediately precede the monolith, is often overlooked. 

Compared with an axial flow, the flow with an oblique entry results in the formation of a recirculation zone near the channel entrance and a shear layer between the separation and freestream regions (see Fig.\,\ref{recirc}). Understanding and quantifying these flow features is crucial for accurately predicting the flow regime and pressure drop across the channel.

Several studies have investigated the effect of oblique entry on pressure losses in various conduits. \cite{KW,Persoons,Quadri,Kamal,Moore} In most studies, it is assumed that the additional pressure loss due to the oblique flow entry, $\Delta P_{Obl}$, is proportional to the dynamic head associated with the transverse velocity component, $U_y$, induced by the oblique entry angle, $\alpha$ (as shown in Fig.\,\ref{recirc}):

\begin{equation}
    \Delta P_{Obl} = f(Re, \alpha) \frac{\rho U_y^2}{2}
\end{equation}

\noindent It is useful to consider a dimensionless pressure drop, normalised by the upstream dynamic
pressure $\rho U_1^2/2$, $K_{Obl} = \Delta P_{Obl} / (\rho U_1^2/2)$, providing:

\begin{equation}
    \label{Eq:KW}
    K_{Obl} = f(Re, \alpha) \text{sin}^2\alpha
\end{equation}

\noindent where $f$ is a function of flow parameters, $Re$ is the Reynolds number, and $U_1$ is the mean upstream velocity magnitude (as shown in Fig.\,\ref{recirc}).

Kuchemann and Weber \cite{KW} were among the first to propose a theoretical model to quantify $\Delta P_{Obl}$. They examined the coolant flow through a finned heat exchanger where the flow entered tubes obliquely. Their derivation assumes that the transverse dynamic head is completely dissipated, i.e. $f$ = 1 in Eq.\,\ref{Eq:KW}. 

At low Reynolds numbers, the flow might not lose all of its oblique kinetic energy upon entering the channel. Instead, it could "creep" around the corner, retaining some of its oblique kinetic energy. This would imply that the correlation developed by Kuchemann and Weber \cite{KW} (referred to as K-W correlation in the rest of this study), which assumes total loss of the transverse dynamic head, may overpredict the actual losses. This has been observed by Persoons et al. \cite{Persoons} who reported only partial dissipation of the dynamic pressure. In their experimental study, Persoons et al. \cite{Persoons} quoted a value of $f$ = 0.459. Computational fluid dynamics (CFD) simulations presented by Haimad \cite{Haimad} also observed only a partial dissipation of the dynamic pressure resulting in a value of $f$ = 0.831. 

Quadri et al.\cite{Quadri} conducted experimental work measuring the pressure drop across a catalyst monolith with axial and oblique upstream flow, achieved with the use of an angled inlet pipe. They used a flow rate range characterised by a Reynolds approach number, $Re_a$ (based on the upstream velocity magnitude, $U_1$), between 250 and 2200. With an inlet angle of $\alpha$ = 60$^\circ$, an additional pressure drop of around 20\% at $Re_a$ = 2000 was observed. The experimental results indicated a power-law dependency of function, $f$, on both $Re_a$ and $\alpha$, leading the authors to propose $f = A Re_a^{n(\alpha)}$, where $A$ is a proportionality constant and $n(\alpha)$ reflects the dependency on the inlet angle:

\begin{equation}
    K_{Obl} = A Re_a^{n(\alpha)} \text{sin}^2 \alpha
    \label{Eq:Quadri}
\end{equation}

\noindent where the values of both $A$ and $n$ depend on $Re_a$ and $\alpha$:

\begin{table}[h] 
    \centering
    \setlength{\tabcolsep}{8pt} % Adjust column spacing
    \begin{tabular}{c c c} 
        & $A$ & $n(\alpha)$ \\  
        30$^\circ \leq \alpha \leq$ 45 $^\circ$ & 0.021 & 0.5 \\ 
        55$^\circ \leq \alpha \leq$ 70 $^\circ$ & 0.18 & 0.24 \\ 
        $\alpha$ = 75$^\circ$ & 0.525 & 0.1\\ 
    \end{tabular} 
\end{table}

The results showed that the non-dimensional oblique pressure loss, $K_{Obl}$, approaches a plateau at higher Reynolds numbers, following an initially steep gradient.

Mat Yamin \cite{Kamal} employed a similar experimental methodology to Quadri et al. \cite{Quadri} but extended the parameter range to higher Reynolds numbers (up to $Re_a$ = 4000). The correlation described Eq.\,\ref{Eq:Quadri} was found to agree well with the experimental results over the broader Reynolds number range, thus extending the correlation validity range.

Thus, most existing correlations for the extra pressure loss caused by oblique entry estimate it as a fraction of the upstream dynamic pressure (Eq.\,\ref{Eq:KW}). However, there is no agreement on the magnitude of the coefficient, $f$, with values varying from 0.459 to 1, or whether it depends on the Reynolds number and oblique entry angle.

Numerical studies of flows in square ducts have also been used to investigate various physical phenomena such as turbulent flow \cite{Coherent,Cornejo,Vidal} and particle transport. \cite{DNS1,particle1,particle2} In such studies, LES is often used due to its advantages over RANS models in capturing transient and complex flow phenomena, including flow separation and transition to turbulence. \cite{LES1,LES2} The advantage of LES is that it resolves large-scale turbulent structures while modelling smaller scales using subgrid-scale models, \cite{Piomelli} therefore combining good modelling accuracy for larger, anisotropic structures, with computationally efficient, averaged turbulence models for smaller, isotropic turbulent eddies. LES has shown promising results in predicting pressure losses in various geometries, such as pipes \cite{LESpipe,LESpipe2} and square channels. \cite{Cornejo} 

The general effects of the oblique flow entry on the flow, such as flow separation with the formation of a shear layer and additional pressure losses, are well-known. However, existing studies disagree not only on the magnitude of these extra pressure losses but also on the effect of the Reynolds number. The flow regime inside the channels at channel Reynolds numbers close to the transition threshold ($\approx$ 2000 and above) has not been clearly defined. The CFD turbulence modelling techniques, such as LES, RANS, or laminar models, are often chosen assuming the flow regime is the same as that for the corresponding axial flows. This assumption overlooks the unique flow features present in flows with an oblique entry, which may influence the turbulence transition threshold and affect the accuracy of the results.

This study aims to gain a deeper understanding of the effect of an oblique flow entry into a square channel on the pressure drop and flow characteristics, and explore the advantages and limitations of LES and RANS for modelling the effect of oblique entry on the flow structure inside the channel. Simulation results are compared with experimental studies \cite{Quadri,Kamal,Persoons} and correlations developed in other studies. \cite{KW,Persoons,Quadri,Moore} The accuracy and limitations of the available correlations are discussed.

The paper is organised as follows. Section II (Methodology) presents details of the assumptions and simulation setup for each of the flow models used (LES, RANS, and laminar flow simulations), and flow parameters. In Section III (Results), subsection A focuses on detailing the salient flow features, including flow recirculation, shear layer formation, secondary flow, and turbulence development. Subsection B discusses the pressure losses due to oblique entry and compares these results with experimental data and existing correlations. The sensitivity of the results to the leading edge geometry and turbulence model choice is also explored. In Section IV, the conclusions and future recommendations are presented.

\section{\label{sec:methodology}Methodology} 
\subsection{\label{sec:geommesh}Geometry \& Mesh}
\begin{figure*}
    \includegraphics[scale=0.39]{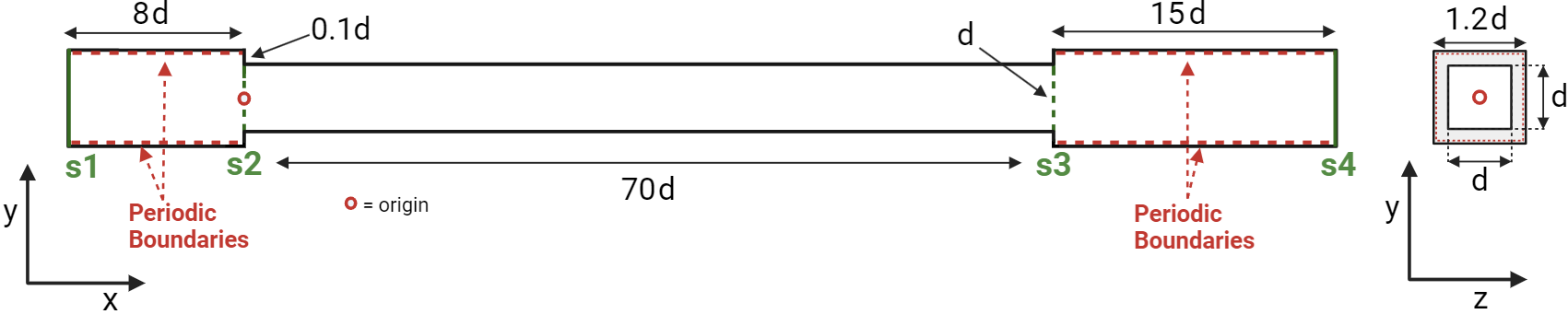}
    \caption{Schematic diagram of the flow domain used for simulations.}
    \label{fig:schematic}
\end{figure*}

Flow in a single square cross-section channel is considered with uniform upstream velocity. In practice, this is representative of flow in a monolithic structure with multiple identical channels, such as heat exchangers or particulate filters (see Fig.\,\ref{recirc}). The channel with width, $d$ = 1 mm, has a wall thickness of 0.2$d$ (Fig.\,\ref{fig:schematic}), which is typical for several applications, for example, automotive catalysts or filter monoliths. Extra inlet and outlet sections of width 1.2$d$ and lengths of 8$d$ and 15$d$ were added upstream and downstream of the channel, respectively, to ensure that flow redistribution induced by contraction and expansion at the channel entrance and exit are properly captured. The simulations demonstrated that the inlet length was sufficiently far from the channel inlet to prevent any effect on the flow inside the channel, however, the development length downstream may extend beyond 15 channel widths. To ensure that the presence of the outlet boundary does not affect the flow in the channel, two extra simulations with an outlet length of 30$d$ were carried out (Sec.\,\ref{effectofoutlet}).

A Cartesian coordinate system ($x,y,z$) is used throughout this study, with $x$ being the axial co-ordinate, and the origin set in the centre of the channel entrance cross-section (indicated by the red circle in Fig.\,\ref{fig:schematic}).

The domain was discretised using a polyhedral mesh with prism cell layers at the walls. The prism layers had a total thickness of 8$\times$10$^{-5}$ m, with the number of layers ranging between 8 and 10 (Table\,\ref{table:meshs}). This configuration ensured y$^+ \leq 1$ for most of the boundary layer along the channel. Table\,\ref{table:meshs} presents a summary of the four meshes utilised in the mesh quality study, detailing the properties of each mesh. The quality of the meshes and results of the mesh independence study are discussed in more detail in Sec.\,\ref{MeshResults}.

\begin{table}
    \arrayrulecolor{black}
    \caption{\label{table:meshs}Properties of the meshes used for flow simulations.}
    \begin{ruledtabular}
    \begin{tabular}{c r r}

    Mesh \# & Cell Count & \makecell{No. of Prism \\ Layers} \\
    \hline
    1 &  679719 & 8 \\
    2 & 1687961 & 8 \\
    3 & 8857225 & 8 \\
    4 & 32814169 & 10 \\
    \end{tabular}
    \end{ruledtabular}
\end{table}

To study the effect of a rounded edge on the flow structure and pressure loss, three additional geometries were created with a radius, $r$ = 0.01$d$, 0.025$d$, and 0.05$d$, applied on the leading edge. The properties of the meshes were kept the same as those for the sharp edge geometry.

\subsection{\label{FlowParams}Flow Parameters}

To study the effect of shear induced by the flow separation on the flow regime, values of the channel Reynolds number, $Re_c$, between 500 and 3000 were used. The channel Reynolds number is defined as:

\begin{equation}
    \label{Eq:Rec}
    Re_c = \frac{\rho U_c d}{\mu}
\end{equation} 

\noindent where the density and viscosity of air, at 25 $^\circ$C, are $\rho$ = 1.184 kg/m$^3$, and $\mu$ = 1.83$\times$10$^{-5}$ Pa$\times$s, respectively. The axial velocity in the channel, $U_c$, averaged over the channel cross-section, is related to the axial velocity upstream of the channel, $U_x$, by the contraction ratio, (1/$\phi$):

\begin{equation}
U_c = U_x \times \frac{1}{\phi}
\label{chanvel}
\end{equation}

In Eq.\,\ref{chanvel}, $\phi$ is the monolith's open area fraction, $\phi = d^2/d_i^2$ $\approx$ 0.695 for the channel geometry used in this study. Here, $d_i$ = 1.2$d$, is the inlet width (see Fig.\,\ref{recirc}).

When the inlet angle, $\alpha$ (Fig.\,\ref{recirc}), increases, the velocity magnitude at the inlet must also increase to maintain the same axial velocity and channel Reynolds number:

\begin{equation}
U_1 = \frac{U_x}{\text{cos}(\alpha)}
\end{equation}

The range of entry angles, $\alpha$, selected for this study spans from $0^\circ$ to $70^\circ$. This range adequately covers the entry angles typically encountered in various applications.

Table\,\ref{tab:sims} provides a summary of all the simulations conducted in this study. Given the extensive number of simulations performed, only those most relevant to the key aspects of the analysis are included in this paper to ensure clarity in the presentation of results.

\begin{table*}
    \centering
    \caption{Summary of conducted simulations.}
    \label{tab:sims}
    \begin{ruledtabular}
    \begin{tabular}{r c c r}
        \textbf{Channel Edge Radius, r (\textmu m)} & \textbf{Flow Model} & \textbf{Entry Angles ($\alpha$ in $^\circ$)} & \textbf{Reynolds Number ($Re_c$)} \\
        \hline
        0 & LES                & 0 - 70                       & 500 - 3000 \\
        0 & RANS               & 0, 15, 30, 50, 60 & 500 - 3000 \\
        0 & Laminar Unsteady   & 0, 15, 30, 45, 50, 60 & 500 - 3000 \\
        0 & Laminar Steady     & 0 - 70                       & 500 - 3000 \\
        10              & LES                & 0, 40                          & 1000 - 3000 \\
        25              & LES                & 0, 40                          & 1000 - 3000 \\
        50              & LES                & 0, 40                          & 1000 - 3000 \\
    \end{tabular}
    \end{ruledtabular}
\end{table*}

\subsection{Pressure Drop}
To assess the losses in the channel, the static pressure drop between the inlet boundary (s1) and the channel outlet (s3) as shown in Fig.\,\ref{fig:schematic}, was monitored and time-averaged over 2 residence times of the flow across the entire domain. As the focus of the study is the oblique entry effect, we use this value as it accounts for the pressure losses incurred within the channel and just before the channel (contraction), but discounts losses caused by fluid expansion upon exiting the channel. To estimate the irreversible pressure loss, the difference between the total pressure at cross-sections s1 and s3 is calculated. The dynamic pressure is estimated using the air density and average velocities at the respective cross-section - that is $U_1$ at the inlet and $U_c$ for the channel. The difference then makes up the non-recoverable pressure drop between cross-sections s1 and s3:

\begin{equation}
\Delta P = \big(P_{s1} + \frac{1}{2}\rho U^2_1\big) - \big(P_{s3} + \frac{1}{2}\rho U^2_c \big)
\label{DPoblique}
\end{equation}

In Eq.\,\ref{DPoblique}, $P_{s1}$ and $P_{s3}$ is the static pressure at cross-sections s1 and s3. 

By subtracting the total pressure drop associated with axial flow ($\alpha$ = 0$^\circ$), we can quantify the specific additional pressure losses attributable to the oblique flow entry, $\Delta P_{Obl}$:

\begin{equation}
\Delta P_{\text{Obl}} = \Delta P_\alpha - \Delta P_{(\alpha=0^\circ)}
\label{Eq:PObl}
\end{equation}

In Eq.\,\ref{Eq:PObl}, $\Delta P_{\alpha}$ is the total pressure drop at angle $\alpha$, and $\Delta P_{(\alpha = 0^\circ)}$ is the pressure drop for axial entry. These oblique losses, $\Delta P_{Obl}$, are non-dimensionalised by the upstream dynamic pressure, $\rho U_1^2 /2$. 

The same definition of the non-recoverable pressure loss (Eq.\,\ref{DPoblique} and\,\ref{Eq:PObl}) and non-dimensional oblique pressure loss was used by Quadri et al. \cite{Quadri} and Mat Yamin, \cite{Kamal} which facilitates comparison with their experimental data. To compare the oblique pressure drop ($\Delta P_{Obl}$) calculated from simulations with the measured $\Delta P_{Obl}$ in experiments involving channels of different lengths, it is necessary to evaluate the pressure drop from the inlet (cross-section s1 in Fig.\,\ref{fig:schematic}) to a specific cross-section at length, $L$, along the channel.

\subsection{\label{flowmodels} Flow Models}
To study the complex, unsteady flow with flow separation and shear, LES and unsteady RANS flow models are used, implemented in the Star CCM+ commercial CFD package. For low Reynolds numbers and angles, the flow may be laminar, therefore in addition to LES and RANS several laminar flow simulations were also conducted (see Table\,\ref{tab:sims} for the full list).

In all simulations, the working fluid used is air, and the flow is assumed to be incompressible, and isothermal. For all simulations, except the steady laminar cases, an adaptive time step was employed to ensure a maximum Courant-Friedrichs-Lewy (CFL) number of less than one:

\begin{equation}
    \text{CFL} = \frac{u\Delta t}{\Delta} \leq 1
    \label{CFL}
\end{equation}

Here, the velocity, $u$, is the local velocity magnitude, $\Delta t$ is the timestep, and $\Delta$ is the cell length. With this criterion, the fluid does not travel more than one cell length per time step. Further information regarding the CFL number is provided by Courant et al. \cite{CFL}

For both RANS and LES simulations, second-order accurate numerical schemes were employed for spatial discretisation to ensure a balance between computational efficiency and solution accuracy. Time integration in LES was performed using a second-order implicit scheme to accurately resolve the unsteady turbulent structures. Convergence was assessed based on the residuals, which were observed to drop to the order of $10^{-9}$ for both RANS and LES. These measures ensured the reliability of the numerical solutions while adequately resolving the complex flow dynamics.

\subsubsection {LES Setup}

The LES approach segregates the turbulent flow structures (eddies) into those that are sufficiently large to be resolved on the mesh, and the remaining, smaller, eddies which are modelled by a sub-grid scale turbulence model. \cite{LES} The mesh cell size acts as a filter sorting the eddies by size. Consequently, a more refined mesh results in a smaller fraction of kinetic energy being modelled. Filtering by the mesh cell length ($\Delta$) and assuming constant density and incompressible flow, the governing continuity and momentum equations are expressed as follows \cite{LES}:

\begin{equation}
    \frac{\partial \bar{u}_i}{\partial x_i} = 0
    \label{LESEqs}
\end{equation}
    
\begin{equation}
    \frac{\partial \bar{u}_i}{\partial t} + \bar{u}_j \frac{\partial \bar{u}_i}{\partial x_j} = -\frac{1}{\rho} \frac{\partial \bar{P}}{\partial x_i} + \nu \frac{\partial^2 \bar{u}_i}{\partial x_j \partial x_j} - \frac{\partial \tau_{ij}}{\partial x_j}
    \label{LESEqs2}
\end{equation}
    
The overbar in Eq.\,\ref{LESEqs2} denotes the resolved velocity components, $u_i$, and resolved pressure, $P$. In these equations, $\nu$ stands for the kinematic viscosity $(\mu/\rho)$, $t$ is time, $\rho$ is density, and $x_i$ represents the three Cartesian coordinate components for $i$ = 1, 2, 3. The un-resolved turbulent kinetic energy is accounted for in the stress tensor, $\tau_{ij}$, modelled using the sub-grid Dynamic Smagorinsky Model. \cite{Germano} The expression for $\tau_{ij}$ in Eq.\,\ref{LESEqs2} is given by:

\begin{equation}
\tau_{ij} = \frac{2}{3} \rho k_{\text{sgs}} \delta_{ij} - 2C_s k_{sgs}^{1/2} \Delta \bar{S}_{ij}
\label{sgEq}
\end{equation}

Here, $S_{ij}$ is the strain rate tensor. The coefficient, $C_s$, in Eq.\,\ref{sgEq}, is computed locally, by employing a test filter twice the size of the cell (see Germano et al. \cite{Germano}). The subgrid kinetic energy, $k_{sgs}$, is calculated with the transport equation as shown by Smagorinsky. \cite{LES} 

\subsubsection{RANS Setup}

LES is considered a superior modelling tool for capturing flows with shear; however, RANS models are frequently used because of their computational efficiency. To compare the accuracy of different models for the flow in the considered range of Reynolds numbers, a RANS model ($k-\omega$ SST) has been used for several configurations. The RANS governing equations closely resemble Eqs.\,\ref{LESEqs} and\,\ref{LESEqs2}, with the overbar denoting temporally filtered variables rather than spatially filtered ones.

The Gamma transition model was employed to capture the laminar to turbulent transition in regions where the flow characteristics change near the boundary layer.

Full details of the flow model are provided by Menter. \cite{Menter}

\subsubsection{Laminar Setup}

Both the LES and RANS models account for turbulence and therefore perform favourably when turbulence is present. Since some simulations are conducted at Reynolds numbers associated with laminar flow transitioning to turbulence, these models may artificially induce turbulence under conditions where it would not develop in reality. To assess whether this might be the case, both unsteady and steady laminar flow models have also been employed. These models solve the steady and unsteady Navier-Stokes equations, inherently accounting for viscous effects without having a stress tensor to model turbulent contributions.

\subsection{\label{sec:boundary}Boundary Conditions}
At the inlet (cross-section s1 in Fig.\,\ref{fig:schematic}) the flow velocity is specified. The inlet flow direction is changed to simulate an oblique flow entry, as indicated by the velocity vector $U_1$ in Fig.\,\ref{recirc}. For LES and laminar simulations, a flat velocity profile without synthetic turbulence was specified at the inlet. Although the flow could be turbulent upstream of the monolith in applications, the focus of this study is on the flow features caused by an oblique flow entry into the square channels. This agrees with the experimental setup used by Mat Yamin \cite{Kamal} and Quadri et al. \cite{Quadri} where measures were taken to ensure uniform, laminar inlet flow where possible. 

Surrounding the inlet and outlet sections are four orthogonal walls set to translational periodic boundaries (Fig.\,\ref{fig:schematic}). This ensures that the flow far upstream and downstream of the channel is uniform and unidirectional. All walls of the channel are solid smooth walls with no-slip conditions imposed. At the outlet (cross-section s4 in Fig.\,\ref{fig:schematic}), the pressure is set to be equal to atmospheric pressure.

In $k-\omega$ turbulence modelling, setting zero turbulence at the inlet may affect the numerical stability of the simulations (because $\omega = \epsilon/k$). Therefore, a small non-zero value of 1\% and a turbulent viscosity ratio of 10 were used to provide a realistic baseline level of turbulence and ensure stable initialisation of the turbulence model. The turbulence fully decayed by the time it reached the channel inlet. While the wall $y^+$ was below 1 for most of the channel, the $k-\omega$ SST model's 'all $y^+$' wall treatment was used. This approach adapts to the mesh resolution, applying wall functions where needed to capture near-wall turbulence accurately.

\subsection{\label{MeshQualityMethod}LES Mesh Resolution Assessment}
On an LES mesh, the quality can be evaluated by several methods. Pope \cite{Pope} suggested that LES performs favourably when 80\% of the turbulent kinetic energy is resolved while the remaining 20\% is modelled. However, as some studies \cite{Celik,Gou} have demonstrated, defining a ratio of resolved to modelled kinetic energy leads to singularities and anomalous errors in the near-wall and laminar regions, where the kinetic energies are small or zero. As an alternative, a quality parameter has been suggested by Celik et al. \cite{Celik} based on the ratio of effective turbulent viscosity to laminar viscosity:

\begin{equation}
\text{LES}_{IQ_\nu} = \frac{1}{1 + \alpha_\nu \left(\frac{\nu_{\text{t,eff}}}{\nu}\right)^n}
\label{celikeq}
\end{equation}

The variables $\alpha_\nu$, and $n$ have been derived by Celik et al. \cite{Celik} and are 0.05 and 0.53, respectively. The effective turbulent viscosity, $\nu_{\text{t,eff}}$, accounts for numerical dissipation by incorporating a numerical viscosity, $\nu_{\text{num}}$:

\begin{equation}
\nu_{t,\text{eff}} = \nu_t + \nu_{\text{num}}
\end{equation}

\begin{equation}
\nu_{\text{num}} = \text{sgn}(k_{\text{num}}) C_\nu \Delta \sqrt{k_{\text{num}}}
\label{vnum}
\end{equation}

\begin{equation}
k_{\text{num}} = C_n \left(\frac{h}{\Delta}\right)^2 k_{\text{sgs}}
\label{eqknum}
\end{equation}

In Eqs.\,\ref{vnum} and\,\ref{eqknum},\, $C_n$ = 1 and $C_\nu$ = 0.17, are model coefficients that control the magnitude of numerical dissipation, and $h$ represents the characteristic element size of the mesh, often taken as the cubic root of the cell volume. The parameter $\Delta$ denotes the cell length, which is typically defined as the local grid spacing.

Celik et al. \cite{Celik} state that $\text{LES}_{IQ\nu}$ > 0.8 implies the mesh is sufficiently resolved for LES. This threshold indicates that the model adequately captures and resolves the appropriate amount of kinetic energy, aligning with Pope's hypothesis that at least 80\% of the kinetic energy is resolved and the remaining 20\% is modelled. Moreover, the authors state that $\text{LES}_{IQ_{\nu}}$ > 0.95 corresponds to the accuracy level of a Direct Numerical Simulation (DNS). 
\label{MeshResults}

To evaluate the quality of the mesh used for all models, several meshes with different levels of refinement were used (Table\,\ref{table:meshs}). Both global and local flow properties were analysed to assess if the corresponding simulations capture most of the important flow features.

\begin{figure}
    \includegraphics[scale=0.57]{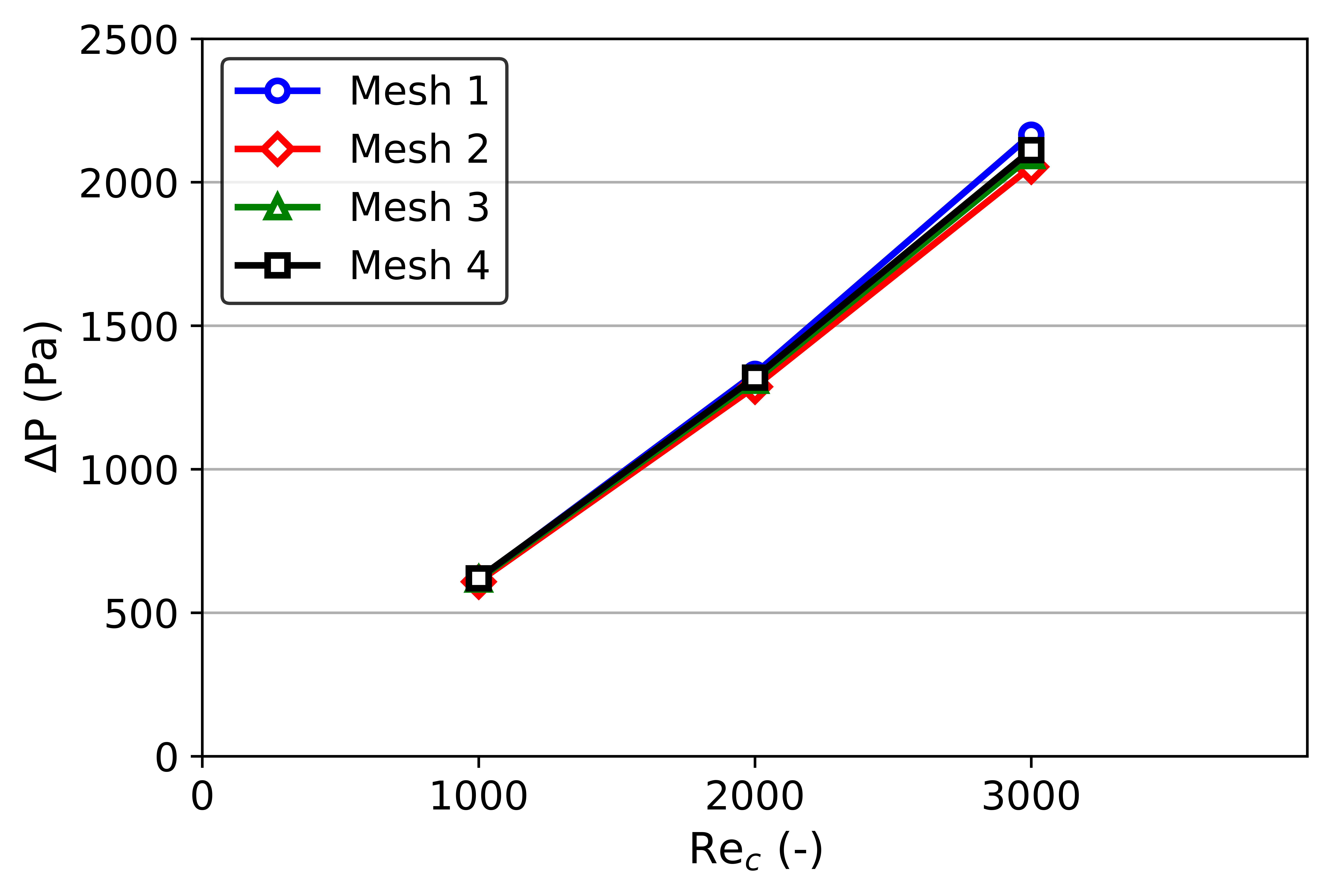 }
    \caption{Comparison of the dimensional static pressure drop between s1 and s3 for the four meshes, $\alpha$ = 0$^\circ$.}
    \label{axialMS}
\end{figure}

\begin{figure}
    \includegraphics[scale=0.57]{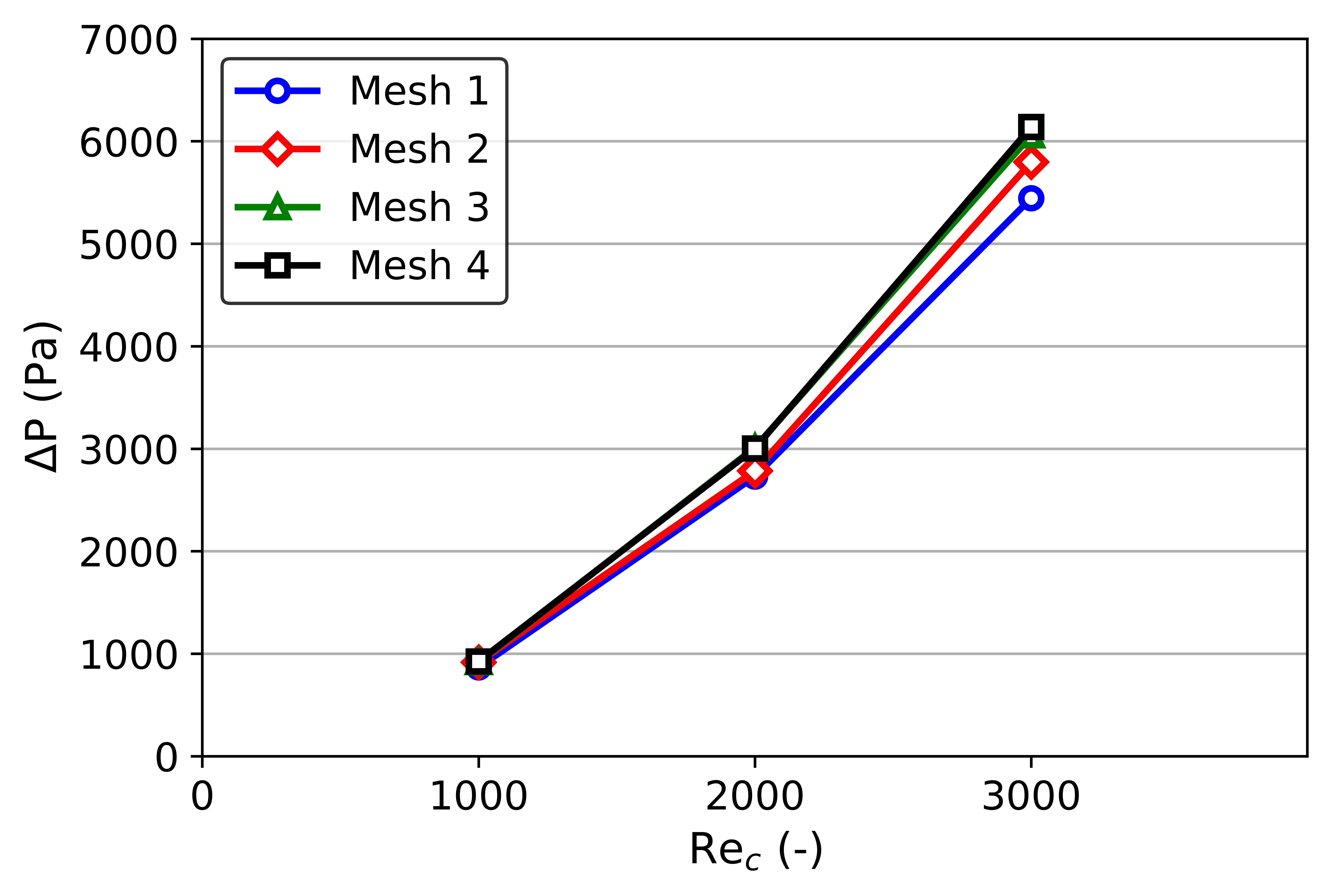 }
    \caption{Comparison of the dimensional static pressure drop between s1 and s3 for the four meshes, $\alpha$ = 60$^\circ$.}
    \label{60MS}
\end{figure}

The dimensional static pressure drop calculated between sections s1 and s3 for three distinct entry angles, $\alpha$ = 0$^\circ$, 30$^\circ$, and 60$^\circ$, is presented in Figs.\,\ref{axialMS} to\,\ref{60MS}. Three values of Reynolds number, $Re_c$ = 1000, 2000, 3000, are used for each mesh.

\begin{figure}
    \includegraphics[scale=0.57]{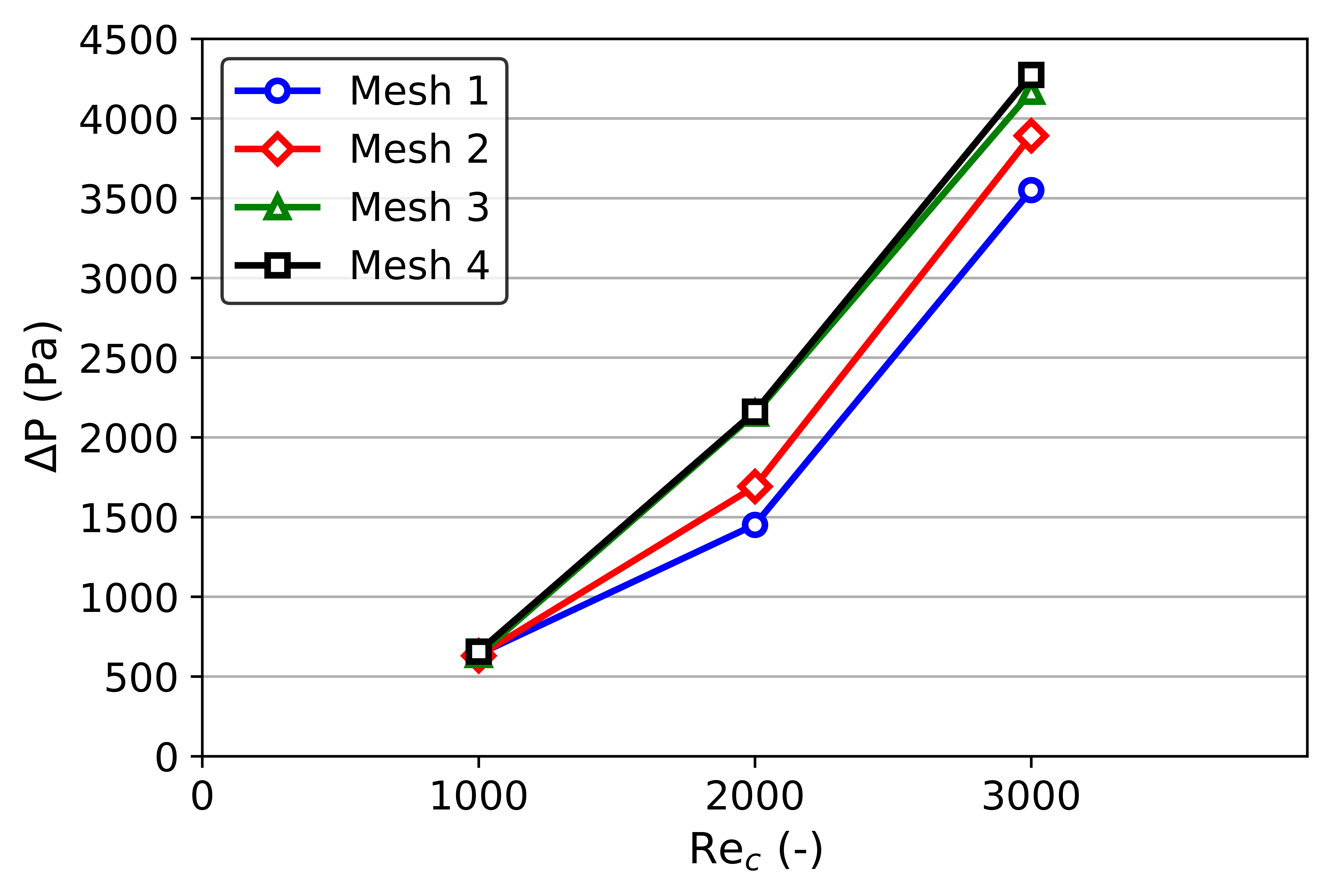 }
    \caption{Comparison of the dimensional static pressure drop between s1 and s3 for the four meshes, $\alpha$ = 30$^\circ$.}
    \label{30MS}
\end{figure}

The results for the axial entry (Fig.\,\ref{axialMS}) indicate a good consistency among the four meshes' pressure drop predictions, with a minor difference observed at $Re_c$ = 3000. The maximum difference is approximately 6\% between the finest and coarsest meshes (mesh 4 and mesh 1, respectively). For $\alpha$ = 30$^\circ$ (Fig.\,\ref{30MS}), the differences in pressure values are higher; however, the results for the two finer meshes, namely mesh 3 and mesh 4, are within 4\%. This trend persists for $\alpha$ = 60$^\circ$ (Fig.\,\ref{60MS}). Notably, the agreement between the meshes improves from $\alpha$ = 30$^\circ$ to $\alpha$ = 60$^\circ$. This suggests that a more refined mesh might be needed to capture some of the complex flow features, particularly where the transition from laminar to turbulent flow is expected. Further discussion of these phenomena is presented in subsequent sections. Based on the results shown in Figs.\,\ref{axialMS} to\,\ref{60MS}, it can be reasonably inferred that the pressure drop becomes independent of the mesh configuration for meshes 3 and 4.

For an LES model, a mesh independence study of a global parameter such as channel pressure drop is not conclusive for demonstrating model/mesh validity. Pressure drop in the channel is a global flow characteristic, and should not be used as the sole criterion for assessment of the resolution of complex flow features.

\begin{figure}
    \includegraphics[scale=0.3]{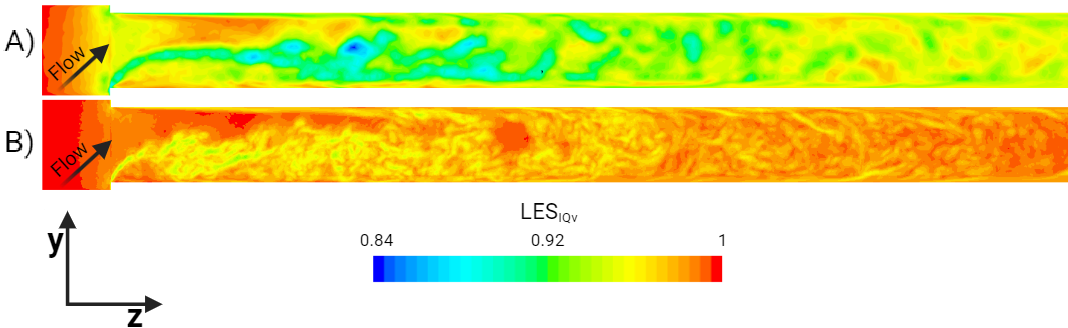 }
    \caption{Instantaneous $\text{LES}_{IQ_{\nu}}$ near the entrance of the channel (XY plane, $z$ = 0), for $Re_c$ = 3000 and $\alpha$ = 60$^\circ$. A) Mesh 2, B) Mesh 4. }
    \label{LESIQ}
\end{figure}

To assess the mesh resolution in the whole domain, the Celik index, $\text{LES}_{IQ_{\nu}}$ (Eq.\,\ref{celikeq}), was calculated for all meshes in a configuration that was expected to be turbulent, with an oblique angle of $\alpha$ = 60$^\circ$ and $Re_c$ = 3000. Fig.\,\ref{LESIQ} shows a contour plot of $\text{LES}_{IQ_{\nu}}$ in the XY plane for mesh 2 and 4 near the entrance of the channel, where the recirculation region and shearing layer are present.

In Fig.\,\ref{LESIQ} the lowest values of $\text{LES}_{IQ_{\nu}}$ are approximately 0.84 and 0.92 for meshes 2 and 4, respectively, and are within the shear layer. This region typically exhibits heightened turbulence and vorticity, indicating a potential requirement for further mesh refinement. Values exceeding 0.8, following Celik's analysis, still indicate reasonable mesh quality. The minimum $\text{LES}_{IQ_{\nu}}$ values across the entire domain for each mesh at $\alpha$ = 60$^\circ$ and $Re_c$ = 3000 are shown in Table\,\ref{LESIQvalues}. $\text{LES}_{IQ_{\nu}}$ values for simulations with lower entry angles or Reynolds numbers, were found to be closer to the values typically considered near DNS resolution (i.e. > 0.95). Nevertheless, mesh 3 was employed across all simulations to ensure consistency in our approach.

\begin{table}
    \centering
    \caption{The minimum values for $\text{LES}_{IQ_{\nu}}$ for all meshes for $Re_c$ = 3000, $\alpha$ = 60$^\circ$.}
    \label{LESIQvalues}
    \begin{ruledtabular}
    \begin{tabular}{cc}
    \textbf{Mesh \#} & \textbf{Minimum $\text{LES}_{IQ_\nu}$}  \\
    \hline
    \centering
    1 & 0.81 \\
    2 & 0.82 \\
    3 & 0.88 \\
    4 & 0.90 \\
    \end{tabular}
    \end{ruledtabular}
\end{table}

\subsection{\label{effectofoutlet}Effect of Outlet Length on Pressure Distribution within the Domain}

\begin{figure}
\includegraphics[scale=0.575]{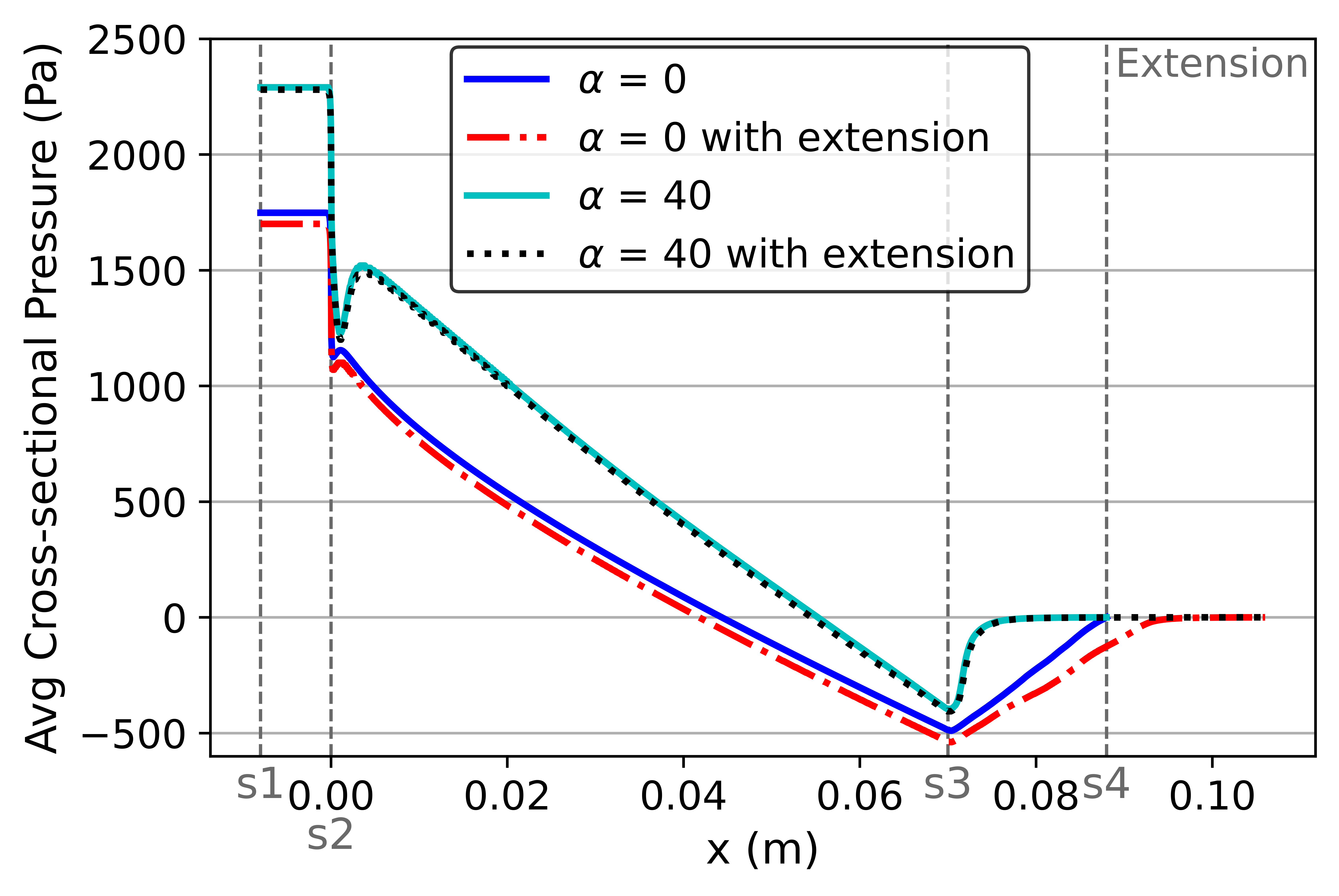}
\caption{Comparison of pressure distribution in the domain with outlet lengths of $L$ = 15$d$ and $L$ = 30$d$. $Re_c$ = 2000, $\alpha$ = 0$^\circ$ and 40$^\circ$.}
\label{Ext}
\end{figure}

Two additional simulations were conducted with an extended outlet length of 30$d$ to establish if a 15$d$ outlet length was sufficient for the flow inside the channel to not be affected by the outlet conditions downstream. One was conducted under conditions expected to produce laminar flow, and another under conditions expected to produce turbulent flow, $\alpha$ = 0$^\circ$ and 40$^\circ$ at a channel Reynolds number, $Re_c$ = 2000. The results (Fig.\,\ref{Ext}) show the time-averaged mean static pressure at cross-sections perpendicular to the channel along the length of the flow domain. Fig.\,\ref{Ext} shows that the longer outlet length does not alter the pressure distribution within the channel or at the inlet for either turbulent or laminar cases. As LES modelling is computationally expensive, a shorter outlet section, of length 15$d$, was chosen for subsequent simulations. 

Fig.\,\ref{Ext} shows that using a shorter outlet does not alter pressure distribution inside the channel, which is the main focus of this study. The fixed pressure outlet boundary condition results in a shift of the whole pressure distribution along the channel by a constant value. To make comparison between different cases easier, in all subsequent sections the pressure plotted is the modified pressure:

\begin{equation}
    P_{\text{mod}} = P - P_{(x=L)}
\end{equation}

This is equivalent to assuming that the average pressure at the outlet is atmospheric, and does not affect the calculations of the pressure losses. The subscript 'mod' is dropped in subsequent sections for convenience. 

\section{\label{sec:Results}Results}
In all subsequent discussions, LES simulation results are used unless stated otherwise.

\subsection{\label{FlowStructure}Flow Structure}

Unlike the case for axial flow entry, where a small vena contracta effect is observed, \cite{Dullien} oblique flow entry introduces several new complex flow features. These include the formation of a large recirculation region with a shear layer, secondary flow caused by corner vortices, and, in some cases, transition to turbulence. This has a significant impact on pressure drop, which is important for many applications.

\subsubsection{\label{RecircShear}Recirculation Region and Shear Layer}

\begin{figure}
    \includegraphics[scale=0.38]{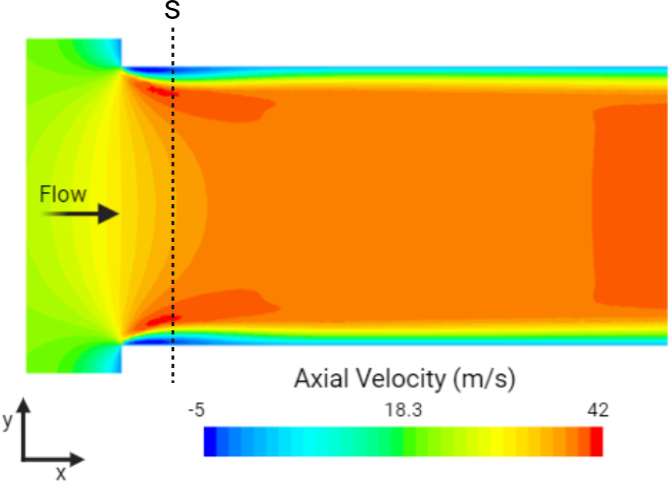 }
    \caption{Axial velocity distribution at the entrance of the channel (XY plane, $z$ = 0) for $Re_c$ = 3000, $\alpha$ = 0$^\circ$.}
    \label{vena}
\end{figure}

In the case of an axial entry, the flow path contracts as it enters the channel. This causes the formation of a vena contracta immediately post-entrance (see Fig.\,\ref{vena}). For contraction ratios typical to monoliths and catalytic converters ($\approx$ 0.75 - 0.9), the effect is not very pronounced, and the separation zones formed at the channel entrance are small. Pressure losses linked to the vena contracta have been studied in the context of automotive filter applications. \cite{KonstInert, Tim} Typically, these losses are estimated using the Borda-Carnot equation \cite{Dullien} or using conservation laws. \cite{Tim}

\begin{figure}
    \includegraphics[scale=0.24]{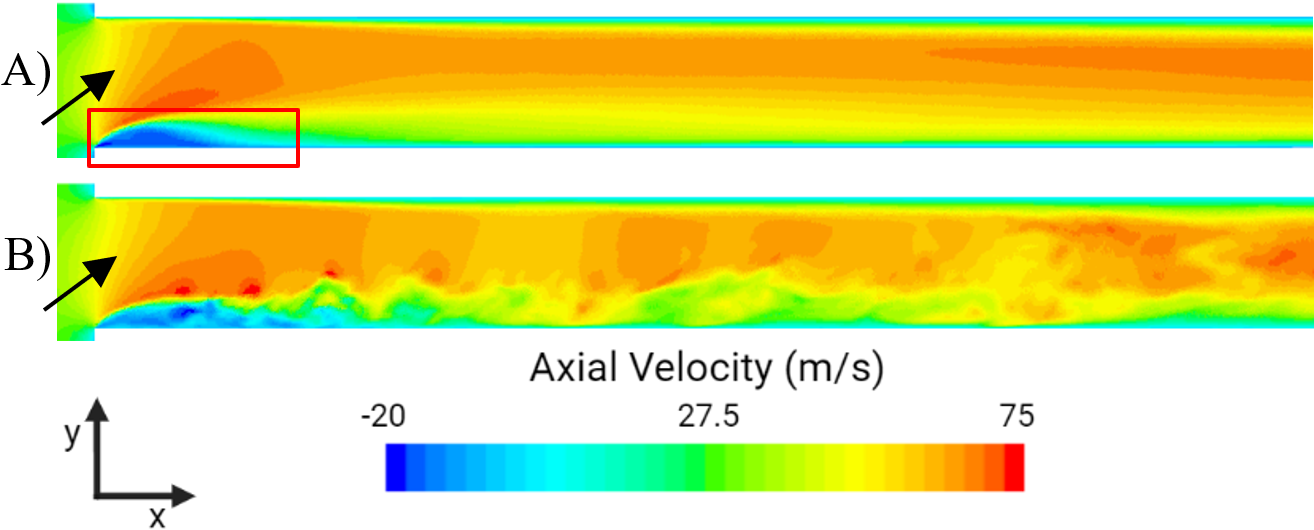 }
    \caption{Axial velocity distribution at the entrance of the channel (XY plane, $z$ = 0) for $Re_c$ = 3000, $\alpha$ = 30$^\circ$. A) Time-averaged (recirculation region position highlighted by a red box) B) Instantaneous}
    \label{30deg3000}
\end{figure}

When the flow approaches the channel at an oblique angle, the symmetry of the vena contracta disappears. A greater angle results in a more pronounced flow separation from the bottom channel wall and a reduction in the flow separation from the upper wall of the channel (Fig.\,\ref{recirc}). The recirculation region, seen within the red box in Fig.\,\ref{30deg3000}A for $Re_c$ = 3000 and $\alpha$ = 30$^\circ$, shows negative axial velocities of up to 20 m/s in magnitude in the separation region, with the freestream velocities of up to 70 m/s. This means that a shear layer is present at the separation zone boundary. High vorticity in the shear layer leads to a flow characterised by unsteady behaviour, causing the shedding of vortical structures into the core downstream, and generation of turbulence, as shown in Fig.\,\ref{30deg3000}B.

\begin{figure}
    \includegraphics[scale=0.57]{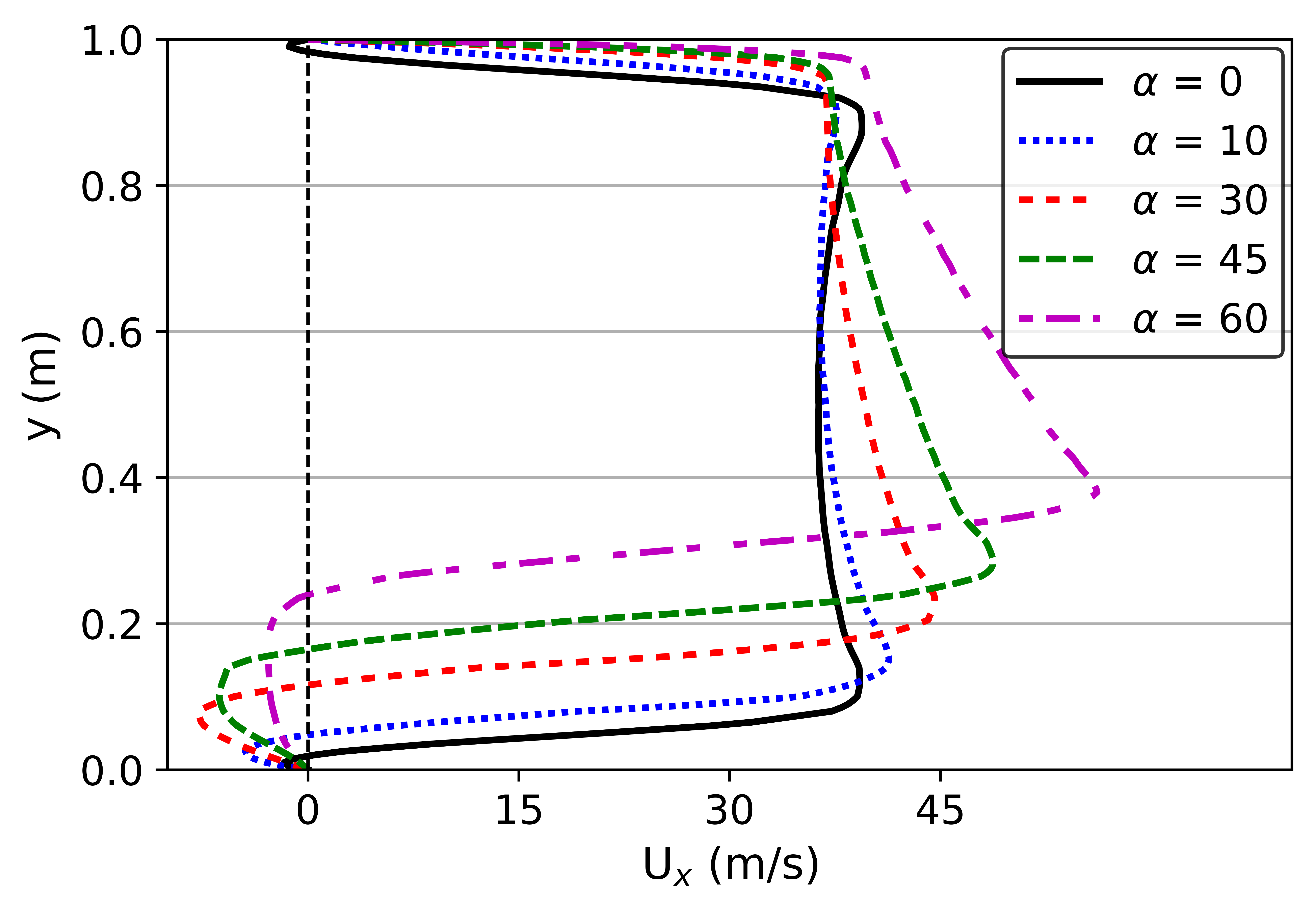 }
    \caption{Time-averaged velocity profiles at the line $z$ = 0, $x$ = 0.05$d$, i.e. 0.05$d$ downstream of the channel entrance for $Re_c$ = 2000 and for different entry angles, $\alpha$.}
    \label{velprofs}
\end{figure}

\begin{figure}
    \includegraphics[scale=0.265]{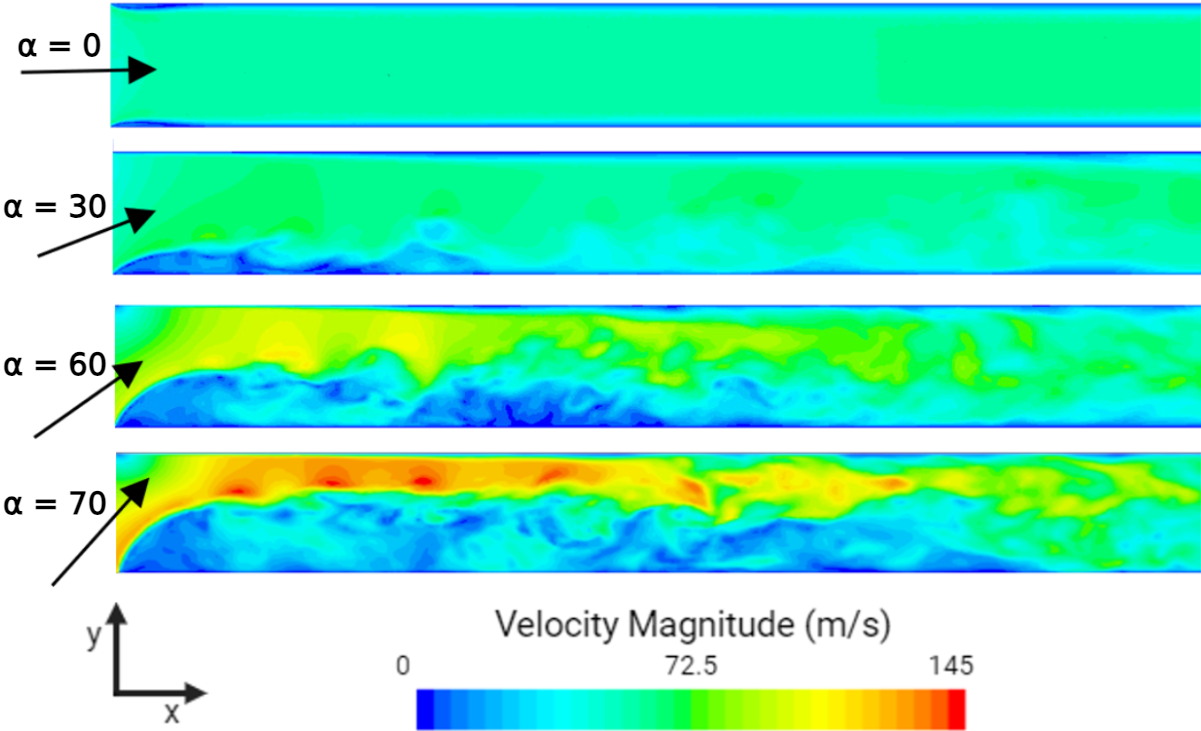 }
    \caption{Instantaneous velocity magnitude at the entrance of the channel (XY plane, $z$ = 0) for $Re_c$ = 3000, $\alpha$ = 0$^\circ$, 30$^\circ$, 60$^\circ$, 70$^\circ$.}
    \label{increasingalpha}
\end{figure}

To illustrate the development of the recirculation zone with increasing entry angle, Fig.\,\ref{velprofs} shows the time-averaged axial velocity profiles at a line $z$ = 0$d$, $x$ = 0.05$d$ downstream of the channel entrance (line S in Fig.\,\ref{vena}), within the recirculation region, for varying entry angles and $Re_c$ = 2000. With an oblique entry angle, the size of the recirculation region increases, leading to two primary effects. Initially, the recirculation bubble narrows the channel entrance significantly, constricting the flow and increasing the bulk velocity. Consequently, a greater shearing stress is generated between the recirculation region and the bulk flow. This causes the formation and the shedding of coherent structures that develop downstream, something that is not seen for the cases with axial flow entry (Fig.\,\ref{increasingalpha}). Furthermore, when flow enters a channel at an oblique angle, the upper surface does not immediately develop a boundary layer as it would in axial flow. This could be caused by the flow approaching the wall surface at an angle, rather than moving parallel to it, delaying the boundary layer formation until further downstream. This effect becomes more pronounced with a greater oblique entry angle.

\begin{figure}
    \includegraphics[scale=0.222]{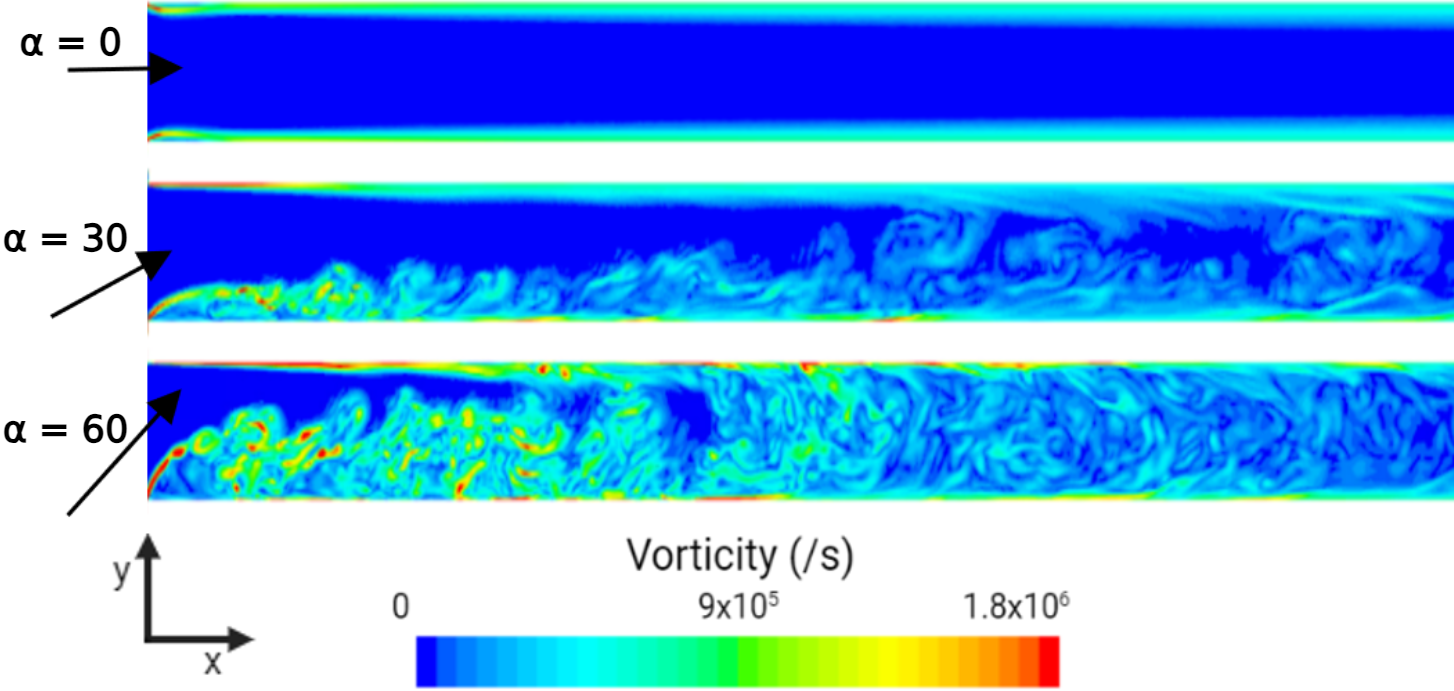 }
    \caption{Instantaneous vorticity vector magnitude at the entrance of the channel (XY plane, z= 0) for $Re_c$ = 3000 and $\alpha$ = 0$^\circ$, 30$^\circ$ and 60$^\circ$.}
    \label{vorticity}
\end{figure}

Fig.\,\ref{vorticity} shows the vorticity magnitude in the XY plane for three different entry angles. When the entry flow is axial ($\alpha$ = 0$^\circ$), vorticity appears mainly near the boundaries, peaking near the vena contracta but dissipating rapidly downstream of the separation zone, with minimal impact on the flow structure. However, at higher entry angles and Reynolds numbers, vorticity magnitude increases notably within the shearing layer. As a result, vortical structures are formed, shedding, and persisting downstream, promoting mixing, and influencing a transition from laminar to turbulent flow throughout the channel. 

\subsubsection{\label{2ndFlow}Secondary Flow}
 
Formation of secondary vortices in non-circular ducts has been observed in many experimental and numerical studies, \cite{secondary1,secondary2,secondary3} and is attributed to the transverse turbulent stress gradients. These vortices play a significant role in shaping the overall flow structure and enhancing mixing within the channel. \cite{Vidal,particle1} At lower Reynolds numbers, the secondary flows are minor in magnitude compared to the primary flow, offering little contribution to mixing and pressure losses. As the Reynolds number increases, the prominence of corner vortices becomes greater, exerting more of an influence on the mixing dynamics and turbulence generation. \cite{Yao}

\begin{figure}
    \centering
    \includegraphics[scale=0.29]{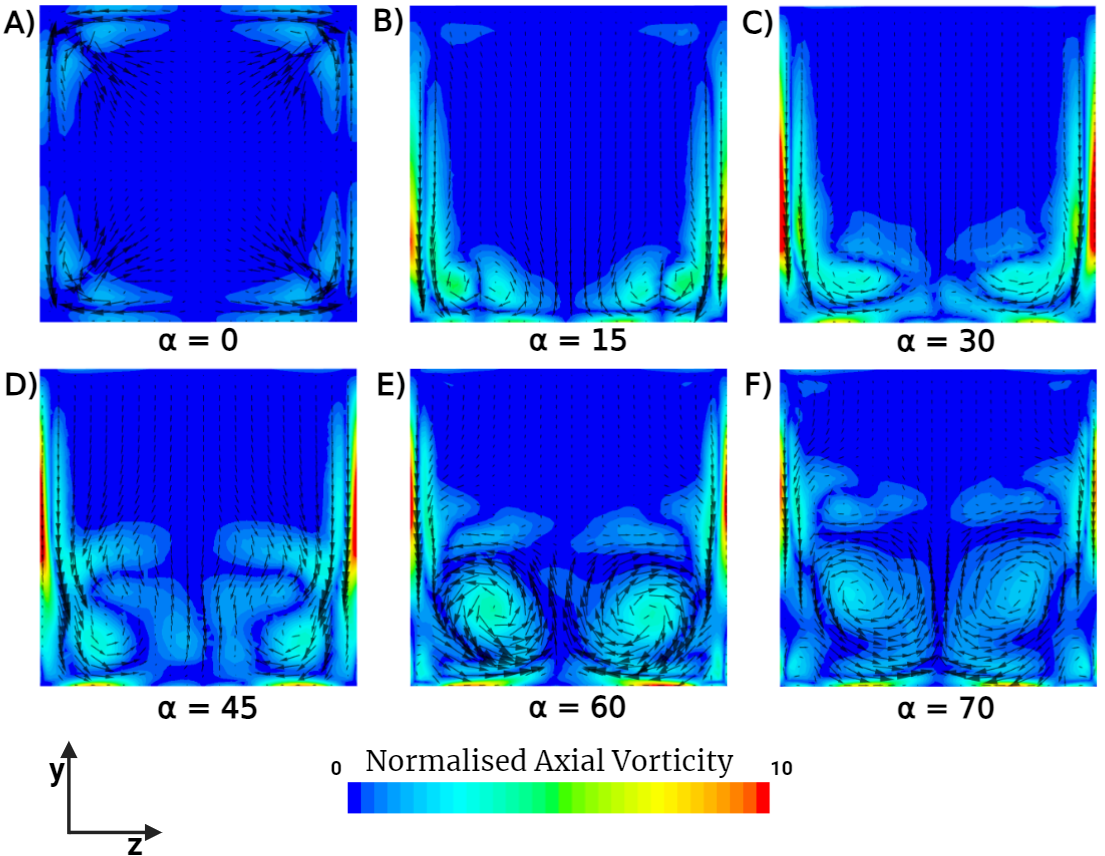 }
    \caption{Time-averaged, non-dimensional axial vorticity 1.6$d$ downstream of the channel entrance (ZY plane, $x$ = 1.6$d$), with superimposed velocity vectors to indicate flow direction in the corners, for $Re_c$ = 2000. A) $\alpha$ = 0$^\circ$, B) $\alpha$ = 15$^\circ$, C) $\alpha$ = 30$^\circ$, D) $\alpha$ = 45$^\circ$, E) $\alpha$ = 60$^\circ$, F) $\alpha$ = 70$^\circ$.}
    \label{cornervorts}
\end{figure}

\begin{figure}
    \centering
    \includegraphics[scale=0.26]{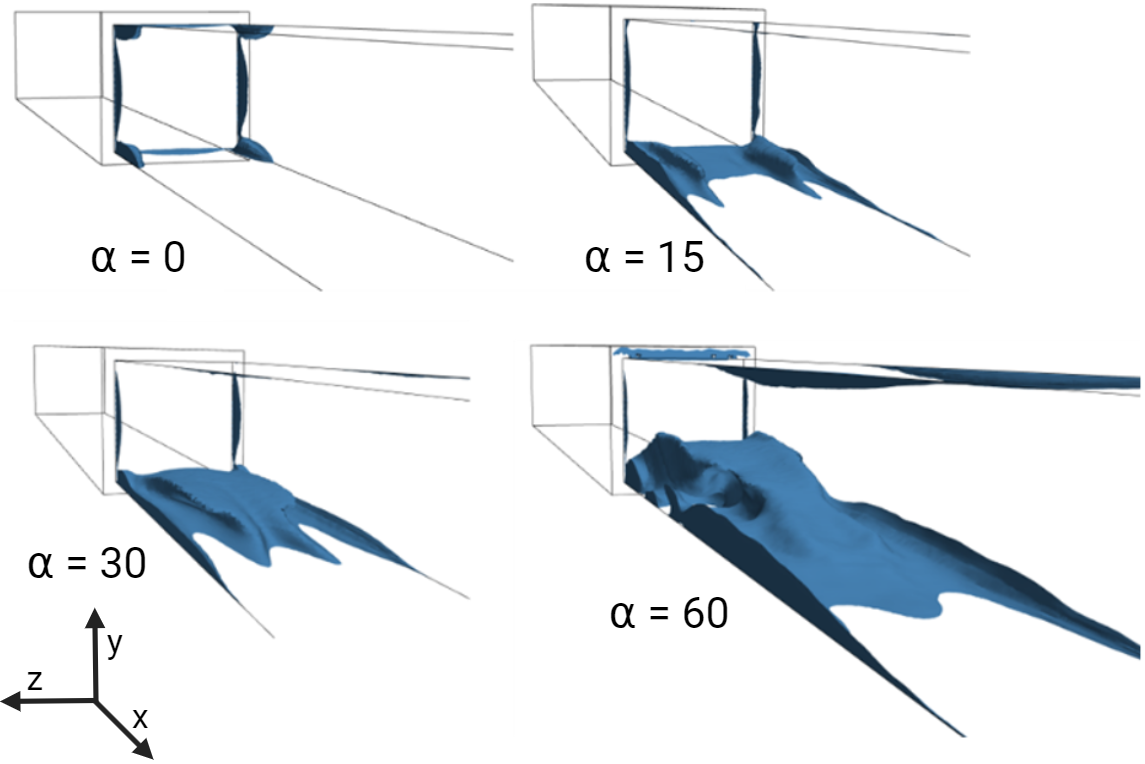 }
    \caption{Isosurfaces of time-averaged, axial velocity, $u$ = -0.1 m/s, depicting the regions with reverse flow for $Re_c$ = 2000, $\alpha$ = 0$^\circ$, 15$^\circ$, 30$^\circ$, and 60$^\circ$.}
    \label{isosurfs}
\end{figure}

Fig.\,\ref{cornervorts} shows the time-averaged axial vorticity non-dimensionalised by $U_1/d$ for cross-sections 1.6$d$ downstream of the channel entrance for various entry angles (within the seperation region). The time-averaged tangential velocity vectors are superimposed on the contour plots to show the direction of the flow in the corners. For axial flow entry, the secondary flow structures in all four corners are mostly identical (Fig.\,\ref{cornervorts}A, $\alpha$ = 0$^\circ$) and agree well with those observed in other studies. \cite{secondary1,secondary3} With an oblique flow entry, the secondary flow in the upper corners disappears and the lower vortices grow in magnitude (Fig.\,\ref{cornervorts}B-F). The lower corner vortices then join the recirculation region with substantial energy, increasing the magnitude of the reverse flow and size of the shearing layers. Subsequently, this secondary flow influences the shape of the recirculation regions, which, rather than a simple recirculation bubble, have a complex but mostly symmetrical shape with respect to the midplane $z$ = 0 (Fig.\,\ref{isosurfs}). 

\subsubsection{\label{exit}Turbulence Development}
\begin{figure}
    \includegraphics[scale=0.575]{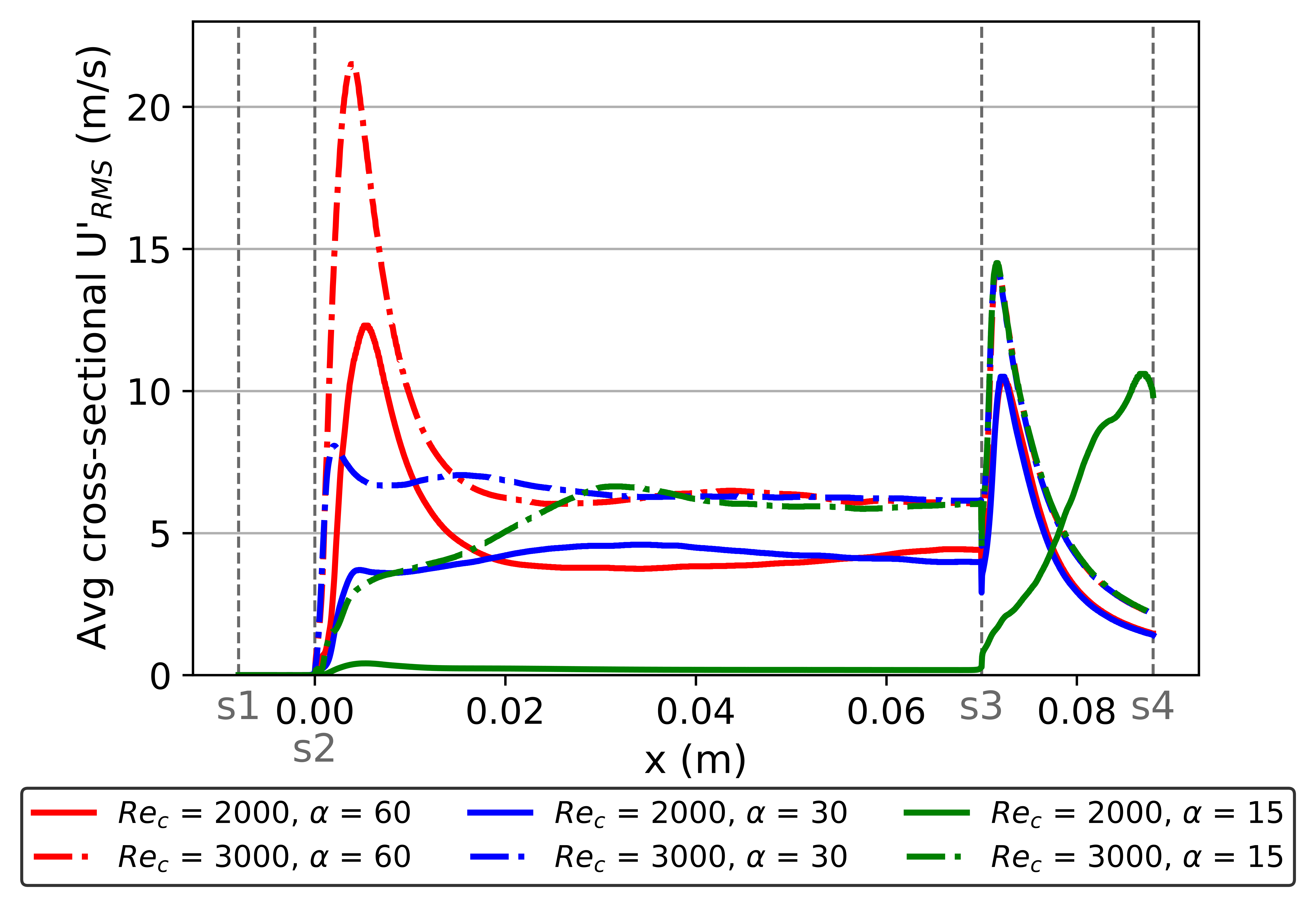 }
    \caption{Distribution of resolved velocity fluctuations averaged in $yz$ cross-sections along the channel for $Re_c$ = 2000 and 3000 and $\alpha$ = 15$^\circ$, 30$^\circ$, 60$^\circ$. The inlet (s1), channel inlet (s2), channel outlet (s3), and outlet (s4) sections are marked by grey vertical lines.}
    \label{Kalong}
\end{figure}

The discussed flow features (flow acceleration due to separation, shear layer, and secondary flow formation) contribute to the transition of the flow to turbulence with increasing Reynolds number and/or oblique entry angle. Fig.\,\ref{Kalong} shows the average cross-sectional distribution of resolved velocity fluctuations, defined as $U'_{RMS} = \sqrt{u'^2 + v'^2 + w'^2}$, along the channel for a range of Reynolds numbers and approach angles. 

For all simulations with axial inflow ($\alpha$ = 0$^\circ$), and oblique inflow simulations with $Re_c$ < 1000, the flow regime is laminar and the recirculation/vena contracta region generates weak fluctuations which dissipate back to zero downstream in the channel. At a certain combination of $Re_c$ and $\alpha$, the velocity fluctuations generated around the recirculation region promote the transition of the flow to turbulence that does not dissipate after the recirculation region disappears. For example, at $Re_c = 2000$ and $\alpha$ = 30$^\circ$, the velocity fluctuation magnitude peaks in the separation region then remains around 5 m/s throughout the channel. Despite no turbulence being seeded at the inlet, the shear layer in the separation region promotes the development of velocity fluctuations within the channel. As expected, increasing $Re_c$ results in higher turbulence at the outlet of the channel.

The velocity fluctuations appearing near the recirculation region, seen in most simulations, tend, mostly, to be partially preserved downstream. This suggests that while the turbulence observed in some simulations is triggered within the recirculation region, it is sustained as the flow develops along the channel, even for moderate $Re_c$ = 2000. At the outlet, between s3 – s4, the increased turbulence levels are due to flow expansion and mixing of jets exiting adjacent channels, but this has a negligible effect on the oblique losses considered here. 

LES provides valuable information on turbulence intensity and isotropy within the monolith channels and at their exit. This information is useful for modelling other processes such as reaction kinetics and particulate collection. Understanding the outlet conditions is important for optimising the geometric design of relevant devices and their downstream components. In more complex, multi-component systems, these outlet conditions can serve as inlet conditions, facilitating better integration and performance across the system. 

The findings suggest that for applications requiring improved mixing, using an oblique entry angle may be beneficial, as it promotes enhanced turbulent mixing. 

\subsection{\label{PressureDrop}Pressure Drop}
For all simulations, the static gauge pressure throughout the domain is time-averaged for two residence times of the full length of the domain. This pressure is then averaged in cross-sections perpendicular to the channel at 0.05$d$ intervals along the whole domain, and the value at $x$ = $L$ is subtracted as discussed in Sec.\,\ref{effectofoutlet}.

\subsubsection{Pressure Drop for Axial Flow Entry Simulations}

\begin{figure}
\includegraphics[scale=0.57]{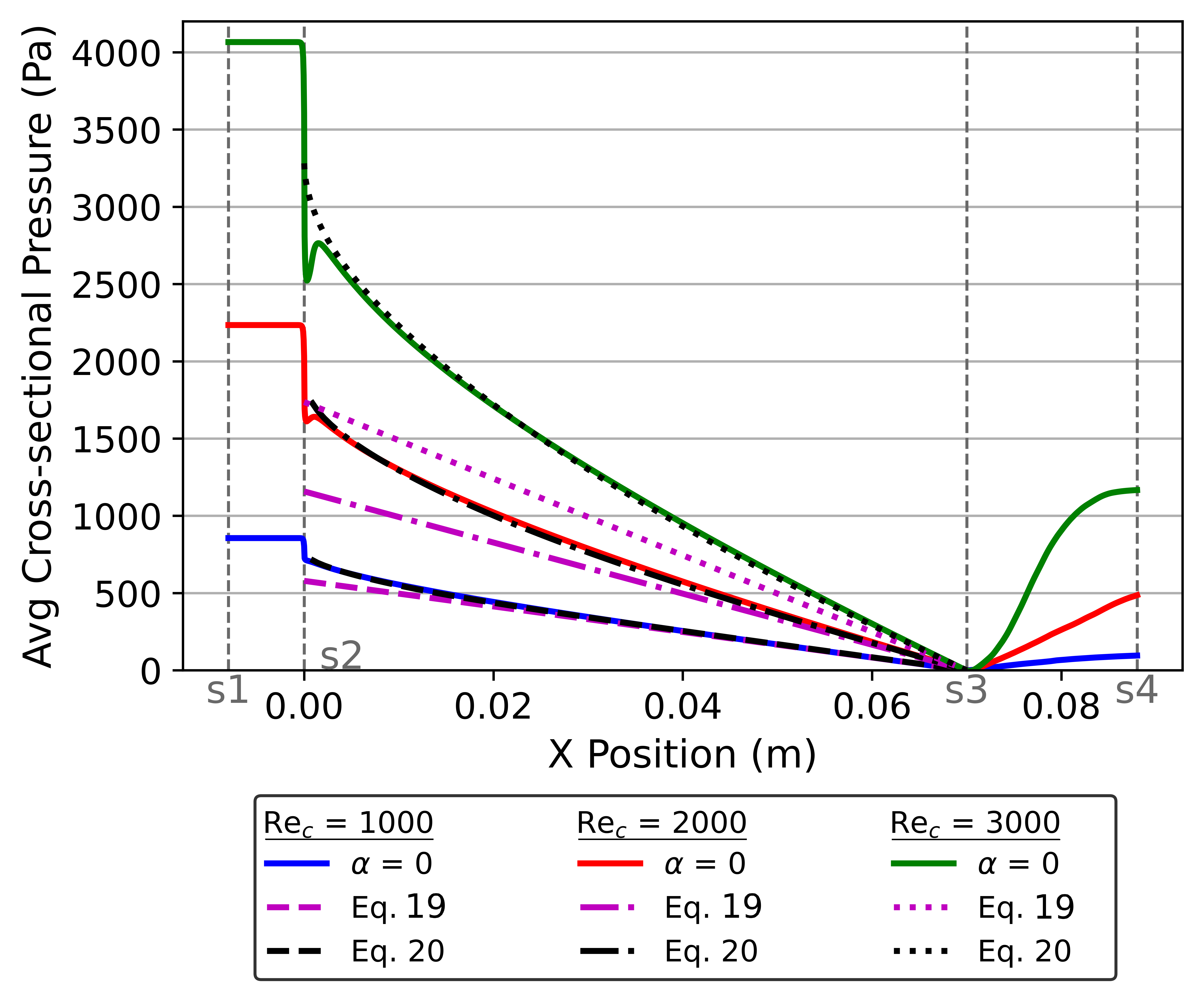 }
\caption{Pressure distribution along the domain for $Re_c$ = 1000, $\alpha$ = 0$^\circ$}
\label{axcombined}
\end{figure}

Fig.\,\ref{axcombined} shows the pressure distribution in the domain for axial entry ($\alpha$ = 0$^\circ$) simulations with $Re_c$ = 1000, 2000, and 3000. To assess the importance of the boundary layer development, and the effects of the oblique entry, the results are compared with the established correlations for frictional losses in a square channel with laminar and turbulent flow. For laminar flow, the fully developed, frictional losses are estimated from the Darcy-Weisbach equation \cite{DarcyWeisFric}: 

\begin{equation}
\Delta P_{Darcy} = f \cdot \bigg(\frac{L}{d}\bigg) \cdot \bigg( \frac{\rho U_c^2}{2}\bigg)
\label{darcyforoblique}
\end{equation}

\noindent where, $L$ is the length of the channel, $d$ is the hydraulic diameter/width of the channel, $U_c$ is the axial velocity in the channel, $\rho$ is the density of air (here $\rho$ = 1.1842 kg/m$^3$), and $f$, is the Darcy friction factor. For fully developed flow in a square channel \cite{OCJones} the friction factor is defined as:

\begin{equation}
    f = \frac{56.908}{Re_c}
    \label{fsq}
\end{equation}

The channel length considered here is generally shorter than the development length for laminar flow, therefore it is not expected that the pressure will have a constant gradient in the axial direction. Shah and London's \cite{ShahLondon} correlation calculates a modified friction factor, $f_{SL}$, to account for laminar developing flow:

\begin{equation}
    f_{SL}Re_c(x^+) =4\bigg( \frac{3.44}{\sqrt{x^+}}+\frac{f \cdot Re_c + \frac{K_\infty}{4x^+}-\frac{3.44}{\sqrt{x^+}}}{1+Cx^{+^{-2}}} \bigg)
\label{shahoblique}
\end{equation}

\noindent where, $x^+$ is the dimensionless axial distance from the entrance in the direction of the flow defined as:

\begin{equation}
    x^+ = \frac{x}{dRe_c}
\end{equation}

The coefficients, $C$ and $K_{\infty}$, were found empirically by Shah \& London. \cite{ShahLondon} For a square cross-section channel $C$ = 0.00029 and $K_{\infty}$ = 1.43. The multiplication by 4 in Eq. \ref{shahoblique} is due to the original form of the equation being for the Fanning friction factor which is a quarter of the Darcy friction factor used here.

As expected, for axial entry ($\alpha$ = 0$^\circ$), the fully developed flow correlation underpredicts pressure losses at the entrance to the channel (Fig.\,\ref{axcombined}). The developing flow correlation by Shah \& London \cite{ShahLondon} agrees well with the simulation results. The differences observed near the entrance of the channel are due to the contraction of the flow and the resulting vena contracta. 

\begin{figure}
\includegraphics[scale=0.57]{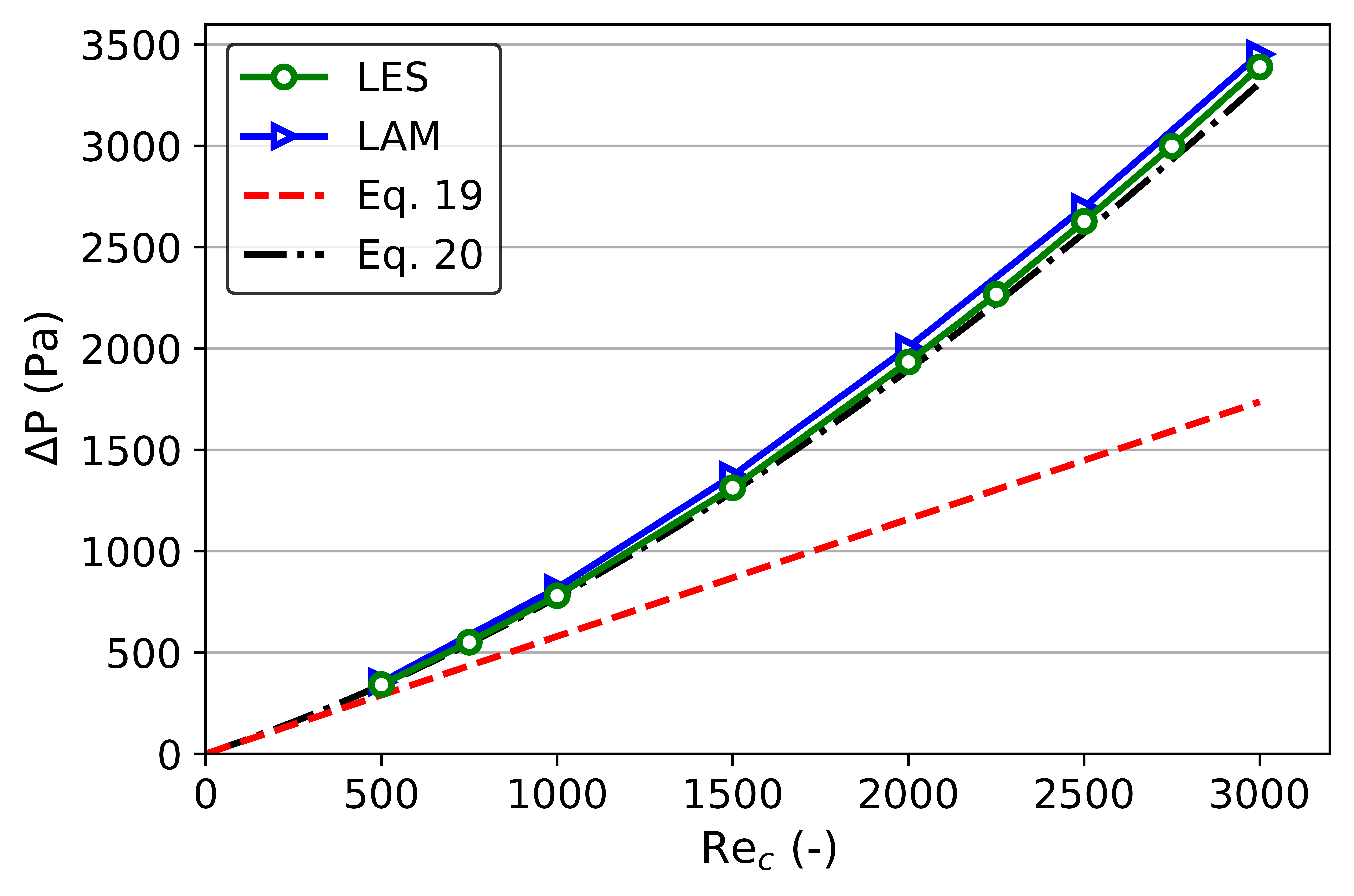 }
\caption{Channel pressure drop, $\Delta P_{(\alpha = 0)}$, vs $Re_c$ for $\alpha$ = 0$^\circ$ (LES stands for Large Eddy Simulation and LAM stands for unsteady laminar flow models).}
\label{axDPvsRe}
\end{figure}

Agreement with Shah \& London's \cite{ShahLondon} correlation is confirmed by comparing the pressure drop in the whole channel (between cross-sections s1 and s3) shown in Fig.\,\ref{axDPvsRe}. As it was evident that the flow was laminar in these cases, a laminar flow model was also used to assess differences between LES and laminar flow models. The difference between unsteady laminar and LES simulations is minor, and both are in good agreement with the developing flow losses in a channel (Eq.\,\ref{shahoblique}) suggesting the flow, even at $Re_c$ = 3000, is laminar and developing for all axial entry simulations (Fig.\,\ref{axDPvsRe}). 

\subsubsection{Pressure Drop for Oblique Flow Entry Simulations}

\begin{figure}
\includegraphics[scale=0.57]{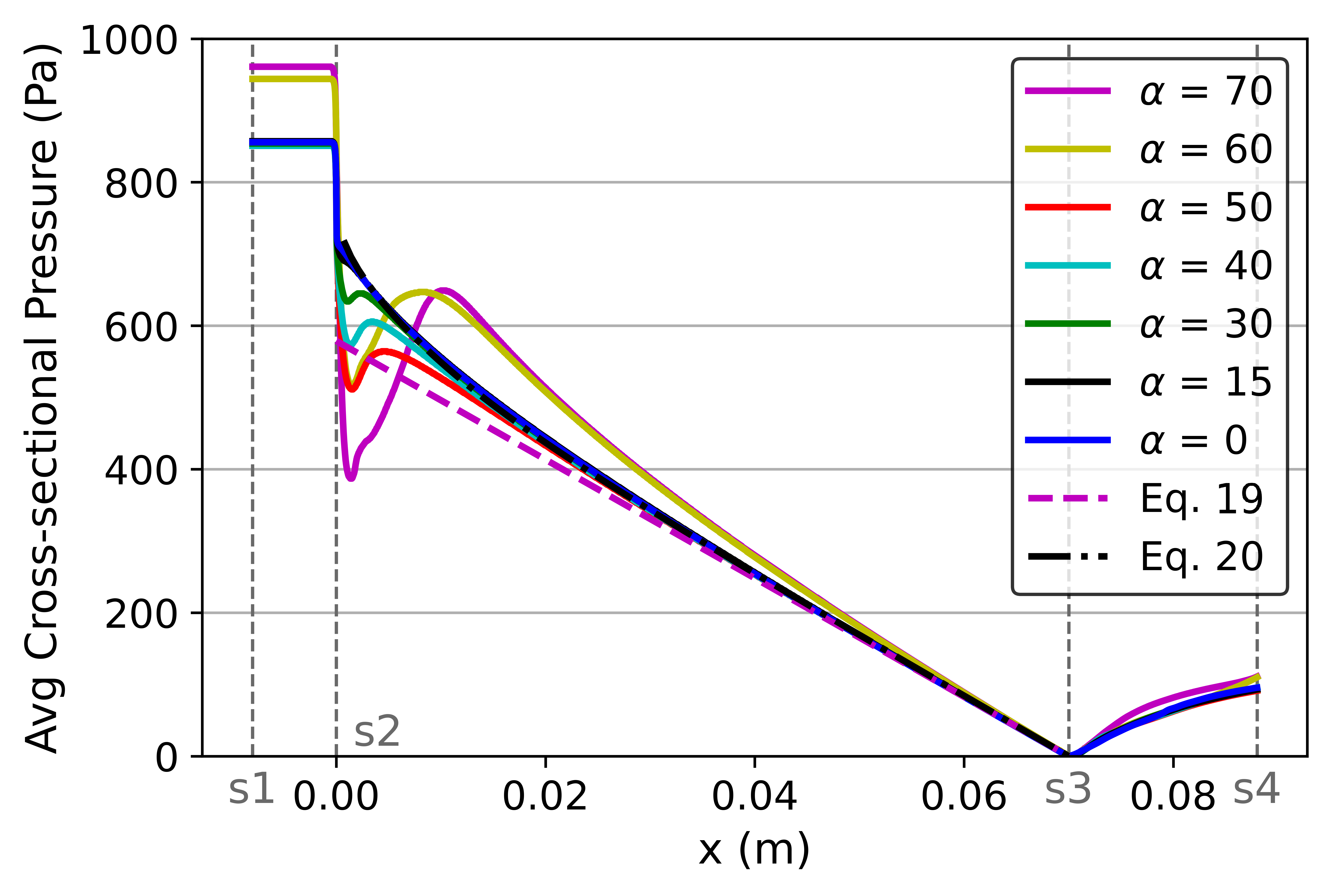 }
\caption{Pressure distribution along the domain for $Re_c$ = 1000, $\alpha$ = 0$^\circ$ - 70$^\circ$.}
\label{oblique1000}
\end{figure}

\begin{figure}
\includegraphics[scale=0.57]{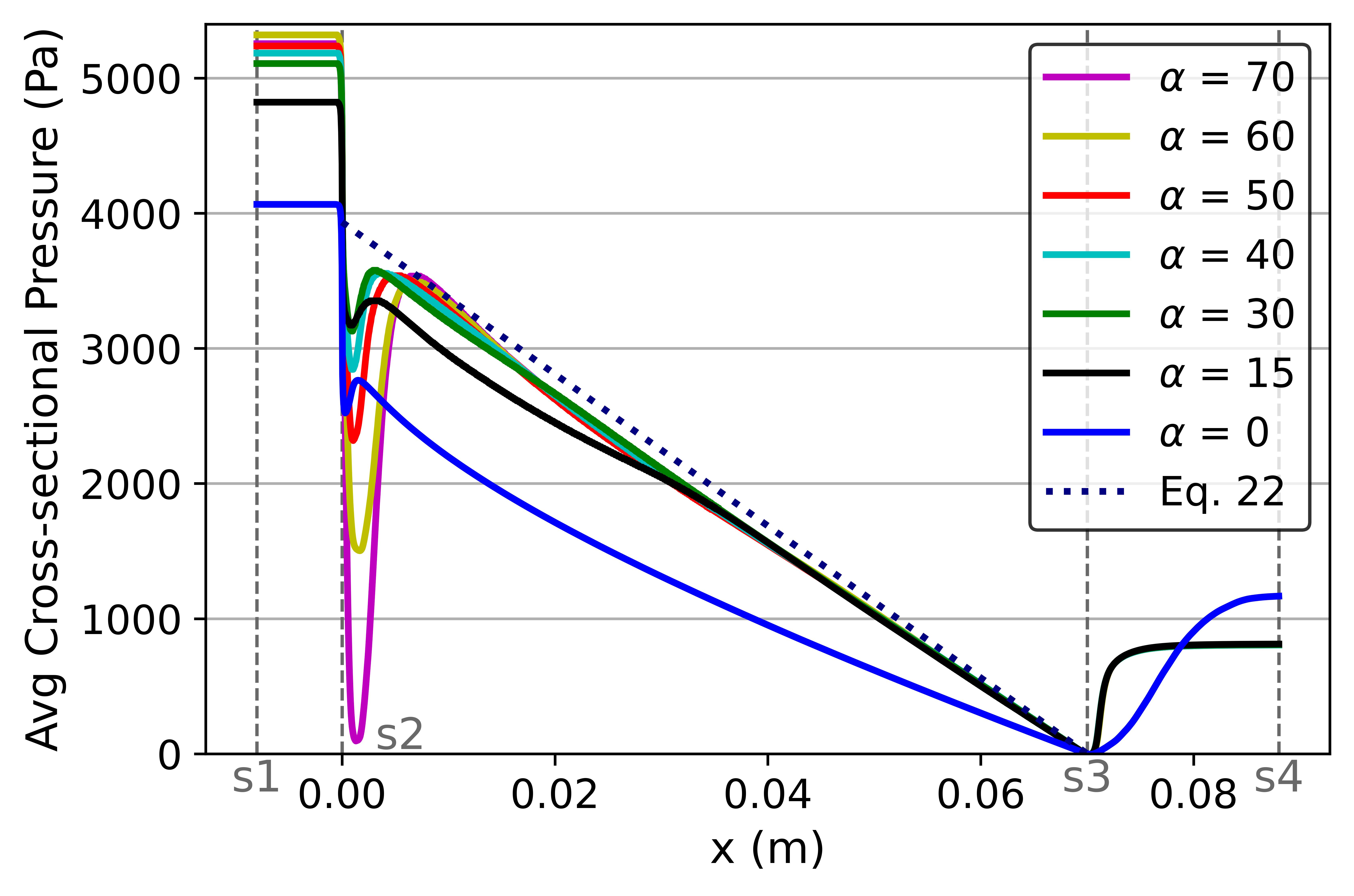 }
\caption{Pressure distribution along the domain for $Re_c$ = 3000, $\alpha$ = 0$^\circ$ - 70$^\circ$.}
\label{oblique3000}
\end{figure}

Figs.\,\ref{oblique1000} and\,\ref{oblique3000} show the pressure distribution along the channel for cases with an oblique flow entry for $Re_c$ = 1000 and 3000. For lower angles, the cross-sectional average pressure drops at the entrance to the channel, but recovers from the contraction and separation of the flow back to the developing flow frictional loss trend observed for axial ($\alpha$ = 0$^\circ$) cases. At a certain angle, depending on the Reynolds number, the pressure distribution along the whole channel changes, with its gradient increasing. This change coincides with the observed transition from developing laminar to turbulent flow (discussed in Sec.\,\ref{FlowStructure}).

In Fig.\,\ref{oblique3000}, frictional losses for fully developed turbulent flow using a turbulent friction factor, derived from Colebrook's equation, \cite{Colebrook} in Eq.\,\ref{darcyforoblique}, have been added:

\begin{equation}
    \frac{1}{\sqrt{f_{\text{Colebrook}}}} = -2 \log_{10} \left( \frac{\epsilon}{d \cdot 3.7} + \frac{2.51}{\text{Re}_c \cdot \sqrt{f_{\text{Colebrook}}}} \right)
\label{eq:colebrook}
\end{equation}

\noindent Here, $\epsilon$ is the roughness of the channel walls ($\epsilon$ = 1 for smooth channel) and $f$ is the Darcy friction factor (Eq. \ref{fsq}).

A reasonable agreement between simulations and fully developed turbulent flow predictions for $Re_c$ = 3000 and $\alpha$ > 0$^\circ$, indicates that the Darcy-Weisbach equation, with a friction factor estimated from Eq. \ref{eq:colebrook}, provides an adequate description of the flow dynamics for the majority of the channel in these cases. 

\begin{figure}
\includegraphics[scale=0.57]{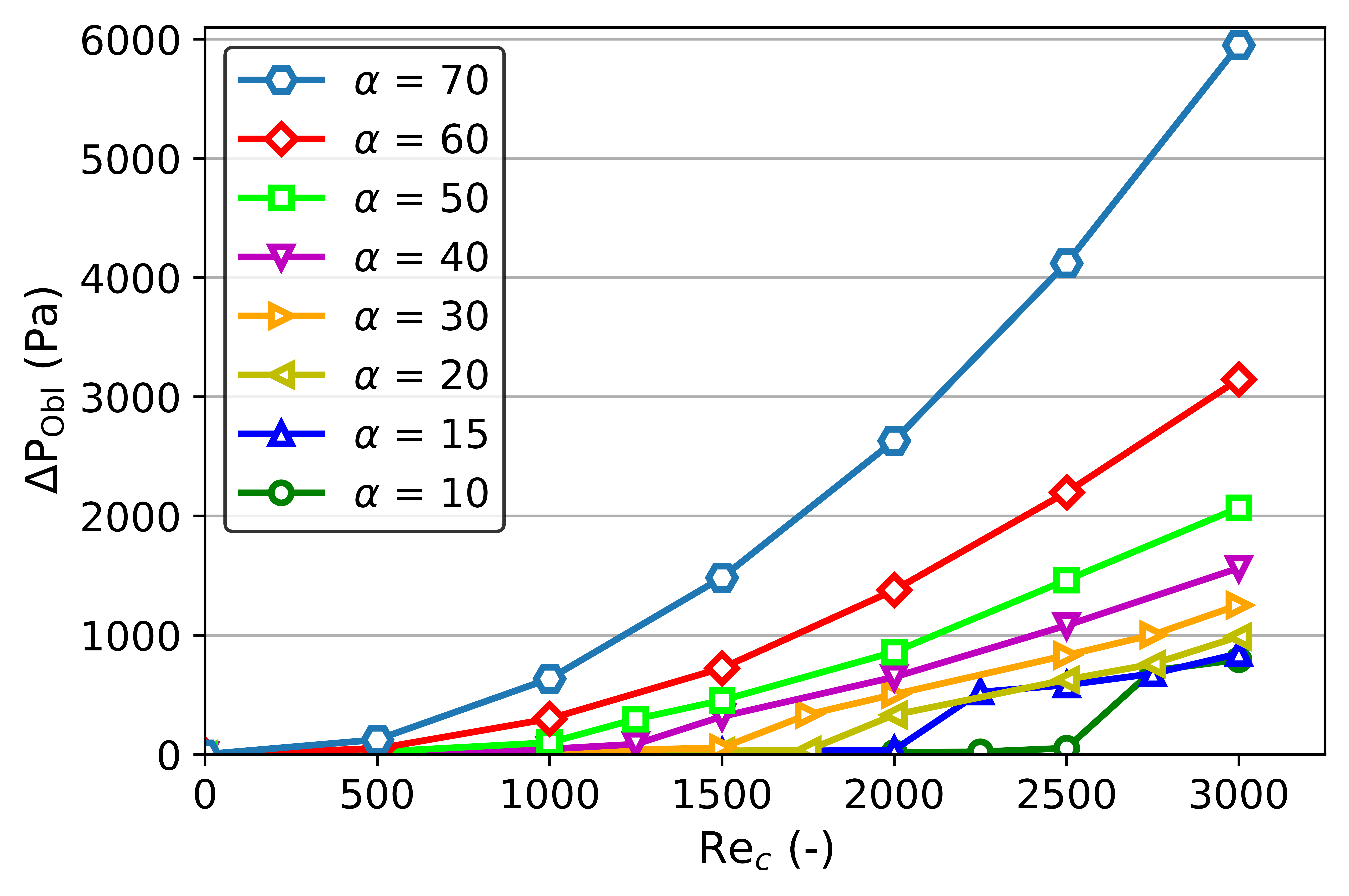 }
\caption{Oblique pressure loss vs $Re_c$ for various entry angles. ($L$ = 70$d$)}
\label{PoblvsRe}
\end{figure}

\begin{figure}
\includegraphics[scale=0.57]{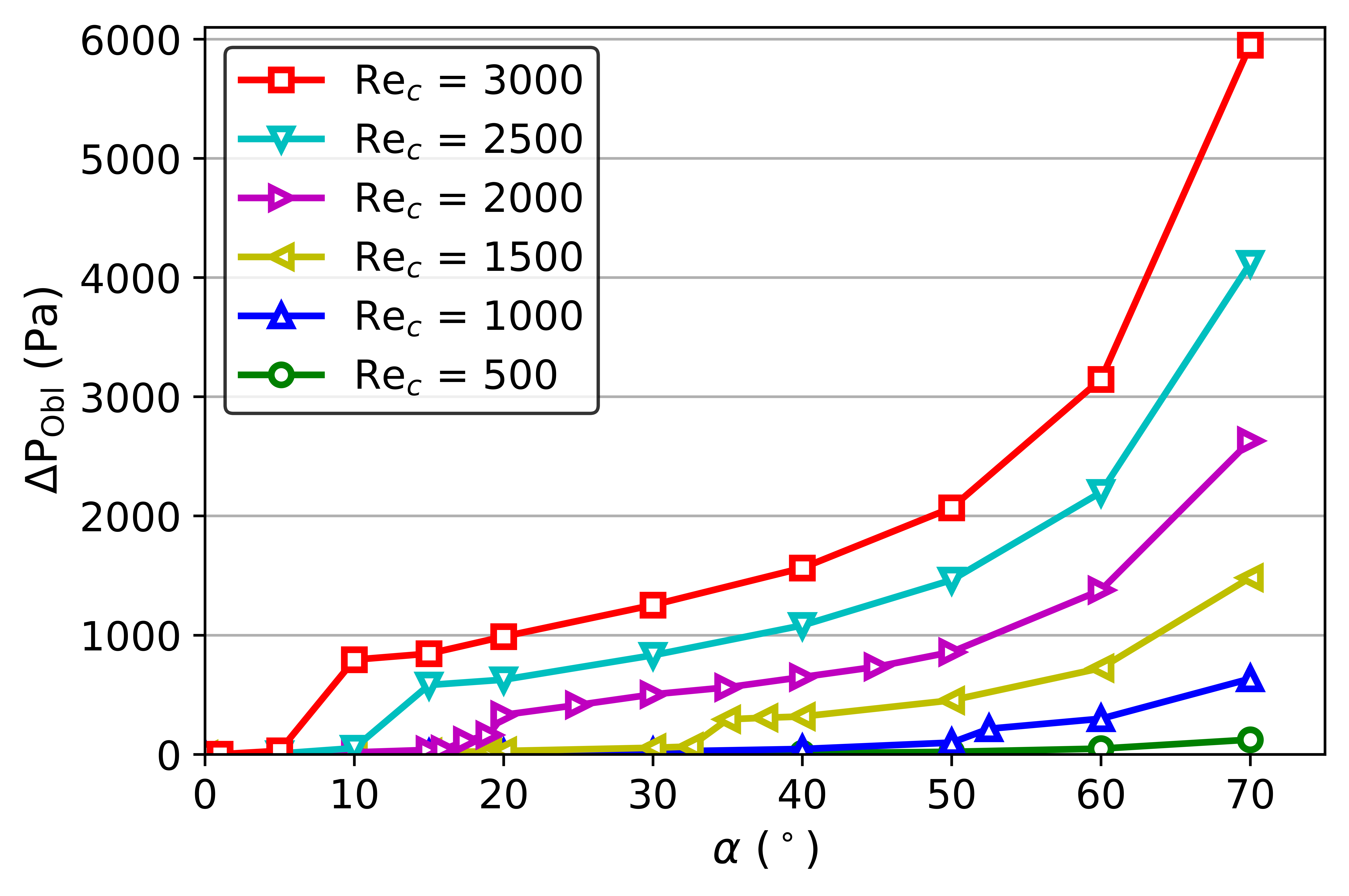 }
\caption{Oblique pressure loss vs entry angle for various channel Reynolds numbers. ($L$ = 70$d$)}
\label{PoblvsAngle}
\end{figure}

In order to assess the contribution of the losses associated with the oblique flow entry, the oblique pressure drop given by Eq.\,\ref{Eq:PObl} is plotted in Figs.\,\ref{PoblvsRe} and\,\ref{PoblvsAngle}. Fig.\,\ref{PoblvsRe} shows the dimensional oblique pressure drop, $\Delta P_{Obl}$ (Eq.\,\ref{DPoblique}), against channel Reynolds number for various entry angles. For low angles and low Reynolds numbers $\Delta P_{Obl}$ is small. At a certain Reynolds number and angle, there is a shift in $\Delta P_{Obl}$. For example, for $\alpha$ = 15$^\circ$, $\Delta P_{Obl}$ remains small until around $Re_c$ = 2000 where the oblique pressure drop contribution becomes significant, exceeding 500 Pa (Fig.\,\ref{PoblvsRe}). For larger angles, e.g. $\alpha$ = 60$^\circ$, the relationship between the Reynolds number and oblique pressure loss is almost parabolic suggesting the oblique losses are significant even at relatively low Reynolds numbers. The same trend is visible when plotting $\Delta P_{Obl}$ against the entry angle (Fig.\,\ref{PoblvsAngle}). For $Re_c$ = 2000, there is a sharp increase in $\Delta P_{Obl}$ between $\alpha$ = 17.5$^\circ$ and $\alpha$ = 20$^\circ$. 

The increase in oblique pressure losses at higher angles and Reynolds numbers is primarily due to increased losses from the transition to turbulence. The definition of $\Delta P_{Obl}$ (Eq.\,\ref{Eq:PObl}) assumes that the flow regime is the same for both the oblique entry case and when $\alpha$ = 0$^\circ$. As a result, friction losses downstream of the inlet are assumed to be cancelled, isolating the contribution from the oblique entry. This assumption does not hold for higher angles and Reynolds numbers, where axial entry flow remains laminar, while oblique entry flow may become transitional or turbulent. Consequently, the assumption made in previous studies, \cite{Quadri,Kamal,Persoons,KW} that $K_{Obl}$ is independent of channel length, is incorrect in such cases, as $K_{Obl}$ also includes the difference between turbulent and laminar friction losses along the entire channel length.

\subsubsection{Comparison with Experimental Data and Existing Correlations}
\label{vsExp}

The results can be compared with correlations and experimental data that exist within the literature. Four common correlations used are detailed in Table\,\ref{table:corrs}. Values of $A$ and $n$ for the correlation by Quadri et al. \cite{Quadri} are shown in Eq. \ref{Eq:Quadri}.

\begin{table}
\caption{\label{table:corrs} $K_{Obl}$ correlations available in the literature for comparison.}
\begin{ruledtabular}
\begin{tabular}{ccccc}
    Source & \makecell{Correlation \\ Definition} & \makecell{Angles \\ ($^\circ$)} & \makecell{Flow \\ Range} & Application \\
\hline
    \makecell{Kuchemann \& \\ Weber \cite{KW}} & $\text{sin}^2\alpha$ & 0 - 90 & Laminar & \makecell{Heat \\ Exchanger} \\
    Persoons et al.\cite{Persoons} & $0.459\text{sin}^2\alpha$ & 0 - 33 & Laminar & \makecell{Catalyst \\ Substrate}\\
    Quadri et al.\cite{Quadri} & $ARe_a^{n(\alpha)}\text{sin}^2\alpha$ & 0 - 90  & Full Range& \makecell{Catalyst \\ Substrate}\\
\end{tabular}
\end{ruledtabular}
\end{table}

\begin{figure}
\includegraphics[scale=0.575]{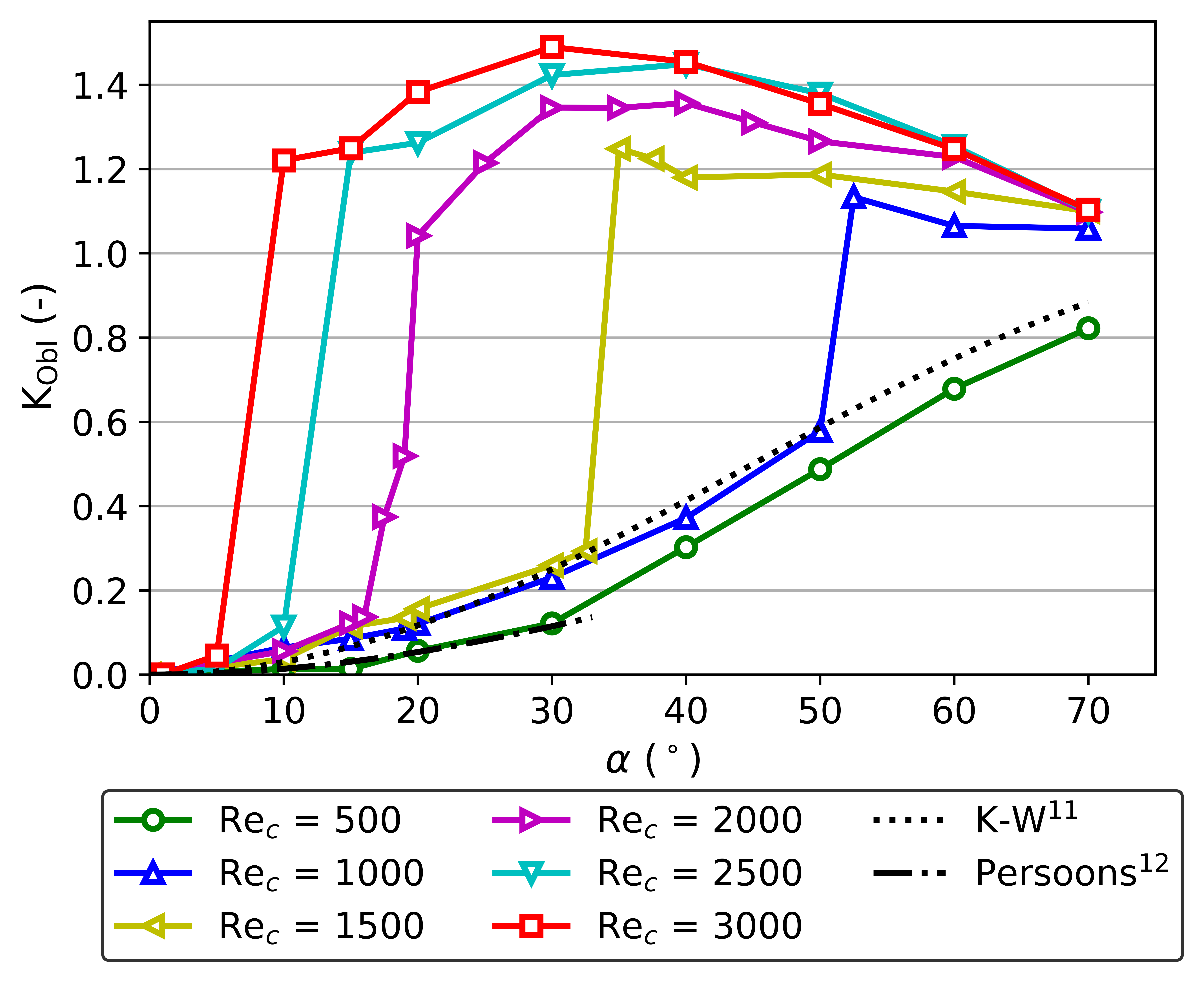 }
\caption{$K_{Obl}$ vs entry angle compared with $\alpha$ dependent correlations. ($L$ = 70$d$)}
\label{KoblKW}
\end{figure}

The results in Fig.\,\ref{KoblKW} show good agreement between predicted non-dimensional pressure loss and the Kuchemann \& Weber (K-W) correlation for most Reynolds numbers for lower values of $\alpha$. However, as the oblique entry angle increases, a pressure drop shift occurs, signalling a transition from laminar to turbulent flow. For example, for $Re_c$ = 1000, up until around $\alpha$ = 50$^\circ$, the simulation results agree well with the K-W correlation. 

The correlation proposed by Persoons et al., \cite{Persoons} which is almost half of the K-W correlation (0.459\(\text{sin}^2 \alpha\)), mostly underpredicts the losses observed in the simulations, apart from $Re_c$ = 500, where there is a good agreement. However, the authors only developed their correlation up to $\alpha$ = 33$^\circ$, so it is not plotted beyond this angle. It is possible that the correlation by Persoons et al. \cite{Persoons} captures the magnitude well for low angles because it considers reduced energy dissipation that aligns better with the viscous-dominated low Reynolds number flow. As the Reynolds number increases to $Re_c$ = 1000, inertial effects start to play a more significant role, where the K-W correlation is more appropriate.

The LES simulations (Fig.\,\ref{KoblKW}) also demonstrate that $K_{Obl}$ changes with the Reynolds number and the assumption that $K_{Obl}$ does not depend on Reynolds number is not valid. This Reynolds number dependency of $K_{Obl}$ has been observed experimentally \cite{Quadri,Kamal}. 

\begin{figure}
\includegraphics[scale=0.575]{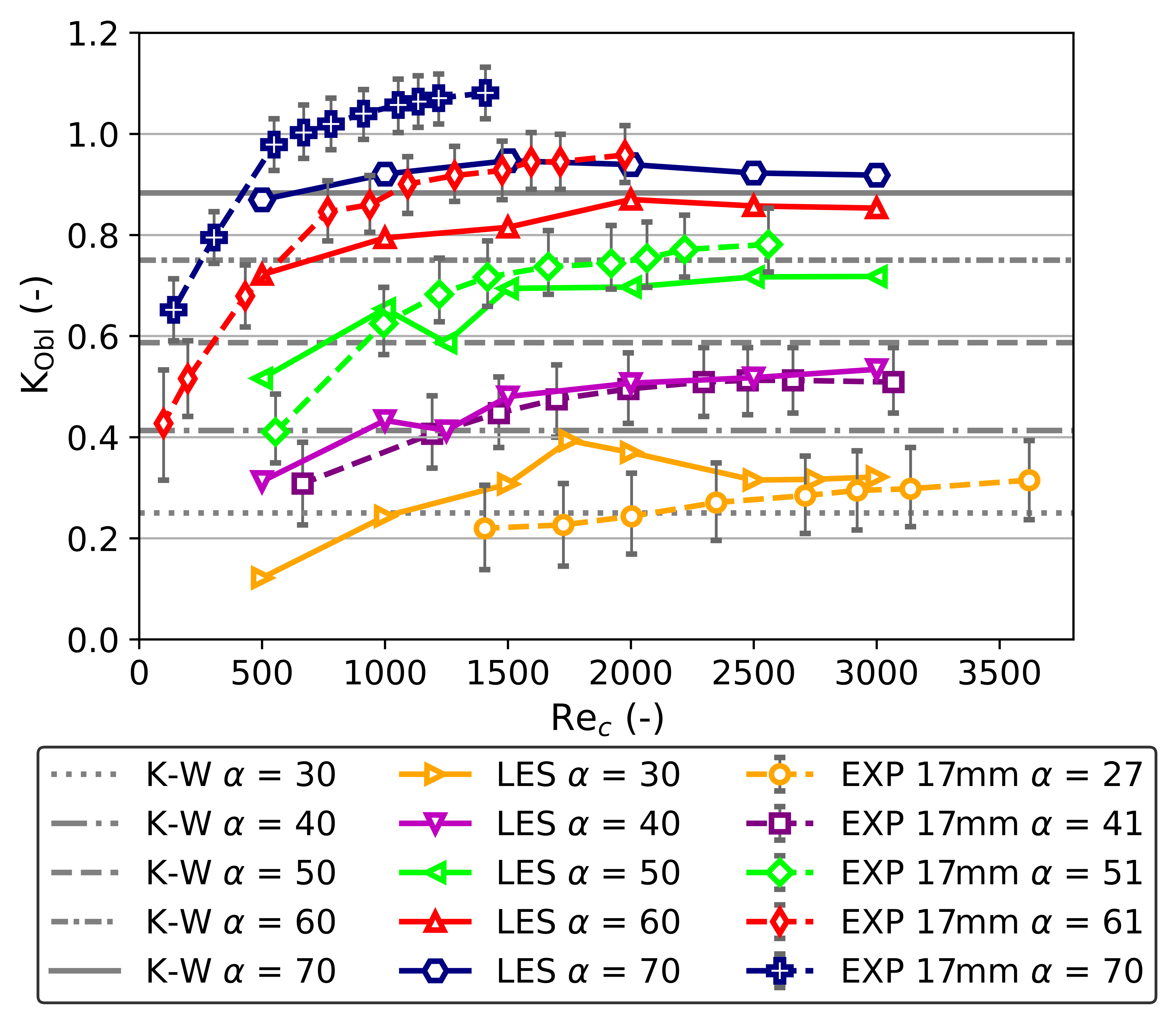}
\caption{$K_{Obl}$ vs $Re_c$ compared with K-W correlation (Eq.\,\ref{Eq:KW}) and the experimental results from Mat Yamin \cite{Kamal} using a 17 mm monolith ($L$ = 17$d$).}
\label{vsKamal17mm}
\end{figure}

\begin{figure}
\includegraphics[scale=0.575]{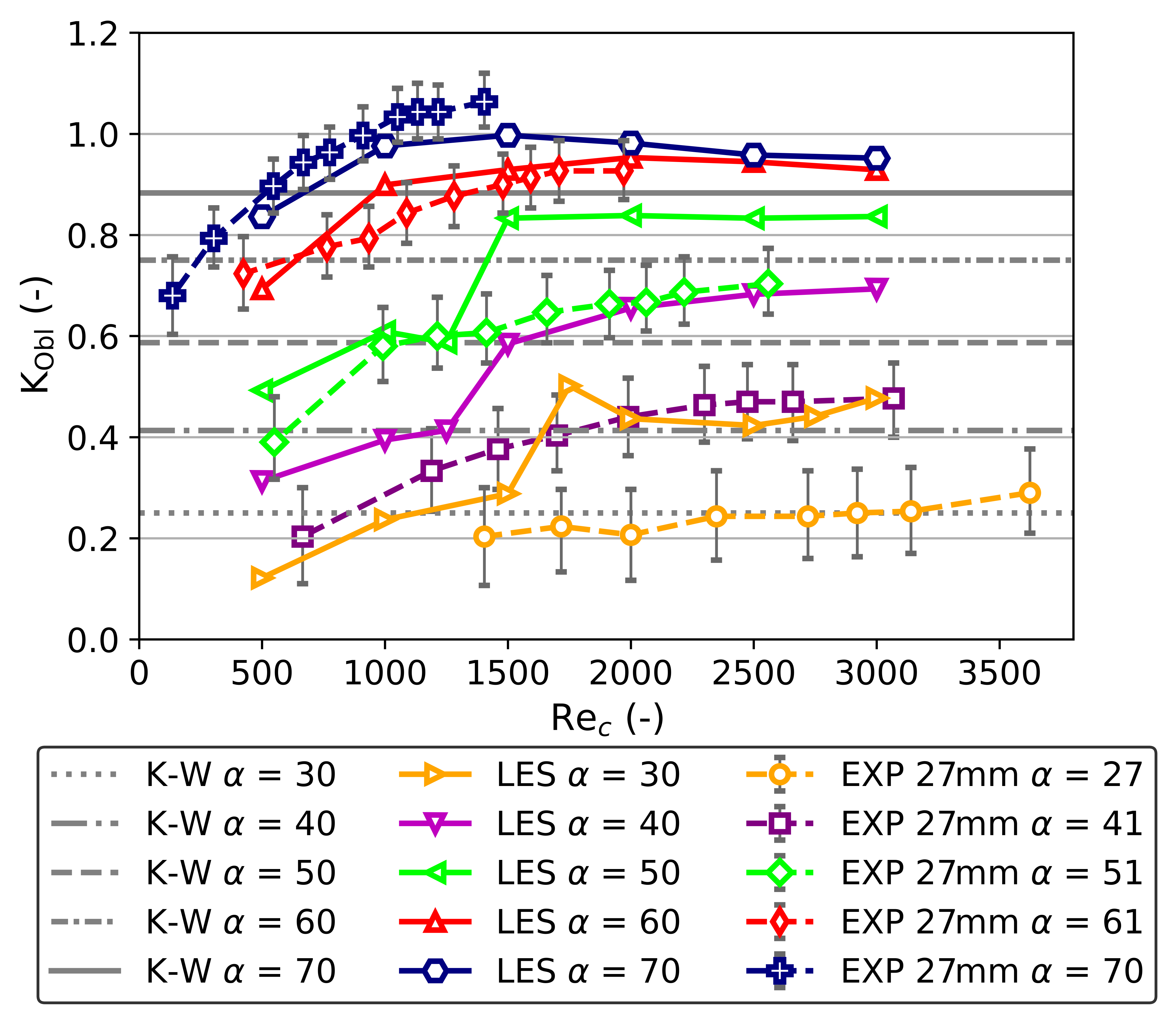 }
\caption{$K_{Obl}$ vs $Re_c$ compared with K-W correlation (Eq.\,\ref{Eq:KW}) and the experimental results from Mat Yamin \cite{Kamal} using a 27 mm monolith ($L$ = 27$d$).}
\label{vsKamal27mm}
\end{figure}

\begin{figure}
\includegraphics[scale=0.575]{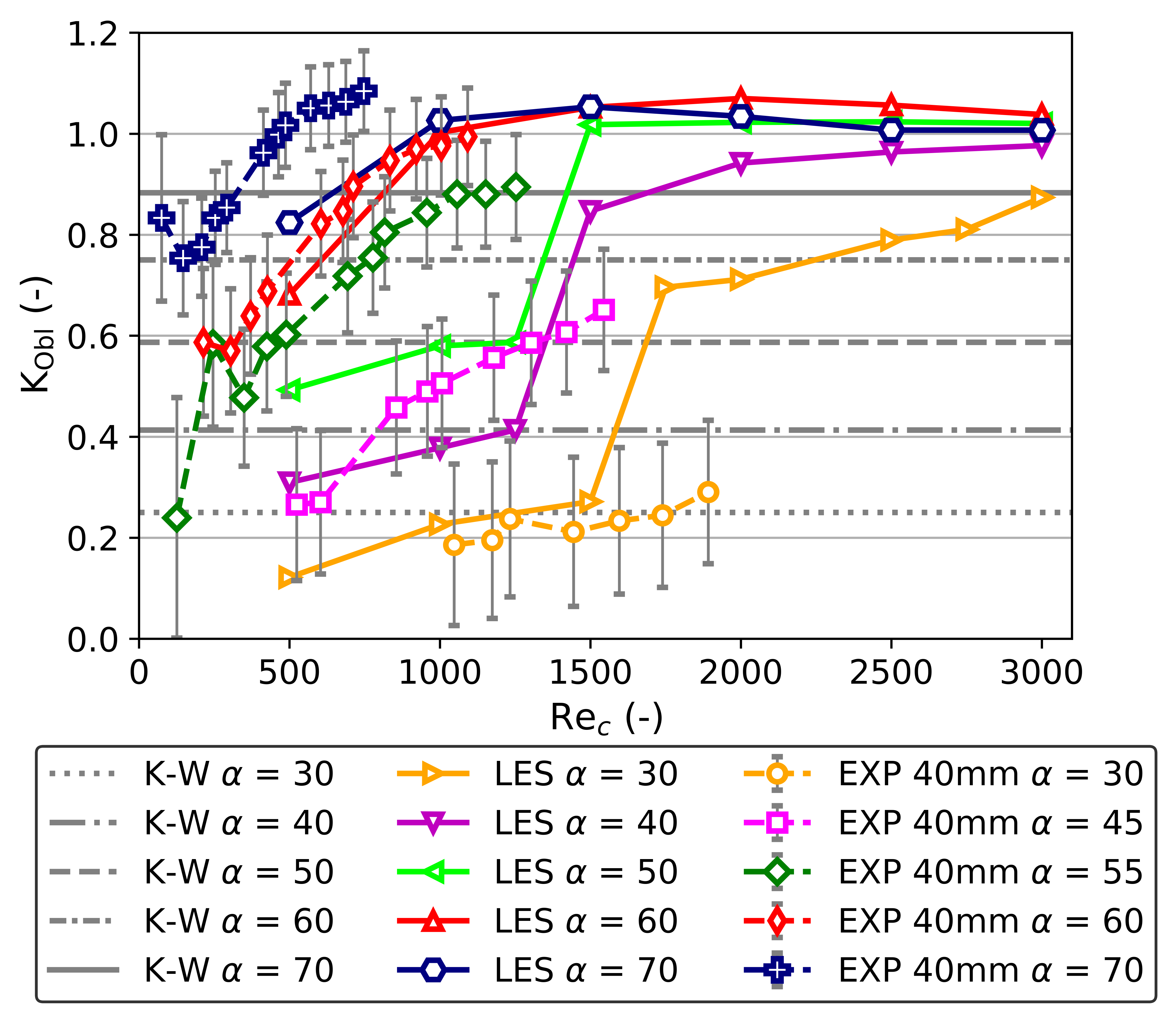 }
    \caption{$K_{Obl}$ vs $Re_c$ compared with K-W correlation (Eq.\,\ref{Eq:KW}) and the experimental results from Quadri et al. \cite{Quadri} using a 40 mm monolith ($L$ = 40$d$).}
    \label{vsQuad40mm}
\end{figure}

\begin{figure}
    \includegraphics[scale=0.575]{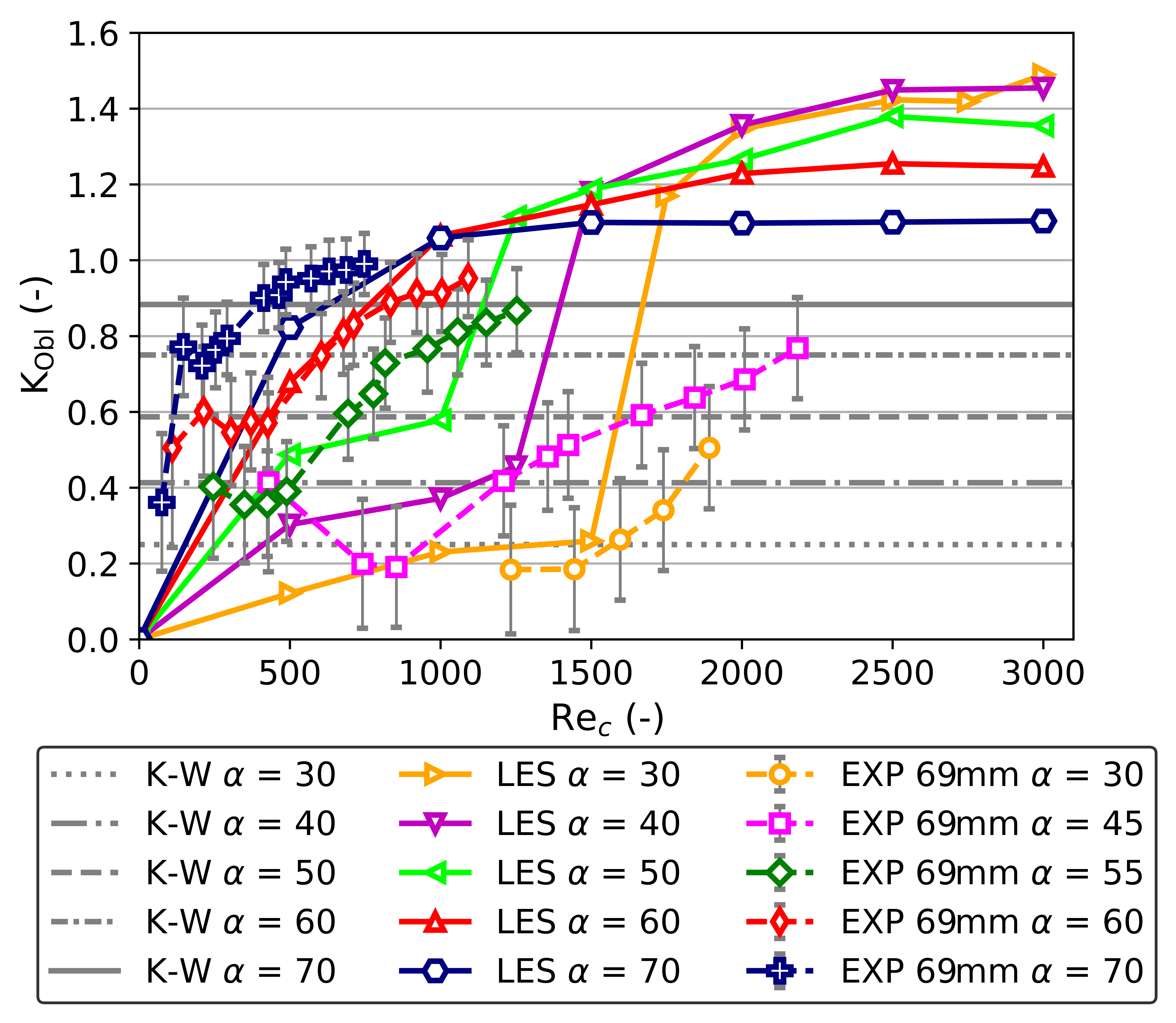 }
    \caption{$K_{Obl}$ vs $Re_c$ compared with K-W correlation (Eq.\,\ref{Eq:KW}) and the experimental results from Quadri et al. \cite{Quadri} using a 69 mm monolith ($L$ = 69$d$).}
    \label{vsQuad69mm}
\end{figure}

In the experimental studies by Quadri et al. \cite{Quadri} and Mat Yamin, \cite{Kamal} the parameter $K_{Obl}$ was calculated for monoliths of various lengths, $L$ = 17 mm, 27 mm, 40 mm, and 69 mm and plotted with experimental error bars (Figs.\,\ref{vsKamal17mm} - \ref{vsQuad69mm}). It was demonstrated in Sec. \ref{vsExp} that, contrary to previous assumptions, $K_{Obl}$ is not always independent of the channel length. Therefore, to enable comparison with simulation results, the pressure drop is evaluated between the domain inlet (s1), and a cross-section located a distance $L$ along the channel, rather than between the domain inlet (s1) and the channel outlet (s3) used in previous discussions. 

For shorter monolith lengths, the agreement between the LES simulation results and the experimental data is generally good, with lower angles showing the best alignment. For example, for $L$ = 17 mm (Fig.\,\ref{vsKamal17mm}), the results for $\alpha$ = 30$^\circ$, 40$^\circ$, and 50$^\circ$ mostly fall within the range of experimental error. As the monolith channel length increases, an increase in $K_{Obl}$ values becomes more apparent and the agreement starts to degrade. For $L$ = 27 mm (Fig.\,\ref{vsKamal27mm}), for $\alpha$ = 30$^\circ$, 40$^\circ$, and 50$^\circ$ between $Re_c$ = 1500 and 2000, a jump in $K_{Obl}$ values is observed in the simulations, which is not present in the experimental data. The agreement improves for higher angles, $\alpha$ = 60$^\circ$ and 70$^\circ$.

Quadri et al. \cite{Quadri} tested longer monoliths, as shown in Figs.\,\ref{vsQuad40mm} and\,\ref{vsQuad69mm}, but in a smaller Reynolds number range. They found that for longer monoliths, the measurement error was greater. Fig.\,\ref{vsQuad40mm} shows good agreement for $\alpha$ = 60$^\circ$, although the error bars for the other angles are large, suggesting that while the agreement may be within experimental error, the uncertainty is significant. A sharp increase in $K_{Obl}$ values in the LES simulations is still evident, but the experiments do not extend to sufficiently high Reynolds numbers to observe whether this increase occurs. For $L$ = 69 mm (Fig.\,\ref{vsQuad69mm}), the results show similar trends, with good agreement at higher angles. At lower angles, such as $\alpha$ = 30$^\circ$ and 45$^\circ$, there is an indication of a sharp upward trend in $K_{Obl}$ in the experiments, but it occurs at higher values of the Reynolds number than predicted by the simulations.

The results confirm that $K_{Obl}$ is dependent on the length of the channel. There is a sharp increase in $K_{Obl}$ that coincides with the coherent structure shedding from the shear layer observed in the simulations, as discussed in Section\,\ref{exit}. While this sharp increase occurs at all values of $L$, its contribution is smaller for shorter channels so it does not considerably increase $K_{Obl}$ values. The agreement for shorter channel lengths is promising, showing good alignment with the LES model and demonstrating strong LES performance, but the agreement is worse for longer monoliths. 

The deviation from the experimental results seen as the monolith length increases could be attributed to several factors, such as differences in the monolith geometry and orientation between simulations and experiments, or model limitations in correctly predicting flow instability and transition. Some of these factors are explored in subsequent sections.

\subsubsection{Effect of Turbulence Model}
It has been shown in Sec. \ref{FlowStructure} that for higher values of Reynolds number and $\alpha$, large vortical structures form within the shear layer. This suggests that the shear layer makes a significant contribution to the total pressure loss, in addition to the turbulent friction downstream of the separation region. It can be argued that the transition to turbulence is a singularity caused by using the LES model, and a laminar flow model should be suitable for modelling the flow, and should agree better with the experiments for a larger range of Reynolds numbers. To check this hypothesis, steady and unsteady laminar flow models were used for several combinations of entry angle and Reynolds number. 

For the oblique cases where the flow transition was observed in the LES simulations, laminar models did not accurately capture the physics. For the steady laminar model, this is unsurprising as turbulence is inherently an unsteady phenomenon. The unsteady model showed poor convergence for transitional flow due to instabilities in the shear layer, resulting in either too large or too small structure shedding compared to LES simulations. For example: for $Re_c$  = 2000, $\alpha$ = 30$^\circ$, the unsteady laminar pressure drop was nearly 2.5 times higher than the experimental value.  Without a sufficiently fine grid or sub-grid turbulence model to capture vortex break-up, the pressure losses were not accurately captured. Therefore, while the steady and unsteady laminar models may be suitable for lower angles and Reynolds numbers, they cannot be used for higher entry angles and Reynolds numbers. Due to the poor performance of the laminar flow models, the full results are not presented here. 

\begin{figure}
\includegraphics[scale=0.575]{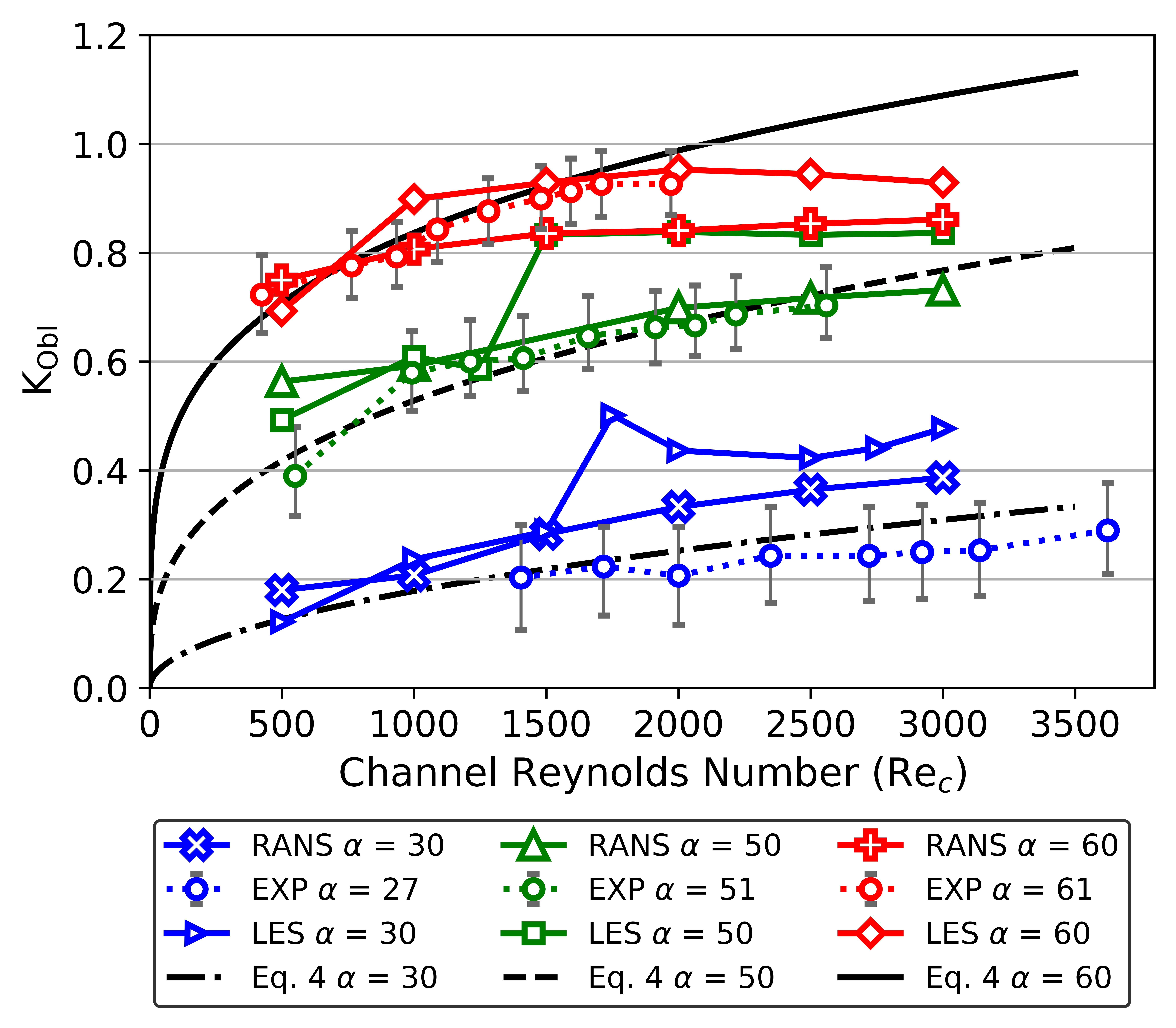 }
\caption{Comparison of LES and RANS flow models with correlation by Quadri et al. \cite{Quadri} (Eq.\,\ref{Eq:Quadri}) and experimental results from Mat Yamin \cite{Kamal} using 27 mm monolith ($L$ = 27$d$)}
\label{LAMvsEXP}
\end{figure}

The suitability of a less computationally expensive RANS model for predicting losses at higher angles and Reynolds numbers was also investigated. A $k-\omega$ SST model, known for its reliable performance in highly separating and transitional flows, \cite{LES2} was selected for this purpose. Fig.\,\ref{LAMvsEXP} presents a comparison between RANS and LES simulations, alongside the correlation proposed by Quadri et al. \cite{Quadri} and experimental results from Mat Yamin. \cite{Kamal}

\begin{figure}
    \includegraphics[scale=0.28]{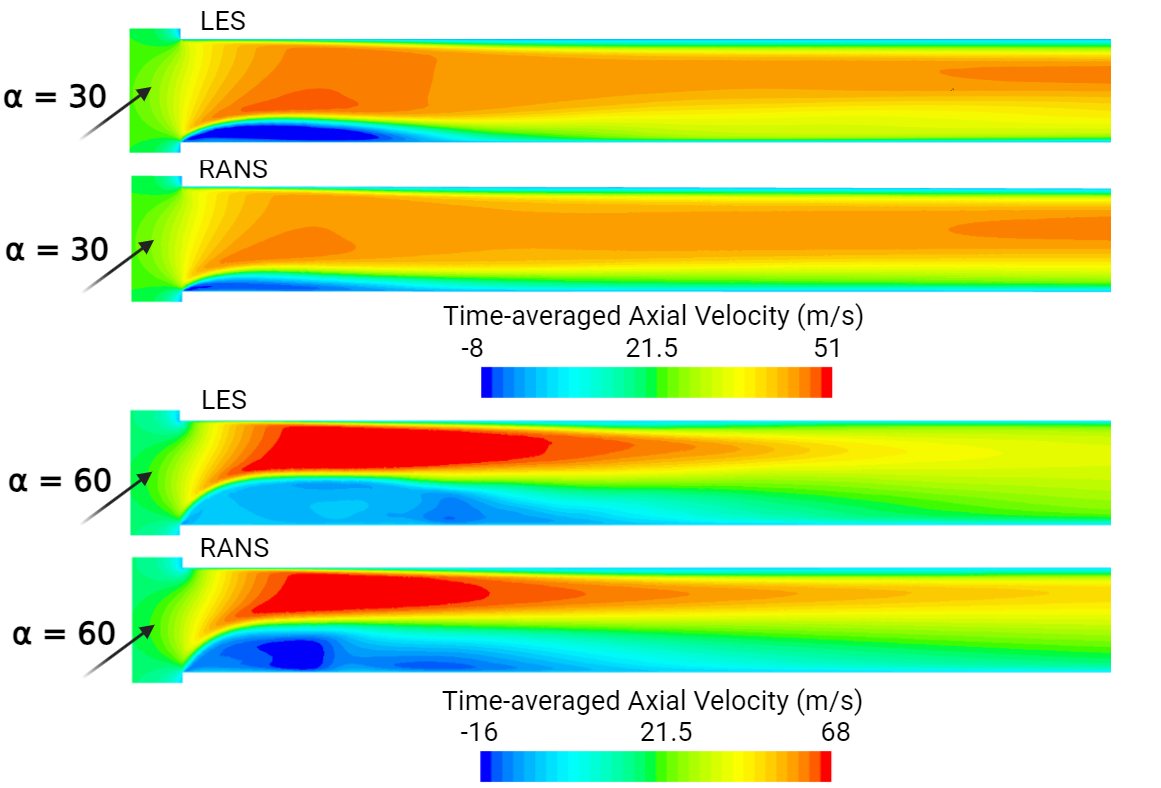 }
    \caption{Time-averaged axial velocity distribution for LES and RANS simulations, $\alpha$ = 30$^\circ$ and 60$^\circ$ (XY Plane).}
    \label{LESvsRANS}
\end{figure}

For $\alpha$ = 30$^\circ$ and 50$^\circ$, RANS simulations are in better agreement with the experiments than LES. The primary losses, mainly due to flow restriction caused by flow separation, are well-captured by the $k-\omega$ SST model, which is finely tuned for such flows. While RANS provides no detailed information about the flow, it accurately captures the separation region structure, delivering pressure drop results closely aligned with experimental data. However, at $\alpha$ = 60$^\circ$, RANS predicts a smaller recirculation region compared to LES (Fig. \ref{LESvsRANS}) and shows less favourable performance against experimental results. 

A more comprehensive analysis of the model differences is outside the scope of this study, but the results indicate that in a large range of parameters, the $k-\omega$ SST model provides a reasonable prediction of the pressure losses in the channel. Using a laminar flow model for flow with an oblique entry, even when $Re_c$ is below the channel flow transition threshold, is not justified.

% \subsubsection{Effect of Inlet Turbulence}

% The presence of turbulence upstream of the monolith in experiments may alter the flow dynamics inside the channel. Although straighteners and nozzles were used in experiments to minimise potential turbulence and promote a laminar/flat velocity profile upstream of the monolith, other unavoidable effects may persist. These effects can be investigated numerically within the CFD model by seeding synthetic turbulence at the inlet boundary of the flow domain (or maybe a baffle - more to follow). To specify some inlet turbulence, a turbulence intensity (\%) and a turbulent eddy length scale need to be defined. In the experiments, the intensity and size of the turbulence are not measured or estimated, so these parameters can be arbitrary. For this reason, three intensities were specified: 1\%, 10\%, and 15\%. The length scale of the turbulence was chosen based on 2.5\% of the channel diameter (these need changing). The same mesh and LES solver settings were used throughout.

\subsubsection{Effect of Rounded Walls}
\label{sec:rounded}
The geometry of the leading edge at the channel entrance plays a crucial role in determining the behaviour of fluid flow around it. \cite{Dullien} A rounded edge allows for a gradual change in the flow direction, reducing the adverse pressure gradient and enabling the boundary layer to remain attached to the surface for a longer distance. Reducing flow separation reduces pressure losses as the separation leads to the formation of a shear layer, shedding of coherent structures, increased turbulence, and energy dissipation.

\begin{figure}
\includegraphics[scale=0.3]{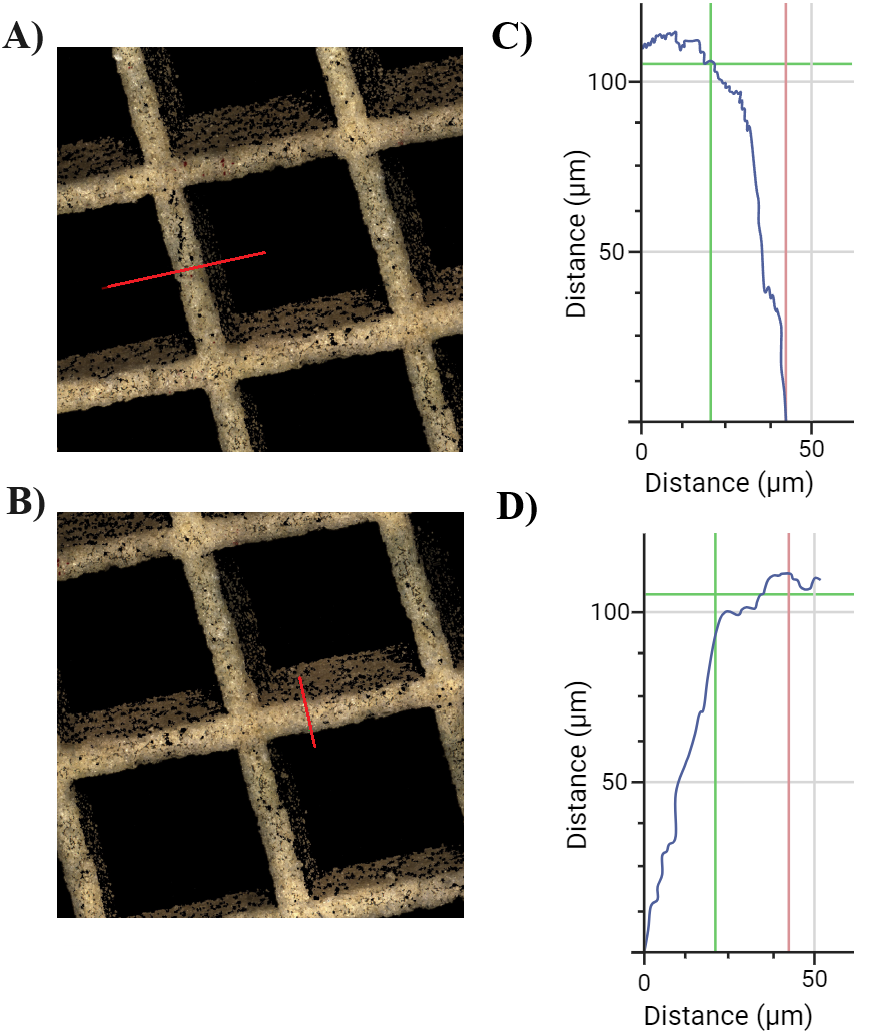}
\caption{A) Microscope images of a catalytic converter substrate front face. B) Plots of the tomography measurements along the red line drawn in the images.}
\label{AME}
\end{figure}

\begin{figure}
\includegraphics[scale=0.25]{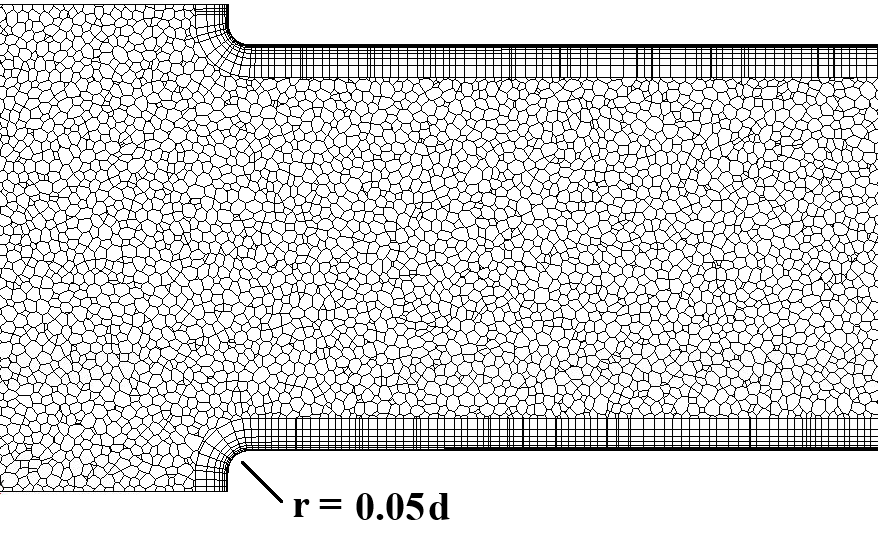}
\caption{Mesh at the entrance of the channel with rounding radius $r$ = 0.05$d$ (XY Plane)}
\label{fillet}
\end{figure}

To estimate the properties of the channel edge encountered in applications, a core from a cordierite catalytic converter (similar to those used by Mat Yamin \cite{Kamal} and Quadri et al. \cite{Quadri}) was inspected under a focus variation microscope. Fig.\,\ref{AME} depicts a section of the monolith front face (Fig.\,\ref{AME} A,B), and the leading edge profile of two monolith walls which show a rounded edge at the entrance to the monolith channel. The measured edge profiles (Fig.\,\ref{AME} C,D) show that for this monolith with a channel width of 1.12 mm the radius is approximately 0.035 mm i.e. 2.92\% of the channel width. To study the effect of the rounding of the leading edge on the channel pressure losses, three new geometries were produced with the leading edge rounded with a radii of $r$ = 0.01$d$, 0.025$d$, and 0.05$d$ (an example of the rounded geometries mesh is shown in Fig.\,\ref{fillet}). All mesh parameters and solver settings for the simulations were kept the same as for zero-radius simulations.

\begin{figure}
\includegraphics[scale=0.575]{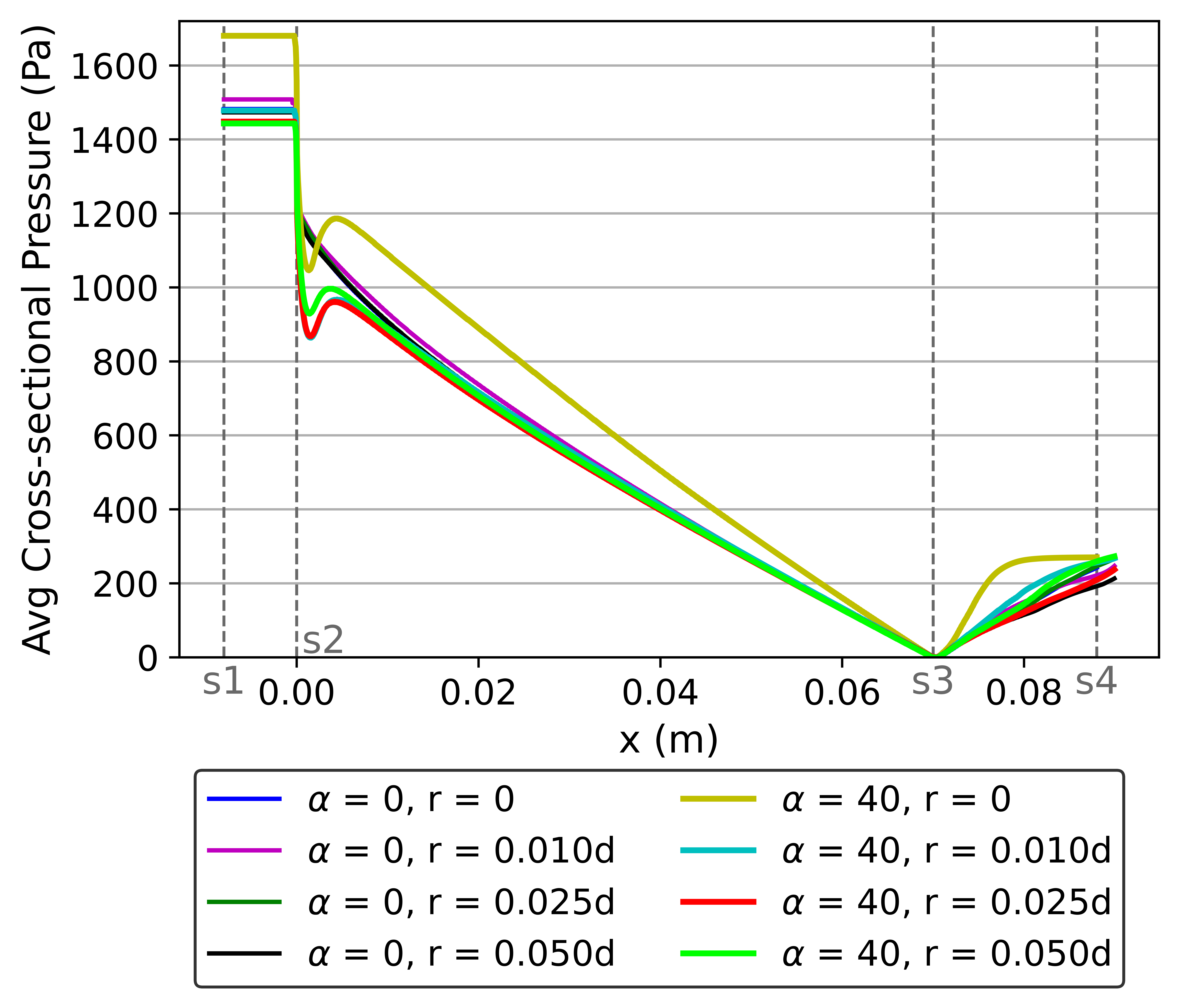 }
\caption{Pressure distribution along the channel for $Re_c$ = 1500 and $\alpha$ = 0$^\circ$ and 40$^\circ$, with rounding radii of $r$ = 0, $r$ = 0.01$d$, 0.025$d$, and 0.05$d$ edge.}
\label{RoundedPD}
\end{figure}

The pressure distribution for the rounded wall cases for $Re_c$ = 1500, $\alpha$ = 0$^\circ$ and $\alpha$ = 40$^\circ$ is shown in Fig.\,\ref{RoundedPD}. As expected, the rounded edge leads to lower pressure losses for oblique entry compared to sharp-edged geometry; for normal entry rounding has little effect. For the $\alpha$ = 40$^\circ$, $r$ = 0 case, there is a noticeable increase in the gradient of the pressure distribution curve compared to the $\alpha$ = 0$^\circ$ case. However, for all cases with rounded radii, the pressure gradient remains consistent with the axial flow scenario ($\alpha$ = 0$^\circ$). 

\begin{figure}
    \includegraphics[scale=0.575]{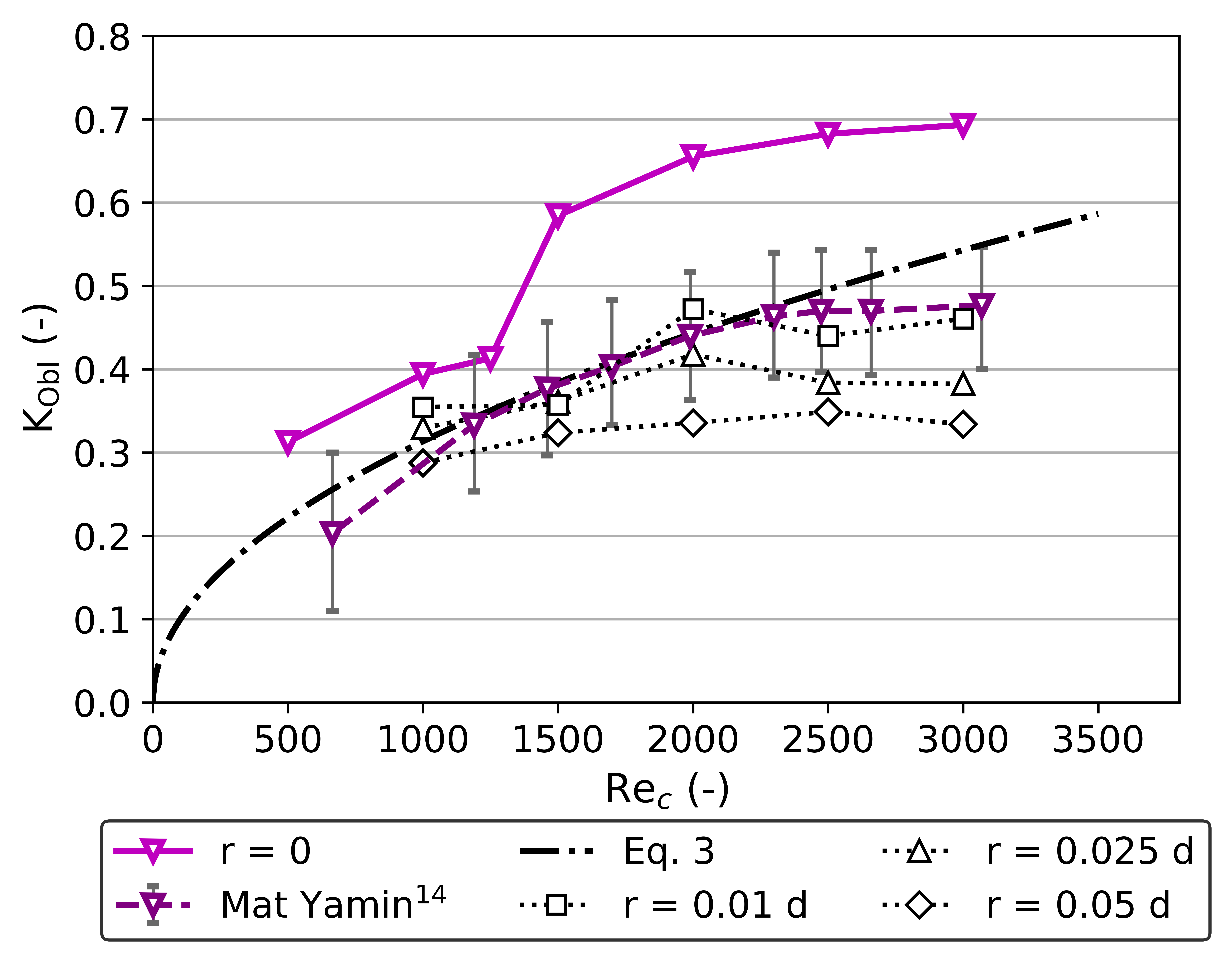 }
    \caption{$K_{Obl}$ for $\alpha$ = 45$^\circ$ vs $Re_c$ for rounded wall simulations compared with experimental results from Mat Yamin \cite{Kamal} for 27 mm monolith, and predictive correlation from Quadri et al. \cite{Quadri} ($L$ = 27$d$)}
    \label{RoundedKObl}
\end{figure}

With increasing Reynolds number, a shift in gradient was still observed, with or without a rounded edge, coinciding with an observed change in flow regime (as in Fig. \ref{RoundedKObl}). This indicates that the presence of a rounded edge delays the transition to turbulence in the channel so higher Reynolds numbers or oblique entry angles are required to observe turbulent flow in the channel.

Fig.\,\ref{RoundedKObl} shows $K_{Obl}$ vs $Re_c$ for the sharp and rounded edge cases, correlation from Quadri \cite{Quadri}, and experimental data by Mat Yamin \cite{Kamal} for the 27 mm monolith. $K_{Obl}$ is consistently lower for the simulations with the rounded edge than those with a sharp edge (Fig.\,\ref{RoundedKObl}). For the sharp edge case, up to around $Re_c$ = 1500, $K_{Obl}$ has reasonable agreement with the correlation developed by Quadri et al.; \cite{Quadri} this breaks down at higher $Re_c$ due to the transition to turbulence. Therefore, the difference in the geometry of the leading edge between simulations and experiments can explain the difference between the pressure losses.

\begin{figure}
    \includegraphics[scale=0.575]{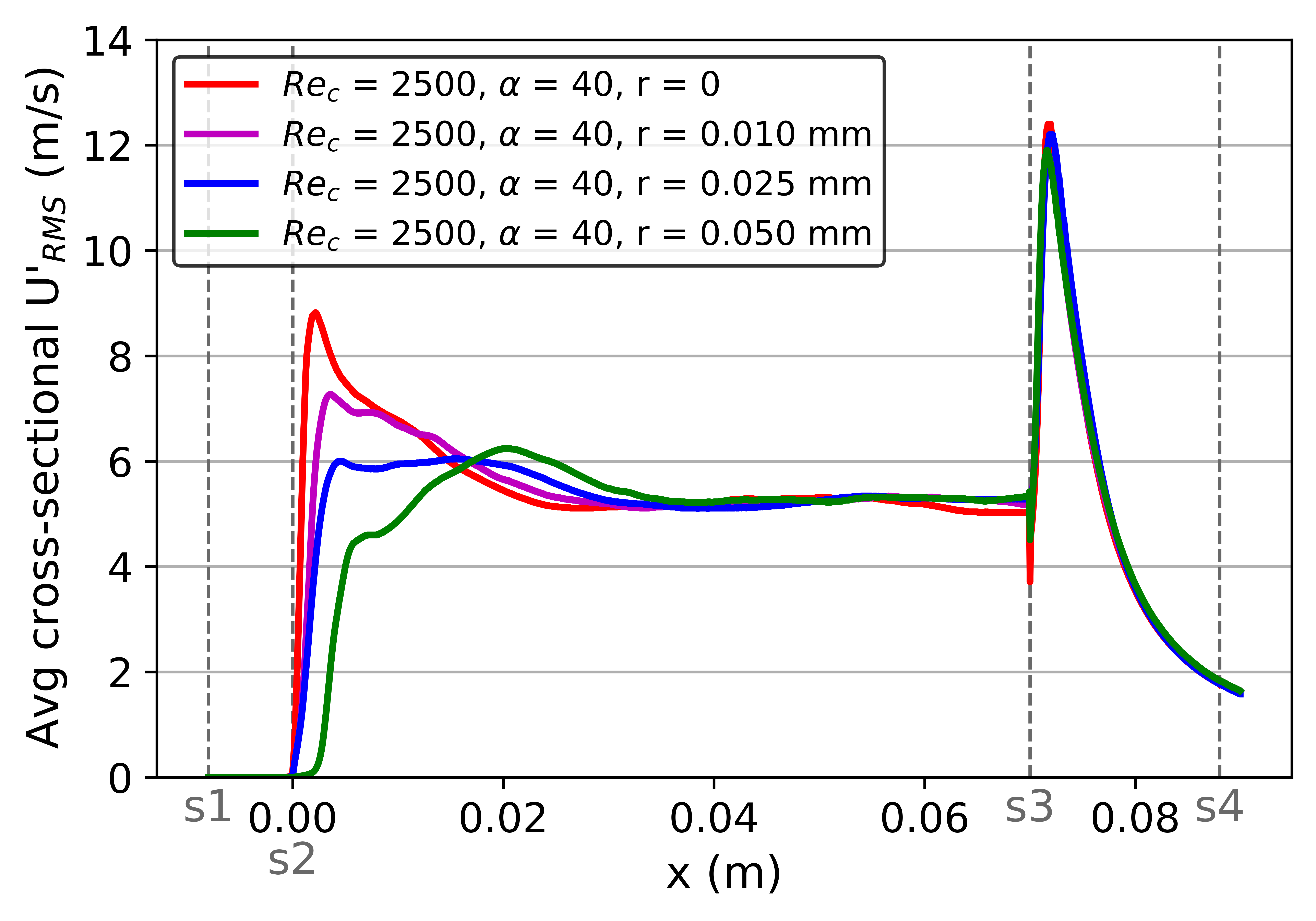 }
    \caption{Distribution of velocity fluctuations along the channel for simulations with a rounded edge, $Re_c$ = 2500 and $\alpha$ = 40$^\circ$.}
    \label{KalongRounded}
\end{figure}

The magnitude of velocity fluctuations along the channel can be used to assess the strength of the structure shedding from the shear layer and downstream turbulence, and the source of associated increase of pressure losses. Fig.\,\ref{KalongRounded} indicates that introducing a rounded leading edge stabilises the shear layer, resulting in fewer and smaller structures being formed and shed. This stabilisation is reflected in the reduced fluctuation magnitude within the recirculation region. However, the downstream velocity fluctuation levels remain consistent at approximately 4 m/s, indicating that while the vortex shedding from the shear layer is mitigated, the flow is turbulent within the boundary layer further along the channel, both with or without rounding.

In summary, the LES simulations show that the presence of a rounded edge delays the onset of turbulence and alters the structure of the separation zone, which results in lower oblique pressure losses. For the case with $r$ = 0.01$d$, the agreement with experiments is within the experimental error. This confirms that the presence of a rounded edge in the experimental setup could be the reason for the differences between measured and predicted pressure losses for higher Reynolds numbers and oblique entry angles.

\section{\label{Conc}Conclusions}

In this work, LES simulations were used to study the effect of oblique flow entry on the flow structure and pressure losses in a square channel. The modelling results were compared with experimental data and existing correlations, as well as with other modelling techniques (RANS and laminar flow simulations), highlighting some of their limitations and care required when setting up the geometry. The insights from the simulations were used to better understand the sources of the extra losses associated with the oblique entry. This is important for a number of applications, in particular in the design of multi-channel devices such as automotive catalysts and filters, or heat exchangers.

The main conclusions from the study are:

\begin{itemize}
    \item Most studies associate the pressure losses from oblique entry solely with the formation of the separation zone at the channel entrance. \cite{Quadri,Kamal} It has been demonstrated that at high incidence angles, the formed shear layer may become unstable and shed coherent structures promoting transition to turbulence, even for $Re_c <$ 2000. This leads to a sharp increase in the pressure losses. This turbulence may be beneficial for applications where enhanced mixing is required.
    \item The LES simulation results agree well with experimental data for low Reynolds numbers, ($Re_c$), for all oblique entry angles, $\alpha$, however, they overestimate the pressure losses at larger $Re_c$ and $\alpha$ - much/all of this overestimation goes if the sharp channel edges are rounded. 
    \item The differences in pressure losses between experiments and simulations at higher Reynolds numbers/entry angles can be explained by the fact that the leading edge of the channel in experiments is not sharp. The rounded leading edge reduces total pressure loss by reducing the shear-driven coherent structure shedding in the shear layer and changing the separation zone structure. Simulations performed with a rounded edge showed better agreement with the experiments. Such sensitivity to the domain geometry highlights the importance of ensuring that all key flow and geometry features are reproduced in simulations, which is often challenging.
    \item LES simulations have shown that the pressure loss, $K_{Obl}$, is dependent on length when the transition to turbulence is triggered by the oblique flow entry. The assumption that subtracting pressure losses at $\alpha$ = 0$^\circ$ would isolate the pressure losses at the channel entrance is not valid when the flow at a non-zero angle is turbulent, while the flow at $\alpha$ = 0$^\circ$ would still be laminar.
    \item The LES model provides valuable insights and can yield highly accurate results with careful understanding and development. For engineering purposes where pressure drop is the key parameter of interest, the RANS $k-\omega$ SST model has been shown to deliver reliable outcomes. Laminar flow models, both steady and unsteady, fail to capture the vortex shedding and dissipation evident in LES and generally are not suitable for modelling flow with oblique entry, even at the values of $Re_c$ usually associated with laminar flow ($Re_c$ = 2000).
\end{itemize}

%\nocite{*}
\bibliography{references}

%merlin.mbs aipnum4-1.bst 2010-07-25 4.21a (PWD, AO, DPC) hacked
%Control: key (0)
%Control: author (8) initials jnrlst
%Control: editor formatted (1) identically to author
%Control: production of article title (0) allowed
%Control: page (1) range
%Control: year (1) truncated
%Control: production of eprint (0) enabled
\providecommand{\noopsort}[1]{}\providecommand{\singleletter}[1]{#1}%
\begin{thebibliography}{42}%
\makeatletter
\providecommand \@ifxundefined [1]{%
 \@ifx{#1\undefined}
}%
\providecommand \@ifnum [1]{%
 \ifnum #1\expandafter \@firstoftwo
 \else \expandafter \@secondoftwo
 \fi
}%
\providecommand \@ifx [1]{%
 \ifx #1\expandafter \@firstoftwo
 \else \expandafter \@secondoftwo
 \fi
}%
\providecommand \natexlab [1]{#1}%
\providecommand \enquote  [1]{``#1''}%
\providecommand \bibnamefont  [1]{#1}%
\providecommand \bibfnamefont [1]{#1}%
\providecommand \citenamefont [1]{#1}%
\providecommand \href@noop [0]{\@secondoftwo}%
\providecommand \href [0]{\begingroup \@sanitize@url \@href}%
\providecommand \@href[1]{\@@startlink{#1}\@@href}%
\providecommand \@@href[1]{\endgroup#1\@@endlink}%
\providecommand \@sanitize@url [0]{\catcode `\\12\catcode `\$12\catcode `\&12\catcode `\#12\catcode `\^12\catcode `\_12\catcode `\%12\relax}%
\providecommand \@@startlink[1]{}%
\providecommand \@@endlink[0]{}%
\providecommand \url  [0]{\begingroup\@sanitize@url \@url }%
\providecommand \@url [1]{\endgroup\@href {#1}{\urlprefix }}%
\providecommand \urlprefix  [0]{URL }%
\providecommand \Eprint [0]{\href }%
\providecommand \doibase [0]{http://dx.doi.org/}%
\providecommand \selectlanguage [0]{\@gobble}%
\providecommand \bibinfo  [0]{\@secondoftwo}%
\providecommand \bibfield  [0]{\@secondoftwo}%
\providecommand \translation [1]{[#1]}%
\providecommand \BibitemOpen [0]{}%
\providecommand \bibitemStop [0]{}%
\providecommand \bibitemNoStop [0]{.\EOS\space}%
\providecommand \EOS [0]{\spacefactor3000\relax}%
\providecommand \BibitemShut  [1]{\csname bibitem#1\endcsname}%
\let\auto@bib@innerbib\@empty
%</preamble>
\bibitem [{\citenamefont {Williams}(2001)}]{Williams}%
  \BibitemOpen
  \bibfield  {author} {\bibinfo {author} {\bibfnamefont {J.~L.}\ \bibnamefont {Williams}},\ }\bibfield  {title} {\enquote {\bibinfo {title} {Monolith structures, materials, properties and uses},}\ }\href {\doibase https://doi.org/10.1016/S0920-5861(01)00348-0} {\bibfield  {journal} {\bibinfo  {journal} {Catal. Today}\ }\textbf {\bibinfo {volume} {69}},\ \bibinfo {pages} {3--9} (\bibinfo {year} {2001})}\BibitemShut {NoStop}%
\bibitem [{\citenamefont {Koltsakis}\ \emph {et~al.}(2013)\citenamefont {Koltsakis}, \citenamefont {Onoufrios}, \citenamefont {Depcik},\ and\ \citenamefont {Ragone}}]{Koltsakis1}%
  \BibitemOpen
  \bibfield  {author} {\bibinfo {author} {\bibfnamefont {G.}~\bibnamefont {Koltsakis}}, \bibinfo {author} {\bibfnamefont {H.}~\bibnamefont {Onoufrios}}, \bibinfo {author} {\bibfnamefont {C.}~\bibnamefont {Depcik}}, \ and\ \bibinfo {author} {\bibfnamefont {J.~C.}\ \bibnamefont {Ragone}},\ }\bibfield  {title} {\enquote {\bibinfo {title} {Catalyzed diesel particulate filter modelling},}\ }\href {\doibase 10.1515/revce-2012-0008} {\bibfield  {journal} {\bibinfo  {journal} {Rev. Chem. Eng.}\ }\textbf {\bibinfo {volume} {29}},\ \bibinfo {pages} {161} (\bibinfo {year} {2013})}\BibitemShut {NoStop}%
\bibitem [{\citenamefont {Fakheri}(2014)}]{Fakheri}%
  \BibitemOpen
  \bibfield  {author} {\bibinfo {author} {\bibfnamefont {A.}~\bibnamefont {Fakheri}},\ }\bibfield  {title} {\enquote {\bibinfo {title} {Efficiency analysis of heat exchangers and heat exchange networks},}\ }\href {\doibase 10.1016/j.ijheatmasstransfer.2014.04.027} {\bibfield  {journal} {\bibinfo  {journal} {Int. J. Heat Mass Transfer}\ }\textbf {\bibinfo {volume} {76}},\ \bibinfo {pages} {99--104} (\bibinfo {year} {2014})}\BibitemShut {NoStop}%
\bibitem [{\citenamefont {Dullien}(1992)}]{Dullien}%
  \BibitemOpen
  \bibfield  {author} {\bibinfo {author} {\bibfnamefont {F.~A.~L.}\ \bibnamefont {Dullien}},\ }\href {\doibase https://doi.org/10.1016/B978-0-12-223651-8.50007-9} {\emph {\bibinfo {title} {Porous Media}}},\ \bibinfo {edition} {2nd}\ ed.\ (\bibinfo  {publisher} {Academic Press},\ \bibinfo {address} {San Diego},\ \bibinfo {year} {1992})\ pp.\ \bibinfo {pages} {5--115}\BibitemShut {NoStop}%
\bibitem [{\citenamefont {Shah}\ and\ \citenamefont {London}(1978)}]{ShahLondon}%
  \BibitemOpen
  \bibfield  {author} {\bibinfo {author} {\bibfnamefont {R.}~\bibnamefont {Shah}}\ and\ \bibinfo {author} {\bibfnamefont {A.}~\bibnamefont {London}},\ }\href@noop {} {\emph {\bibinfo {title} {Laminar-flow forced convection in ducts}}}\ (\bibinfo  {publisher} {Academic Press},\ \bibinfo {address} {New York},\ \bibinfo {year} {1978})\ pp.\ \bibinfo {pages} {98--212}\BibitemShut {NoStop}%
\bibitem [{\citenamefont {Bissett}(1984)}]{Bissett}%
  \BibitemOpen
  \bibfield  {author} {\bibinfo {author} {\bibfnamefont {E.~J.}\ \bibnamefont {Bissett}},\ }\bibfield  {title} {\enquote {\bibinfo {title} {Mathematical model of the thermal regeneration of a wall-flow monolith diesel particulate filter},}\ }\href {\doibase https://doi.org/10.1016/0009-2509(84)85084-8} {\bibfield  {journal} {\bibinfo  {journal} {Chem. Eng. Sci.}\ }\textbf {\bibinfo {volume} {39}},\ \bibinfo {pages} {1233--1244} (\bibinfo {year} {1984})}\BibitemShut {NoStop}%
\bibitem [{\citenamefont {Konstandopoulos}\ and\ \citenamefont {Johnson}(1989)}]{KonstOG}%
  \BibitemOpen
  \bibfield  {author} {\bibinfo {author} {\bibfnamefont {A.~G.}\ \bibnamefont {Konstandopoulos}}\ and\ \bibinfo {author} {\bibfnamefont {J.~H.}\ \bibnamefont {Johnson}},\ }\bibfield  {title} {\enquote {\bibinfo {title} {Wall-flow diesel particulate filters—their pressure drop and collection efficiency},}\ }\href {\doibase https://doi.org/10.4271/890405} {\bibfield  {journal} {\bibinfo  {journal} {SAE Tech. Paper 890405}\ } (\bibinfo {year} {1989}),\ https://doi.org/10.4271/890405}\BibitemShut {NoStop}%
\bibitem [{\citenamefont {Watling}\ \emph {et~al.}(2017)\citenamefont {Watling}, \citenamefont {Ravenscroft}, \citenamefont {Cleeton}, \citenamefont {Rees},\ and\ \citenamefont {Wilkins}}]{Tim}%
  \BibitemOpen
  \bibfield  {author} {\bibinfo {author} {\bibfnamefont {T.}~\bibnamefont {Watling}}, \bibinfo {author} {\bibfnamefont {M.}~\bibnamefont {Ravenscroft}}, \bibinfo {author} {\bibfnamefont {J.}~\bibnamefont {Cleeton}}, \bibinfo {author} {\bibfnamefont {I.}~\bibnamefont {Rees}}, \ and\ \bibinfo {author} {\bibfnamefont {D.}~\bibnamefont {Wilkins}},\ }\bibfield  {title} {\enquote {\bibinfo {title} {Development of a particulate filter model for the prediction of backpressure: Improved momentum balance and entrance and exit effect equations},}\ }\href {\doibase 10.4271/2017-01-0974} {\bibfield  {journal} {\bibinfo  {journal} {SAE Int. J. Eng.}\ }\textbf {\bibinfo {volume} {10}},\ \bibinfo {pages} {1765--1794} (\bibinfo {year} {2017})}\BibitemShut {NoStop}%
\bibitem [{\citenamefont {Masoudi}, \citenamefont {Heibel},\ and\ \citenamefont {Then}(2000)}]{Masoudi}%
  \BibitemOpen
  \bibfield  {author} {\bibinfo {author} {\bibfnamefont {M.}~\bibnamefont {Masoudi}}, \bibinfo {author} {\bibfnamefont {A.}~\bibnamefont {Heibel}}, \ and\ \bibinfo {author} {\bibfnamefont {P.}~\bibnamefont {Then}},\ }\bibfield  {title} {\enquote {\bibinfo {title} {Predicting pressure drop of wall-flow diesel particulate filters - theory and experiment},}\ }\href {\doibase 10.4271/2000-01-0184} {\bibfield  {journal} {\bibinfo  {journal} {SAE Tech. Paper 2000-01-0184}\ } (\bibinfo {year} {2000}),\ 10.4271/2000-01-0184}\BibitemShut {NoStop}%
\bibitem [{\citenamefont {Haralampous}\ \emph {et~al.}(2004)\citenamefont {Haralampous}, \citenamefont {Kandylas}, \citenamefont {Koltsakis},\ and\ \citenamefont {Samaras}}]{Haral}%
  \BibitemOpen
  \bibfield  {author} {\bibinfo {author} {\bibfnamefont {O.~A.}\ \bibnamefont {Haralampous}}, \bibinfo {author} {\bibfnamefont {I.~P.}\ \bibnamefont {Kandylas}}, \bibinfo {author} {\bibfnamefont {G.~C.}\ \bibnamefont {Koltsakis}}, \ and\ \bibinfo {author} {\bibfnamefont {Z.~C.}\ \bibnamefont {Samaras}},\ }\bibfield  {title} {\enquote {\bibinfo {title} {Diesel particulate filter pressure drop part 1: Modelling and experimental validation},}\ }\href {\doibase 10.1243/146808704773564550} {\bibfield  {journal} {\bibinfo  {journal} {Int. J. Eng. Res.}\ }\textbf {\bibinfo {volume} {5}},\ \bibinfo {pages} {149--162} (\bibinfo {year} {2004})}\BibitemShut {NoStop}%
\bibitem [{\citenamefont {Kuchemann}\ and\ \citenamefont {Weber}(1953)}]{KW}%
  \BibitemOpen
  \bibfield  {author} {\bibinfo {author} {\bibfnamefont {D.}~\bibnamefont {Kuchemann}}\ and\ \bibinfo {author} {\bibfnamefont {J.}~\bibnamefont {Weber}},\ }\href@noop {} {\emph {\bibinfo {title} {Aerodynamics of Propulsion}}}\ (\bibinfo  {publisher} {McGraw Hill},\ \bibinfo {address} {New York},\ \bibinfo {year} {1953})\BibitemShut {NoStop}%
\bibitem [{\citenamefont {Persoons}, \citenamefont {Vanierschot},\ and\ \citenamefont {Van~den Bulck}(2008)}]{Persoons}%
  \BibitemOpen
  \bibfield  {author} {\bibinfo {author} {\bibfnamefont {T.}~\bibnamefont {Persoons}}, \bibinfo {author} {\bibfnamefont {M.}~\bibnamefont {Vanierschot}}, \ and\ \bibinfo {author} {\bibfnamefont {E.}~\bibnamefont {Van~den Bulck}},\ }\bibfield  {title} {\enquote {\bibinfo {title} {Oblique inlet pressure loss for swirling flow entering a catalyst substrate},}\ }\href {\doibase 10.1016/j.expthermflusci.2008.02.002} {\bibfield  {journal} {\bibinfo  {journal} {Exp. Therm. Fluid Sci.}\ }\textbf {\bibinfo {volume} {32}},\ \bibinfo {pages} {1222--1231} (\bibinfo {year} {2008})}\BibitemShut {NoStop}%
\bibitem [{\citenamefont {Quadri}, \citenamefont {Benjamin},\ and\ \citenamefont {Roberts}(2009)}]{Quadri}%
  \BibitemOpen
  \bibfield  {author} {\bibinfo {author} {\bibfnamefont {S.~S.}\ \bibnamefont {Quadri}}, \bibinfo {author} {\bibfnamefont {S.~F.}\ \bibnamefont {Benjamin}}, \ and\ \bibinfo {author} {\bibfnamefont {C.~A.}\ \bibnamefont {Roberts}},\ }\bibfield  {title} {\enquote {\bibinfo {title} {An experimental investigation of oblique entry pressure losses in automotive catalytic converters},}\ }\href {\doibase 10.1243/09544062JMES1565} {\bibfield  {journal} {\bibinfo  {journal} {Proc. Inst. Mech. Eng., Part C: J. Mech. Eng. Sci.}\ }\textbf {\bibinfo {volume} {223}},\ \bibinfo {pages} {2561--2569} (\bibinfo {year} {2009})}\BibitemShut {NoStop}%
\bibitem [{\citenamefont {Mat~Yamin}(2012)}]{Kamal}%
  \BibitemOpen
  \bibfield  {author} {\bibinfo {author} {\bibfnamefont {A.~K.}\ \bibnamefont {Mat~Yamin}},\ }\emph {\bibinfo {title} {Pulsating Flow Studies in a Planar Wide-angled Diffuser Upstream of Automotive Catalyst Monoliths}},\ \href {https://pureportal.coventry.ac.uk/en/studentTheses/pulsating-flow-studies-in-a-planar-wide-angled-diffuser-upstream-} {\bibinfo {type} {Ph.d thesis}},\ \bibinfo  {school} {Coventry University}, \bibinfo {address} {Coventry, West Midlands} (\bibinfo {year} {2012})\BibitemShut {NoStop}%
\bibitem [{\citenamefont {Moore}\ and\ \citenamefont {Torrence}(1984)}]{Moore}%
  \BibitemOpen
  \bibfield  {author} {\bibinfo {author} {\bibfnamefont {F.~K.}\ \bibnamefont {Moore}}\ and\ \bibinfo {author} {\bibfnamefont {K.~E.}\ \bibnamefont {Torrence}},\ }\bibfield  {title} {\enquote {\bibinfo {title} {Air flow in dry natural-drought cooling towers subject to wind},}\ }\href@noop {} {\bibfield  {journal} {\bibinfo  {journal} {Cornell Energy Rep.}\ } (\bibinfo {year} {1984})}\BibitemShut {NoStop}%
\bibitem [{\citenamefont {Haimad}(1997)}]{Haimad}%
  \BibitemOpen
  \bibfield  {author} {\bibinfo {author} {\bibfnamefont {N.}~\bibnamefont {Haimad}},\ }\emph {\bibinfo {title} {A theoretical and experimental investigation of the flow performance of automotive catalytic converters}},\ \href@noop {} {Ph.D. thesis},\ \bibinfo  {school} {Coventry University}, \bibinfo {address} {Coventry, West Midlands} (\bibinfo {year} {1997})\BibitemShut {NoStop}%
\bibitem [{\citenamefont {Atzori}\ \emph {et~al.}(2019)\citenamefont {Atzori}, \citenamefont {Vinuesa}, \citenamefont {Lozano-Dur'an},\ and\ \citenamefont {Schlatter}}]{Coherent}%
  \BibitemOpen
  \bibfield  {author} {\bibinfo {author} {\bibfnamefont {M.}~\bibnamefont {Atzori}}, \bibinfo {author} {\bibfnamefont {R.}~\bibnamefont {Vinuesa}}, \bibinfo {author} {\bibfnamefont {A.}~\bibnamefont {Lozano-Dur'an}}, \ and\ \bibinfo {author} {\bibfnamefont {P.}~\bibnamefont {Schlatter}},\ }\bibfield  {title} {\enquote {\bibinfo {title} {Coherent structures and secondary flow in turbulent square duct},}\ }\href {\doibase arxiv.org/abs/1906.00886v1} {\bibfield  {journal} {\bibinfo  {journal} {arXiv: Fluid Dynamics}\ } (\bibinfo {year} {2019}),\ arxiv.org/abs/1906.00886v1}\BibitemShut {NoStop}%
\bibitem [{\citenamefont {Cornejo}\ \emph {et~al.}(2020)\citenamefont {Cornejo}, \citenamefont {Nikrityuk}, \citenamefont {Lange},\ and\ \citenamefont {Hayes}}]{Cornejo}%
  \BibitemOpen
  \bibfield  {author} {\bibinfo {author} {\bibfnamefont {I.}~\bibnamefont {Cornejo}}, \bibinfo {author} {\bibfnamefont {P.}~\bibnamefont {Nikrityuk}}, \bibinfo {author} {\bibfnamefont {C.}~\bibnamefont {Lange}}, \ and\ \bibinfo {author} {\bibfnamefont {R.~E.}\ \bibnamefont {Hayes}},\ }\bibfield  {title} {\enquote {\bibinfo {title} {Influence of upstream turbulence on the pressure drop inside a monolith},}\ }\href {\doibase 10.1016/j.cep.2019.107735} {\bibfield  {journal} {\bibinfo  {journal} {Chem. Eng. Process}\ }\textbf {\bibinfo {volume} {147}},\ \bibinfo {pages} {107735} (\bibinfo {year} {2020})}\BibitemShut {NoStop}%
\bibitem [{\citenamefont {Vidal}\ \emph {et~al.}(2017)\citenamefont {Vidal}, \citenamefont {Vinuesa}, \citenamefont {Schlatter},\ and\ \citenamefont {Nagib}}]{Vidal}%
  \BibitemOpen
  \bibfield  {author} {\bibinfo {author} {\bibfnamefont {A.}~\bibnamefont {Vidal}}, \bibinfo {author} {\bibfnamefont {R.}~\bibnamefont {Vinuesa}}, \bibinfo {author} {\bibfnamefont {P.}~\bibnamefont {Schlatter}}, \ and\ \bibinfo {author} {\bibfnamefont {H.}~\bibnamefont {Nagib}},\ }\bibfield  {title} {\enquote {\bibinfo {title} {Influence of corner geometry on the secondary flow in turbulent square ducts},}\ }\href {\doibase 10.1016/j.ijheatfluidflow.2017.09.011} {\bibfield  {journal} {\bibinfo  {journal} {Int. J. Heat Fluid Flow}\ }\textbf {\bibinfo {volume} {67}},\ \bibinfo {pages} {94--103} (\bibinfo {year} {2017})}\BibitemShut {NoStop}%
\bibitem [{\citenamefont {Wang}\ \emph {et~al.}(2024)\citenamefont {Wang}, \citenamefont {Mortimer}, \citenamefont {Fairweather}, \citenamefont {Ma},\ and\ \citenamefont {Zhen}}]{DNS1}%
  \BibitemOpen
  \bibfield  {author} {\bibinfo {author} {\bibfnamefont {Y.}~\bibnamefont {Wang}}, \bibinfo {author} {\bibfnamefont {L.~F.}\ \bibnamefont {Mortimer}}, \bibinfo {author} {\bibfnamefont {M.}~\bibnamefont {Fairweather}}, \bibinfo {author} {\bibfnamefont {W.}~\bibnamefont {Ma}}, \ and\ \bibinfo {author} {\bibfnamefont {Y.}~\bibnamefont {Zhen}},\ }\bibfield  {title} {\enquote {\bibinfo {title} {{Particle transport in turbulent square duct flows with a free surface}},}\ }\href@noop {} {\bibfield  {journal} {\bibinfo  {journal} {Phys. Fluids}\ }\textbf {\bibinfo {volume} {36}},\ \bibinfo {pages} {013340} (\bibinfo {year} {2024})}\BibitemShut {NoStop}%
\bibitem [{\citenamefont {Wang}, \citenamefont {Zhao},\ and\ \citenamefont {Yao}(2019{\natexlab{a}})}]{particle1}%
  \BibitemOpen
  \bibfield  {author} {\bibinfo {author} {\bibfnamefont {Y.}~\bibnamefont {Wang}}, \bibinfo {author} {\bibfnamefont {Y.}~\bibnamefont {Zhao}}, \ and\ \bibinfo {author} {\bibfnamefont {J.}~\bibnamefont {Yao}},\ }\bibfield  {title} {\enquote {\bibinfo {title} {Large eddy simulation of particle deposition and resuspension in turbulent duct flows},}\ }\href {\doibase 10.1016/j.apt.2019.01.012} {\bibfield  {journal} {\bibinfo  {journal} {Adv. Powder Tech.}\ }\textbf {\bibinfo {volume} {30}},\ \bibinfo {pages} {656--671} (\bibinfo {year} {2019}{\natexlab{a}})}\BibitemShut {NoStop}%
\bibitem [{\citenamefont {Wang}, \citenamefont {Zhao},\ and\ \citenamefont {Yao}(2019{\natexlab{b}})}]{particle2}%
  \BibitemOpen
  \bibfield  {author} {\bibinfo {author} {\bibfnamefont {Y.}~\bibnamefont {Wang}}, \bibinfo {author} {\bibfnamefont {Y.}~\bibnamefont {Zhao}}, \ and\ \bibinfo {author} {\bibfnamefont {J.}~\bibnamefont {Yao}},\ }\bibfield  {title} {\enquote {\bibinfo {title} {Particle dispersion in turbulent, square open duct flows of high reynolds number},}\ }\href {\doibase https://doi.org/10.1016/j.powtec.2019.05.085} {\bibfield  {journal} {\bibinfo  {journal} {Powder Tech.}\ }\textbf {\bibinfo {volume} {354}},\ \bibinfo {pages} {92--107} (\bibinfo {year} {2019}{\natexlab{b}})}\BibitemShut {NoStop}%
\bibitem [{\citenamefont {Xiao}\ and\ \citenamefont {Fu}(2016)}]{LES1}%
  \BibitemOpen
  \bibfield  {author} {\bibinfo {author} {\bibfnamefont {Z.}~\bibnamefont {Xiao}}\ and\ \bibinfo {author} {\bibfnamefont {S.}~\bibnamefont {Fu}},\ }\bibfield  {title} {\enquote {\bibinfo {title} {Simulations of complex turbulent flows with rans-les hybrid approaches},}\ }\href@noop {} {\bibfield  {journal} {\bibinfo  {journal} {Fluid-Struct.-Sound Interact. Control}\ ,\ \bibinfo {pages} {271--281}} (\bibinfo {year} {2016})}\BibitemShut {NoStop}%
\bibitem [{\citenamefont {Marchioli}\ \emph {et~al.}(2024)\citenamefont {Marchioli}, \citenamefont {Garc{\'i}a-Villalba}, \citenamefont {Salvetti},\ and\ \citenamefont {Schlatter}}]{LES2}%
  \BibitemOpen
  \bibfield  {author} {\bibinfo {author} {\bibfnamefont {C.}~\bibnamefont {Marchioli}}, \bibinfo {author} {\bibfnamefont {M.}~\bibnamefont {Garc{\'i}a-Villalba}}, \bibinfo {author} {\bibfnamefont {M.~V.}\ \bibnamefont {Salvetti}}, \ and\ \bibinfo {author} {\bibfnamefont {P.}~\bibnamefont {Schlatter}},\ }\bibfield  {title} {\enquote {\bibinfo {title} {Advances in direct and large-eddy simulations},}\ }\href {\doibase 10.1007/s10494-023-00524-0} {\bibfield  {journal} {\bibinfo  {journal} {Flow, Turb. Comb.}\ }\textbf {\bibinfo {volume} {112}},\ \bibinfo {pages} {1--2} (\bibinfo {year} {2024})}\BibitemShut {NoStop}%
\bibitem [{\citenamefont {Piomelli}\ and\ \citenamefont {Chasnov}(1996)}]{Piomelli}%
  \BibitemOpen
  \bibfield  {author} {\bibinfo {author} {\bibfnamefont {U.}~\bibnamefont {Piomelli}}\ and\ \bibinfo {author} {\bibfnamefont {J.~R.}\ \bibnamefont {Chasnov}},\ }\href {\doibase 10.1007/978-94-015-8666-5_7} {\emph {\bibinfo {title} {Turbulence and Transition Modelling}}}\ (\bibinfo  {publisher} {Springer Netherlands},\ \bibinfo {address} {Dordrecht},\ \bibinfo {year} {1996})\ pp.\ \bibinfo {pages} {269--336}\BibitemShut {NoStop}%
\bibitem [{\citenamefont {Garg}\ \emph {et~al.}(2024)\citenamefont {Garg}, \citenamefont {Wang}, \citenamefont {Andersson},\ and\ \citenamefont {Fureby}}]{LESpipe}%
  \BibitemOpen
  \bibfield  {author} {\bibinfo {author} {\bibfnamefont {H.}~\bibnamefont {Garg}}, \bibinfo {author} {\bibfnamefont {L.}~\bibnamefont {Wang}}, \bibinfo {author} {\bibfnamefont {M.}~\bibnamefont {Andersson}}, \ and\ \bibinfo {author} {\bibfnamefont {C.}~\bibnamefont {Fureby}},\ }\bibfield  {title} {\enquote {\bibinfo {title} {{Large eddy simulations of turbulent pipe flows at moderate Reynolds numbers}},}\ }\href {\doibase 10.1063/5.0201967} {\bibfield  {journal} {\bibinfo  {journal} {Phys. Fluids}\ }\textbf {\bibinfo {volume} {36}},\ \bibinfo {pages} {045138} (\bibinfo {year} {2024})}\BibitemShut {NoStop}%
\bibitem [{\citenamefont {Holgate}\ \emph {et~al.}(2019)\citenamefont {Holgate}, \citenamefont {Skillen}, \citenamefont {Craft},\ and\ \citenamefont {Revell}}]{LESpipe2}%
  \BibitemOpen
  \bibfield  {author} {\bibinfo {author} {\bibfnamefont {J.}~\bibnamefont {Holgate}}, \bibinfo {author} {\bibfnamefont {A.}~\bibnamefont {Skillen}}, \bibinfo {author} {\bibfnamefont {T.}~\bibnamefont {Craft}}, \ and\ \bibinfo {author} {\bibfnamefont {A.}~\bibnamefont {Revell}},\ }\bibfield  {title} {\enquote {\bibinfo {title} {A review of embedded large eddy simulation for internal flows},}\ }\href {\doibase 10.1007/s11831-018-9272-5} {\bibfield  {journal} {\bibinfo  {journal} {Arch. Comput. Methods Eng.}\ }\textbf {\bibinfo {volume} {26}},\ \bibinfo {pages} {865--882} (\bibinfo {year} {2019})}\BibitemShut {NoStop}%
\bibitem [{\citenamefont {Courant}, \citenamefont {Friedrichs},\ and\ \citenamefont {Lewy}(1928)}]{CFL}%
  \BibitemOpen
  \bibfield  {author} {\bibinfo {author} {\bibfnamefont {R.}~\bibnamefont {Courant}}, \bibinfo {author} {\bibfnamefont {K.}~\bibnamefont {Friedrichs}}, \ and\ \bibinfo {author} {\bibfnamefont {H.}~\bibnamefont {Lewy}},\ }\bibfield  {title} {\enquote {\bibinfo {title} {Über die partiellen differenzengleichungen der mathematischen physik},}\ }\href {\doibase 10.1007/BF01448839} {\bibfield  {journal} {\bibinfo  {journal} {Math. Ann.}\ }\textbf {\bibinfo {volume} {100}},\ \bibinfo {pages} {32--74} (\bibinfo {year} {1928})}\BibitemShut {NoStop}%
\bibitem [{\citenamefont {Smagorinsky}(1963)}]{LES}%
  \BibitemOpen
  \bibfield  {author} {\bibinfo {author} {\bibfnamefont {J.}~\bibnamefont {Smagorinsky}},\ }\bibfield  {title} {\enquote {\bibinfo {title} {General circulation experiments with the primitive equations: I. the basic experiment},}\ }\href {\doibase 10.1175/1520-0493(1963)091<0099:GCEWTP>2.3.CO;2} {\bibfield  {journal} {\bibinfo  {journal} {Mon. Weather Rev.}\ }\textbf {\bibinfo {volume} {91}},\ \bibinfo {pages} {99 -- 164} (\bibinfo {year} {1963})}\BibitemShut {NoStop}%
\bibitem [{\citenamefont {Germano}\ \emph {et~al.}(1991)\citenamefont {Germano}, \citenamefont {Piomelli}, \citenamefont {Moin},\ and\ \citenamefont {Cabot}}]{Germano}%
  \BibitemOpen
  \bibfield  {author} {\bibinfo {author} {\bibfnamefont {M.}~\bibnamefont {Germano}}, \bibinfo {author} {\bibfnamefont {U.}~\bibnamefont {Piomelli}}, \bibinfo {author} {\bibfnamefont {P.}~\bibnamefont {Moin}}, \ and\ \bibinfo {author} {\bibfnamefont {W.~H.}\ \bibnamefont {Cabot}},\ }\bibfield  {title} {\enquote {\bibinfo {title} {A dynamic subgrid-scale eddy viscosity model},}\ }\href {\doibase 10.1063/1.857955} {\bibfield  {journal} {\bibinfo  {journal} {Phys. Fluids A}\ }\textbf {\bibinfo {volume} {3}},\ \bibinfo {pages} {1760--1765} (\bibinfo {year} {1991})}\BibitemShut {NoStop}%
\bibitem [{\citenamefont {Menter}(1993)}]{Menter}%
  \BibitemOpen
  \bibfield  {author} {\bibinfo {author} {\bibfnamefont {F.}~\bibnamefont {Menter}},\ }\bibfield  {title} {\enquote {\bibinfo {title} {Zonal two equation k-w turbulence models for aerodynamic flows},}\ \ }(\bibinfo  {publisher} {AIAA},\ \bibinfo {year} {1993})\BibitemShut {NoStop}%
\bibitem [{\citenamefont {Pope}(2000)}]{Pope}%
  \BibitemOpen
  \bibfield  {author} {\bibinfo {author} {\bibfnamefont {S.~B.}\ \bibnamefont {Pope}},\ }\href@noop {} {\emph {\bibinfo {title} {Turbulent Flows}}}\ (\bibinfo  {publisher} {Cambridge Univ. Press},\ \bibinfo {address} {Cambridge, UK},\ \bibinfo {year} {2000})\BibitemShut {NoStop}%
\bibitem [{\citenamefont {Celik}, \citenamefont {Cehreli},\ and\ \citenamefont {Yavuz}(2005)}]{Celik}%
  \BibitemOpen
  \bibfield  {author} {\bibinfo {author} {\bibfnamefont {I.~B.}\ \bibnamefont {Celik}}, \bibinfo {author} {\bibfnamefont {Z.~N.}\ \bibnamefont {Cehreli}}, \ and\ \bibinfo {author} {\bibfnamefont {I.}~\bibnamefont {Yavuz}},\ }\bibfield  {title} {\enquote {\bibinfo {title} {Index of resolution quality for large eddy simulations},}\ }\href {\doibase 10.1115/1.1990201} {\bibfield  {journal} {\bibinfo  {journal} {J. Fluids Eng.}\ }\textbf {\bibinfo {volume} {127}},\ \bibinfo {pages} {949--958} (\bibinfo {year} {2005})}\BibitemShut {NoStop}%
\bibitem [{\citenamefont {Gou}, \citenamefont {Su},\ and\ \citenamefont {Yuan}(2018)}]{Gou}%
  \BibitemOpen
  \bibfield  {author} {\bibinfo {author} {\bibfnamefont {J.}~\bibnamefont {Gou}}, \bibinfo {author} {\bibfnamefont {X.}~\bibnamefont {Su}}, \ and\ \bibinfo {author} {\bibfnamefont {X.}~\bibnamefont {Yuan}},\ }\bibfield  {title} {\enquote {\bibinfo {title} {Adaptive mesh refinement method-based large eddy simulation for the flow over circular cylinder at red = 3900},}\ }\href {\doibase 10.1080/10618562.2018.1461845} {\bibfield  {journal} {\bibinfo  {journal} {Int. J. Comput. Fluid Dyn.}\ }\textbf {\bibinfo {volume} {32}},\ \bibinfo {pages} {1--18} (\bibinfo {year} {2018})}\BibitemShut {NoStop}%
\bibitem [{\citenamefont {Konstandopoulos}, \citenamefont {Skaperdas},\ and\ \citenamefont {Masoudi}(2001)}]{KonstInert}%
  \BibitemOpen
  \bibfield  {author} {\bibinfo {author} {\bibfnamefont {A.~G.}\ \bibnamefont {Konstandopoulos}}, \bibinfo {author} {\bibfnamefont {E.}~\bibnamefont {Skaperdas}}, \ and\ \bibinfo {author} {\bibfnamefont {M.}~\bibnamefont {Masoudi}},\ }\bibfield  {title} {\enquote {\bibinfo {title} {Inertial contributions to the pressure drop of diesel particulate filters},}\ }\href {\doibase 10.4271/2001-01-0909} {\bibfield  {journal} {\bibinfo  {journal} {SAE Tech. Paper 2001-01-0909}\ } (\bibinfo {year} {2001}),\ 10.4271/2001-01-0909}\BibitemShut {NoStop}%
\bibitem [{\citenamefont {Bottaro}, \citenamefont {Soueid},\ and\ \citenamefont {Galletti}(2006)}]{secondary1}%
  \BibitemOpen
  \bibfield  {author} {\bibinfo {author} {\bibfnamefont {A.}~\bibnamefont {Bottaro}}, \bibinfo {author} {\bibfnamefont {H.}~\bibnamefont {Soueid}}, \ and\ \bibinfo {author} {\bibfnamefont {B.}~\bibnamefont {Galletti}},\ }\bibfield  {title} {\enquote {\bibinfo {title} {Formation of secondary vortices in turbulent square-duct flow},}\ }\href {\doibase 10.2514/1.17327} {\bibfield  {journal} {\bibinfo  {journal} {AIAA}\ }\textbf {\bibinfo {volume} {44}},\ \bibinfo {pages} {803--811} (\bibinfo {year} {2006})}\BibitemShut {NoStop}%
\bibitem [{\citenamefont {Modesti}\ \emph {et~al.}(2018)\citenamefont {Modesti}, \citenamefont {Pirozzoli}, \citenamefont {Orlandi},\ and\ \citenamefont {Grasso}}]{secondary2}%
  \BibitemOpen
  \bibfield  {author} {\bibinfo {author} {\bibfnamefont {D.}~\bibnamefont {Modesti}}, \bibinfo {author} {\bibfnamefont {S.}~\bibnamefont {Pirozzoli}}, \bibinfo {author} {\bibfnamefont {P.}~\bibnamefont {Orlandi}}, \ and\ \bibinfo {author} {\bibfnamefont {F.}~\bibnamefont {Grasso}},\ }\bibfield  {title} {\enquote {\bibinfo {title} {On the role of secondary motions in turbulent square duct flow},}\ }\href {\doibase 10.1017/jfm.2018.391} {\bibfield  {journal} {\bibinfo  {journal} {J. Fluid Mech.}\ }\textbf {\bibinfo {volume} {847}},\ \bibinfo {pages} {R1} (\bibinfo {year} {2018})}\BibitemShut {NoStop}%
\bibitem [{\citenamefont {Madabhushi}\ and\ \citenamefont {Vanka}(1991)}]{secondary3}%
  \BibitemOpen
  \bibfield  {author} {\bibinfo {author} {\bibfnamefont {R.~K.}\ \bibnamefont {Madabhushi}}\ and\ \bibinfo {author} {\bibfnamefont {S.~P.}\ \bibnamefont {Vanka}},\ }\bibfield  {title} {\enquote {\bibinfo {title} {Large eddy simulation of turbulence‐driven secondary flow in a square duct},}\ }\href {\doibase 10.1063/1.858163} {\bibfield  {journal} {\bibinfo  {journal} {Phys. Fluids A}\ }\textbf {\bibinfo {volume} {3}},\ \bibinfo {pages} {2734--2745} (\bibinfo {year} {1991})}\BibitemShut {NoStop}%
\bibitem [{\citenamefont {Yao}, \citenamefont {Zhao},\ and\ \citenamefont {Fairweather}(2015)}]{Yao}%
  \BibitemOpen
  \bibfield  {author} {\bibinfo {author} {\bibfnamefont {J.}~\bibnamefont {Yao}}, \bibinfo {author} {\bibfnamefont {Y.}~\bibnamefont {Zhao}}, \ and\ \bibinfo {author} {\bibfnamefont {M.}~\bibnamefont {Fairweather}},\ }\bibfield  {title} {\enquote {\bibinfo {title} {Numerical simulation of turbulent flow through a straight square duct},}\ }\href {\doibase 10.1016/j.applthermaleng.2015.08.065} {\bibfield  {journal} {\bibinfo  {journal} {Appl. Therm. Eng.}\ }\textbf {\bibinfo {volume} {91}},\ \bibinfo {pages} {800--811} (\bibinfo {year} {2015})}\BibitemShut {NoStop}%
\bibitem [{\citenamefont {Darcy}\ and\ \citenamefont {Weisbach}(1857)}]{DarcyWeisFric}%
  \BibitemOpen
  \bibfield  {author} {\bibinfo {author} {\bibfnamefont {H.}~\bibnamefont {Darcy}}\ and\ \bibinfo {author} {\bibfnamefont {J.}~\bibnamefont {Weisbach}},\ }\bibfield  {title} {\enquote {\bibinfo {title} {Recherches expérimentales relatives au mouvement de l'eau dans les tuyaux},}\ }\href@noop {} {\bibfield  {journal} {\bibinfo  {journal} {Mémoires de l'Académie des Sciences de l'Institut de France}\ }\textbf {\bibinfo {volume} {9}},\ \bibinfo {pages} {193--706} (\bibinfo {year} {1857})}\BibitemShut {NoStop}%
\bibitem [{\citenamefont {Jones}(1976)}]{OCJones}%
  \BibitemOpen
  \bibfield  {author} {\bibinfo {author} {\bibfnamefont {J.}~\bibnamefont {Jones}, \bibfnamefont {O.~C.}},\ }\bibfield  {title} {\enquote {\bibinfo {title} {An improvement in the calculation of turbulent friction in rectangular ducts},}\ }\href {\doibase 10.1115/1.3448250} {\bibfield  {journal} {\bibinfo  {journal} {J. Fluids Eng.}\ }\textbf {\bibinfo {volume} {98}},\ \bibinfo {pages} {173--180} (\bibinfo {year} {1976})}\BibitemShut {NoStop}%
\bibitem [{\citenamefont {Colebrook}(1939)}]{Colebrook}%
  \BibitemOpen
  \bibfield  {author} {\bibinfo {author} {\bibfnamefont {C.~F.}\ \bibnamefont {Colebrook}},\ }\bibfield  {title} {\enquote {\bibinfo {title} {Turbulent flow in pipes, with particular reference to the transition region between the smooth and rough pipe laws},}\ }\href {\doibase 10.1680/ijoti.1939.13150} {\bibfield  {journal} {\bibinfo  {journal} {J. Inst. Civil Eng.}\ }\textbf {\bibinfo {volume} {11}},\ \bibinfo {pages} {133--156} (\bibinfo {year} {1939})}\BibitemShut {NoStop}%
\end{thebibliography}%

\end{document}